\documentclass[a4paper,12pt,times,numbered,print,index]{Classes/PhDThesisPSnPDF}

\usepackage{pdfpages}

\input{Preamble/preamble}

\title{An Efficient Data Protection Architecture Based on Fragmentation and Encryption}

\author{QIU Han}

\dept{Department of Network and Computer Science}

\university{Telecom-ParisTech}

\supervisor{\textbf{Prof. MEMMI Gerard}}

\degreetitle{Doctor of Philosophy}

\college{Telecom Paristech}

\subject{LaTeX} 

\ifdefineAbstract
\fi

\ifdefineChapter
\fi

\begin{document}

\includepdf{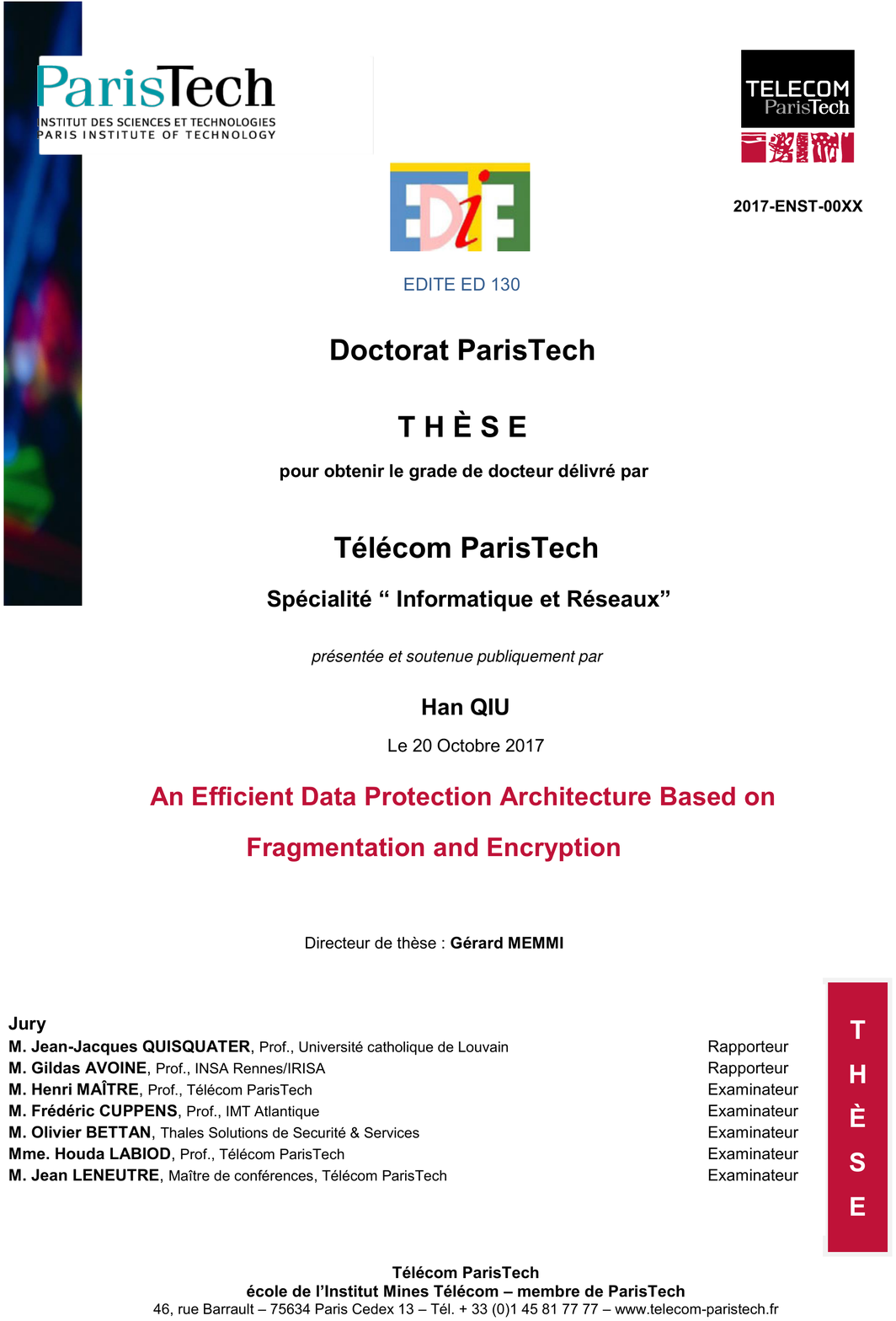} 

\frontmatter



\begin{acknowledgements}

Firstly, I would like to express my most sincere gratitude to my supervisor Prof. Gérard Memmi for the tremendous support of my PhD study and research, for his patience, motivation, immense knowledge, and dedication to students. Thank you for leading me through such a splendid journey. With your leading, the past four years has been a fantastic adventure that will be unique and important in my life.

Besides my advisor, I would like to specially thank my jury member Prof. Henri Maître who contributed to the most important discussions that helped to improve this work and encourage me to widen my research from various perspectives.

I am grateful to all the other jury members: Prof. Frédéric Cuppens, Prof. Houda Labiod, Mr. Olivier Bettan, and Mr. Jean Leneutre for attending my dissertation, and, in particular, the reviewers of the manuscript: Prof. Jean-Jacques Quisquater and Prof. Gildas Avoine, for their important suggestions and comments.

Thanks also to all the colleagues and staffs in Télécom ParisTech for all help and collaboration in technical knowledge and administrative tasks.

Last but not least, I would like to express my gratitude to my family for encouraging and supporting me so many years, and I hope this work will be the pride of you. I also want to thank my girlfriend for being with me and sharing with me for every piece joy and upset in these years. Thanks to my friends I met in France for the memorable moments we shared.

\let\thefootnote\relax\footnote{This thesis is funded by ITEA CAP2 project.}


\end{acknowledgements}

\begin{abstract}

In this thesis, a completely revisited data protection scheme based on selective encryption is presented. First, this new scheme is agnostic in term of data format, second it has a parallel architecture using GPGPU allowing performance to be at least comparable to full encryption algorithms.

Bitmap, as a special uncompressed multimedia format, is addressed as a first use case. Discrete Cosine Transform (DCT) is the first transformation for splitting fragments, getting data protection, and storing data separately on local device and cloud servers.
This work has largely improved the previous published ones for bitmap protection by providing new designs and practical experimentations. General purpose graphic processing unit (GPGPU) is exploited as an accelerator to guarantee the efficiency of the calculation compared with traditional full encryption algorithms. Then, an agnostic selective encryption based on lossless Discrete Wavelet Transform (DWT) is presented. This design, with practical experimentations on different hardware configurations, provides strong level of protection and good performance at the same time plus flexible storage dispersion schemes. Therefore, our agnostic data protection and transmission solution combining fragmentation, encryption, and dispersion is made available for a wide range of end-user applications. Also a complete set of security analysis are deployed to test the level of provided protection.

\end{abstract}



\tableofcontents

\listoffigures

\listoftables



\mainmatter


\chapter{Introduction}  

\ifpdf
    \graphicspath{{Chapter1/Figs/Raster/}{Chapter1/Figs/PDF/}{Chapter1/Figs/}}
\else
    \graphicspath{{Chapter1/Figs/Vector/}{Chapter1/Figs/}}
\fi

\section{Background} 

In the last two decades, digital data has increased in a very large scale in many fields. In 2008, International Data Corporation (IDC) estimated $2.25 \times 10^{21}$ bits of digital information had been created \cite{berman2008got}. This amount would surpass $6 \times 10^{23}$ bits by 2023. More importantly, for the personal users, the latest advances of information technology (IT) including computers, smart phones and tablets make it very easy to generate data to distribute. For example, nowadays 72 hours of videos are uploaded to YouTube in every minute on average \cite{mayer2013big}. Therefore, the data being generated, processed, transmitted and distributed is massive through the Internet.

Both large scale parallel multi-core machines and more efficient and affordable PCs  were built to serve generating, transmitting, storing and computing digital data. One of the most important advance in distributed systems is to link smaller, more affordable servers together to build a large scale computer cluster for data service. The main advantage of Cloud is to offer more scalable, fault-tolerant services with high performance at a low cost compared with one super computer. Moreover, Cloud computing technology can basically provide almost infinite computing and storage resources on demand that can fits both individual users and companies by renting hardware resources remotely on a short-term basis (most commonly, a number of processors by the hour and storage space by the day).  Therefore, cloud users enjoy the variety of cloud services (e.g. Data as a Service (Daas), Software as a Service (SaaS), Platform as a Service (PaaS), Infrastructure as a service (IaaS), etc). 


With both the development of digital data and computer technology, the trends in recent years is to outsource information storage and processing to cloud-based services. Especially the cloud-based data storage services for individual users are gaining popularity. Relying on large free storage space and reliable communication channel, cloud-based service providers like Dropbox, Google Drive are providing individual users almost infinite and low cost storage space.

However, this situation raises a question of the trustworthiness of cloud-based service providers. In fact, many security and privacy incidents are observed in today’s Cloud-based systems. Some of these incidents are listed in \cite{zhou2010security}:

• Steven Warshak stops the government’s repeated secret searches and seizures of his stored email using the federal Stored Communications Act (SCA) in July, 2007.

• A Salesforce.com employee fell victim to a phishing attack and leaked a customer list, which generated further targeted phishing attacks in October 2007.

• Google Docs found a flaw that inadvertently shares users docs in March 2009.

• Epic.com lodged a formal complaint to the FTC against Google for its privacy practices in March 2009. EPIC was successful in an action against Microsoft Passport.

• Yahoo confirmed that at least 500 million user accounts has been stolen from the company’s network in late 2014.

• Equifax announced that 143 million US-based users had their credit history information compromised in 2017.

Most of these incidents are due to human errors. Moreover, the cloud providers themselves cannot be trusted either. In 2013, the PRISM surveillance program \cite{gellman2013us} was exposed. In this program, the NSA has obtained direct access to the systems of Google, Facebook, Apple and other US Internet giants which made privacy of individual users' data vulnerable. This is due to the data that transmitted to the cloud will be handled by the Cloud itself. The situation could be even worse in some specific use cases like outsourcing encryption shown in \citet{xiang2015outsourcing} (the client need to outsource protected images to other users through an insecure channel but does not have sufficiently computational power or energy supply to perform the encryption). Thus, it becomes increasingly important for users to efficiently protect their personal data(texts, images, or videos) independently from the storage or any other application or service providers.

So, in this work, one basic assumption is that Cloud service providers cannot be entirely trusted. We have to assume that one 'curious' or 'malicious' program sits on at least one Cloud server and is able to observe all the data stored in the Cloud and transmitted through the Cloud. In a worse case, all data stored on the Cloud server can be used by this program to sniff the user's privacy by any means of analysis or attack for even a small piece of data. More importantly, the data transmission channel is also not perfectly protected and more threats like crackers could compromise the data on it. And the only trustworthy area is the local machine one end-user has. In this thesis, the design for the data stored in local machines includes encryption algorithms applied so even stealing data from local machine is not a threat. And another basic assumption is there is no such situation that a malicious program stays on end-user's computer and can observe all data in the process.


One reasonable solution is to protect the data locally on an end-user's machine before it is sent to Cloud servers. And this makes encryption naturally become promising. Traditional encryption systems like the standard cipher symmetric key encryption systems (e.g. 3DES, or its successor AES, etc.) work with the assumption that data are sequence of symbols relatively independent (i.i.d.) and of even importance and indeed, that the data must be decrypted with accuracy. This typically does not apply to most of the personal data that are photos and videos: pixels are known to be highly correlated with theirs neighbors and there is well-known strong inter-frame correlation as well. The spatial or temporal redundancies of these multimedia data are not sufficiently exploited by historical encryption methods, as when they are designed, multimedia data with special formats are still rare. For example, users may even tolerate some small level of distortion in some cases when deciphering an image with a moderate requirement on its rendition \cite{krikor2009image}. Another problem is that the traditional encryption methods are not enough to protect: for instance, an image has been encrypted rowwise by means of AES can let element of the structure of an image still understandable (see Fig. 1 of \cite{grangetto2006multimedia}). 

Some other data protection methods like Selective Encryption (SE) have been published in recent decades. The aim at exploiting special redundancies of multimedia data and are based on compression algorithms. SE usually dedicated to image or video protection where they support to automatically separate the image or video into two fragments: a ‘private’ fragment which contains most of the information such that this fragment is sufficient to understand the original image or at least process some exploitation, a second fragment that we call ‘public’ which is supposed to contain a much smaller amount of information such that this fragment is not exploitable. These two fragments are protected using different approaches depending on their respective levels of importance or confidentiality. The state of the art in Selective Encryption methods is showing that all these methods propose to encrypt the private fragment as a small subset of the original content \cite{massoudi2008overview} which in some cases constitutes a lightweight and fast encryption compared with a full encryption. This raises a first issue consisting in determining the optimal private fragment which first, deserves strong protection and secondly, is as small as possible. Then we face a second issue consisting in making sure that the weak level of protection we apply to the public fragment will prevent leaks of useful information. 

Not every SE used image compression transforms, for instance, one very simple answer would be to encrypt the center of the image, leaving the border in clear (see Figure 4 in \cite{sadourny2003proposal} for instance). This simple solution can be considered for lightweight protection, however, it does not address our two issues since the border of the image may leak valuable key information. A more interesting one is to use transformations used in image compression algorithms such as the Discrete Cosine Transform (DCT) (see \cite{krikor2009image} or our own work in \cite{qiu2015fast} for instance) to separate the information in the frequency domain.

Although the SE methods are more suitable for multimedia data in some cases, there are limitations , for instance, most SE methods are specifically related to the format of data (bitmap, jpeg) they are dealing with. Once SE method is designed based on the compression methods or coding technology used, it is dedicated for a specific multimedia content only. More importantly, there are some large volume data transmitted today that are not compressed or cannot be compressed to save storage spaces like an operating system image. It is not efficient to exploit many SE methods according to many different multimedia data formats. Thus, a challenge comes up that if it is possible to design an efficient SE method that can generally fits all kinds of data formats and guarantee not only security but also data integrity.

\section{Motivation} 

As pointed before, outsourcing information storage and processing, cloud-based services for data storage have gained in popularity and today can be considered as mainstream. They attract organizations or enterprises especially individual users who do not want or cannot cope with the cost of a private cloud. Beside the economic factor, both groups of customers subordinate their choice of an adequate cloud provider to other factors, particularly resilience, security, and privacy. 

Hardening data protection using multiple methods rather than ‘just’ encryption is becoming of paramount importance when considering continuous and powerful attacks to spy, alter, or even destroy information. Even if encryption is a great technology rapidly progressing, encryption is ‘just’ not enough to progress with this unsolvable question not mentioning its high computational complexity. In \cite{adrian2015imperfect}, the authors showed how to compromise https sites with 512-bit group; the authors even suggested that 1024-bit encryption could be crypt-analyzed with enough computational power. Cryptograph never like the idea that a cipher can be broken and information can be read given sufficient computational resources \cite{Rambaud2017}, this is nevertheless one of the central design tenets of a number of projects like the Potshards system \cite{storer2009potshards}. Moreover, there remains the difficult question of the management of the encryption key that over time, can be known by too many people, and stolen or lost.

One ultimate purpose and ambition is to look at data protection and privacy from end to end by way of combining fragmentation, encryption, and then dispersion \cite{memmi2015data,memmi2015dataCAP}. This means to derive general schemes and architecture to protect data during their entire life cycle everywhere they go throughout a network of machines where they are being processed, transmitted, and stored. Moreover, it is to offer users choices among various well understood cost effective levels of privacy and security which would come with predictable levels of performance in terms of memory occupation, energy consumption, and processing time. However, in order to provide a practical method for protecting data during their storage, we will set a series of assumptions for the hardware and software environment that is the end-users have a resource limited personal environment like laptops or desktops. Moreover, the execution time has to be comparable to the traditional full encryption algorithms. To verify this point, we will need to setup a benchmark.

Also, the concept of 'Fragmentation' is introduced with a different usage. Normally fragmentation is vastly used for resilience purposes. In \cite{rabin1989efficient}, one of the first results about fragmenting for both fault-tolerance and data protection is found. In \cite{kapusta2016poster}, the authors address this question by using a Reed Solomon error correcting code \cite{reed1960polynomial} to avoid mere duplication. In summary, fragmentation means separating with a more or less complex algorithm data into pieces or fragments for resilience purposes. In this thesis, we redefine the fragmentation as separating a piece of data by considering difference in confidentiality, data nature and space usage, in order to protect the fragments differently according to their level of confidentiality or criticality. For instance, the uncompressed image is containing a lot of redundancy that encryption only a small part of the low frequency coefficients can effectively reduce the image quality. Then these fragments should in turn be stored in different physical locations in a more or less sophisticated manner in order to increase the level of protection for the whole information.

Defining different levels of data importance is based on the thesis that massive amount of data have a non-uniform level of criticality or confidentiality (therefore, a non-uniform need for protection). In fact, non-uniform distribution of data is the basis of compression and only pure white noise is uniformly distributed. Also, as data has not been produced at the same time, they are aging at a non-uniform pace which again relate to the non-uniform level of criticality and a need for a multilevel security system. This makes the idea of combining fragmentation with encryption possible by letting some critical data be separated and strongly encrypted, while some other data less critical be only fragmented and possibly more rapidly encrypted with a weaker encryption algorithm or even in some use cases, let clear.

Last but not least, by definition, fragmentation enables the parallelization of transforming or encrypting pieces of information which lets us expect strong gain in efficiency compared with a full encryption sequentially executed, addressing scalability requirements. Defragmentation could then have to follow a reverse parallel pattern.

Then the other basic assumption is the need for a trusted area. Whatever is the software solution used for protecting data, it is our belief that a complete solution will have to use hardened hardware (a trusted area of one or several machines) at one critical moment or another during the data life cycle. In particular, places where information is being fragmented or defragmented, encrypted or decrypted are particularly critical since the information is gathered in clear during a period of time. Also, places where information is being created, printed out, or visualized by a human end-user have to be trusted and protected from any uninvited observer. A last, reason for considering a trusted area would be to use it as a safe and store ultra-confidential information even as this information is strongly encrypted. This point is widely recognized since a long time and in many publications \cite{fray1986intrusion} or \cite{aggarwal2005two} for instance) or by many industry experts. In fact, most of the trusted area are just relatively more secure than the others while there is a race between the crackers and protection technology. In order to save the endless challenges about whether a storage space is a trusted area, we define in this thesis that the local area is trustworthy compared with the cloud storage space while all data stored locally are still encrypted at application layer by default.

Use cases are important since a specific architecture can comply with a set of use cases but at the same time may very well fail at addressing needs for another group of use cases. Use cases can be defined according to the number of desired authorized participants (one, two, or many), their roles as users or end-users (owner, author (who may not be the owner), read-only user, service provider,…) (aka Alice and Bob), the number and type of attackers (from honest but curious, eavesdropper (aka Eve), to malicious (aka Mallory), insider, man in the middle, coalition of attackers, powerful rogue enterprise,…), the type and location of attacks (at storage, transmission, processing time, …), the size, nature, and format of the data (image, video, text, database, unstructured data,…), the kind of distributed machine environments (one machine to another machine, one personal machine (from a laptop to a mobile device like smart phone or a tablet) to one cloud, a general distributed environment involving several providers,..). We can see by combining these various possibilities that use cases can be very contrasted and their number can be relatively large. 

We consider the use case with relatively simplified situation: an end-user (Alice) wants to save her multimedia data in a public cloud in order to save memory in her private resource-limited environment (be a desktop, a laptop, or even a smart phone), however, for privacy reasons, she does not want putting her entire data either in plain-text or encrypted in the hands one storage provider. The solution is quite straight forward that is to protect the data on the private resource-limited environment that end-users have with all the possible calculation resources to achieve a reasonable performance.

In this thesis, we first present the related work mainly around the notion of Selective Encryption (SE) methods which are designed for specific multimedia contents in Chapter 2. The performance issue and the limitations are given to illustrate weakness of most SE methods. Then in Chapter 3, we introduce the hardware level discussion, mainly the idea of using General Purpose Graphic Unit (GPGPU) which is original for SE methods. Of course GPGPU behave as an accelerator for the methods designed in subsequent chapters but they also have an issue of portability that we will discuss. In Chapter 4, a special use case of bitmap image is considered as the data need to be protected. All design and implementation details are given with benchmark evaluations. In Chapter 5, we upgrade methods of Chapter 4 to fit with agnostic fashion of data by not only design with practical concerns but also parallel implementations partly on a CPU, partly on a GPU. We analyze in details of performance, security, and integrity issues, and describe how our SE methods can be used to safely store public fragments in public storage systems. Then we conclude in Chapter 6 with future works.

\subsection{Benchmark problem} 

Benchmark is critical and is a key rationale \cite{pommer2003selective} for developing protection methods. There are very little research that thoroughly investigate performance of existing specific SE methods in a practical way \cite{khashan2014performance}. One main reason is that some SE works are integrated within the compression or encoding algorithms which authorize authors to simply ignore possible delay caused by the first step of SE methods since they are shared. This unpractical issue is explained in Chapter 2.4.3 within a real end user environment. Moreover, most of the SE methods are not comparable with traditional full encryption algorithms like AES implemented with state of the art hardware or software, or are not considering the huge progression in performance caused by constantly evolving hardware.

In fact, it is easy to assume that encrypting a small part of the data ought to be faster than doing it in full based on the syntax 'selective' encryption. However, the gain in performance is not that obvious, as this approach may adds a pre-processing phase of data analysis and splitting that could lead to overall worse performance than full encryption. We propose to benchmark these methods from an end-user viewpoint: from the moment he is starting the operation of protection to the moment his data is protected (this is an end to end consideration) and compare the method with a full encryption (today, AES)–Of course, this comparison must use similar hardware. 

The need for regularly performing benchmarking is enforced by the fast pace progression of various hardware architecture (particularly GPU architectures) and software implementations of full encryption methods (for instance, there is a clear acceleration from AES to AES-NI \cite{bogdanov2015comb}). These changes of hardware architecture and software algorithms may very well change the ranking of the various methods and ultimately, change the end-user decision. This is why we pay attention to the implementation of these methods and test on different hardware environment. For example, even with the GPU acceleration, we have to recognize that performance for GPU implementation is still not just a simple software coding but, in fact, a particular implementation could reach best in class performance on a given platform but not on another one \cite{dai2007crypto++}.  

It is important to consider performance as a key factor to determine whether the SE method is practical. Also the possible changes caused by hardware upgrade and software optimization still need to be considered and discussed as they may change the whole design. In summary, we are showing the possibility of using the SE methods in a practical way rather than giving an ultimate solution for end-user data protection use case.



\subsection{Security analysis}  
\label{section 1.2.2}

Attack resistance is, in our opinion, another key rationale even if a number of authors accept to present SE as a compromise between security and performance and categorize SE as a lightweight security process. Most of the state of the art papers we have seen in \cite{massoudi2008overview} or later are mostly looking at visual degradation. It is fair to consider only the visual degradation for the image case as it it the most important standard. Just like in Chapter 4 we show the protection for bitmap format is analyzed with mainly the traditional visual degradation. However, for an agnostic SE methods, more requirements are needed including statistical analysis based on frequency analysis, correlation analysis, entropy analysis, differential analysis and whether subject to a possible avalanche effect (resisting to error propagation). This is done in Chapter 5 where we use different file formats to test the design.

In fact, as the design in this thesis is based on the fragmentation for the data, there will be fragments with different security levels and dispersed on different locations. For the most important fragment, the protection method is the traditional full encryption (can be easily replaced with any other protection methods) and the storage place is considered as secure so the security analysis is omitted. The fragments that are transmitted and stored on the Cloud servers are the part that needs security analysis.

In Chapter 5, we present figures for the security analysis for one case and some statistical results in tables for many times repeated tests. As long as different file formats are used, some criteria like PSNR and SSIM just suit for images are not used for other file formats like texts. And some compressed multimedia data is also used for test but just with a general statistical analysis. 

In summary, all core purposes of security analysis is to prove no matter what kind of plain texts are the input, the output cipher texts ought to be as close as possible to the ideal random data. And as the encryption key is introduced in Chapter 5 for the protection of the data stored locally, the sensitivity analysis of the key is also needed to prove the resistance for attacks like chosen plain text attack.

\chapter{Data protection methods}

\ifpdf
    \graphicspath{{Chapter2/Figs/Raster/}{Chapter2/Figs/PDF/}{Chapter2/Figs/}}
\else
    \graphicspath{{Chapter2/Figs/Vector/}{Chapter2/Figs/}}
\fi

In this chapter, firstly, basic introduction of secure storage and secure computation is given. A small test for FHE accelerated by GPGPU is also presented. Then, selective encryption, a special data protection method normally for multimedia data, is introduced and discussed. At last, our selective encryption approach is given.

\section{Secure storage and secure computation}

Three main functions are required to protect digital data during its life cycle: secure storage, secure computing and secure sharing. One of the most promising method for securing computing is Fully Homomorphic Encryption (FHE) which provides full privacy during the whole computing process for the encrypted data. And the most popular technology for data storage and sharing is Cloud computing, which offers several benefits like fast development, pay-for-use and lower costs, scalability, rapid provisioning, greater resiliency, low-cost disaster recovery, and data storage solutions. With over three decades long, outsourcing information storage and processing, cloud-based services for data storage have gained in popularity and today can be considered as mainstream. They attract organizations or enterprises as well as end users who do not want or cannot cope with the cost of a private cloud. 

The cloud offers all these advantages, however, this is not without taking cloud computing needs to move the application data or databases to large data centers, where the operation and management of the data and services are not trustworthy \cite{prism2013news}.

Hardening data protection using multiple methods rather than ‘just’ encryption is becoming of paramount importance when considering continuous and powerful attacks to spy, alter, or even destroy private and confidential information. Even if encryption is a great technology rapidly progressing, encryption is ‘just’ not enough to progress with this unsolvable question not mentioning its high computational complexity. In \cite{adrian2015imperfect}, the author shows how to compromise Diffie-Hellman key exchange (used in https sites) with 512-bit group. It is also shown that 1024-bit encryption could be cryptanalyzed with enough computational power. Cryptographs never like the idea that a cipher can be broken and information can be read given sufficient computational resources \cite{Rambaud2017}, this is nevertheless one of the central design tenets of a number of projects like the Potshards system \cite{storer2009potshards}. Moreover, there remains the difficult question of the management of the encryption key that over time, can be known by too many people, and stolen or lost.

Our purpose and ultimate ambition is to look at data protection and privacy from end to end by way of combining fragmentation, encryption, and then dispersion. This means to derive general schemes and architecture to protect data during their entire life cycle everywhere they go throughout a network of machines where they are being processed, transmitted, and stored. Moreover, it is to offer end users choices among various well understood cost effective levels of privacy and security which would come with predictable levels of performance in terms of memory occupation and processing time. For this thesis, we aim to provide secure data storage scheme for end users with reasonable assumptions that end users will have a resource limited personal environment and will look at a honest but curious third party cloud storage provider with a cost effectiveness additional constraint.

\section{Fully homomorphic encryption}

\subsection{What is FHE}

Fully Homomorphic Encryption (FHE) is a concept asked in 1978 by~\citet{rivest1978data} and answered by~\citet{gentry2009fully}. This concept can be described as “is there a way that delegates processing of your data, without giving away access to it”.We immediately understand the value proposition of such encryption algorithms even before considering outsourcing or public cloud computing since it is about performing computation with encrypted data in perfect security. The trustworthiness question in cloud computing has been discussed for years and today, there is still no perfect solution. FHE could very well be this ‘perfect solution’ only once proven efficient from a performance point of view which depends upon the use case under consideration. 

\begin{figure}[htbp!] 
\centering    
\includegraphics[width=0.8\textwidth]{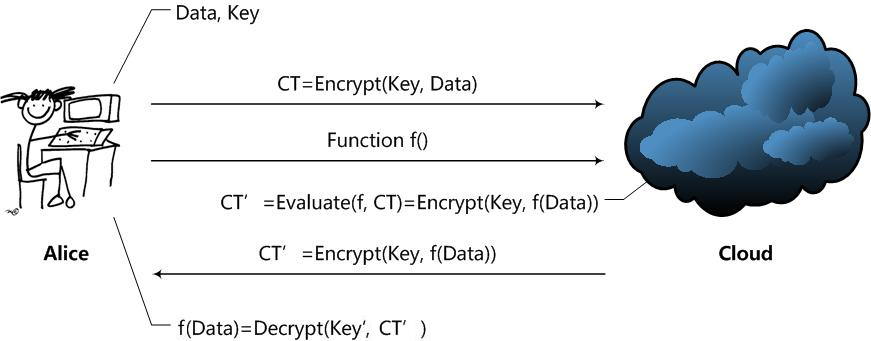}
\caption[General concept of how FHE works.]{General concept of how FHE works.}
\label{fig:fhe1}
\end{figure}

~\figurename~\ref{fig:fhe1} shows how FHE can be used with a cloud server to compute the value of a function \textit{f} for a data \textit{Data} : the user Alice sends the encrypted \textit{Data} (encrypted with Key) and the function \textit{f} (the calculation Alice wants) to the cloud and the cloud will receive only the encrypted \textit{Data} and the function \textit{f}. The property of FHE allows the cloud to perform the computation on encrypted \textit{Data} with the \textit{Evaluate} function (shown in Fig 2.1, compute on ciphertext) without knowing or accessing to \textit{Data}. Alice will be able to decrypt 'evaluated' \textit{Data} with the corresponding Key' and get the result \textit{f(Data)}. This process is basically shown as the equation below:

\begin{equation}
Evaluate(f, Encrypt(Data)) = Encrypt(Evaluate(f, Data))
\end{equation}


Other encryption algorithms are known for having somewhat homomorphic property (see Wikipedia~\cite{wikiFHE}). For instance, RSA is homomorphic with regards to multiplication: If the RSA public key is modulus \textit{m} and exponent \textit{e}, then the encryption of a message \textit{x} is given by: 


\begin{equation}
\varepsilon (x) = x^{e} mod (m)
\end{equation}

The homomorphic property for the multiplication is then:

\begin{equation}
\varepsilon (x_{1})\cdot \varepsilon (x_{2})  = (x_{1}^{e} \cdot x_{2}^{e}) mod (m) = (x_{1}x_{2})^{e}mod (m) = \varepsilon (x_{1}\cdot x_{2})
\end{equation}

which means if the evaluation function is multiply, the RSA has property of homomorphic.

\subsection{Related work of FHE}

Since 2009 when FHE based on ideal lattice was introduced by~\citet{gentry2009fully}, three main branches of FHE schemes have been developed: lattice-based, integer-based and learning-with-errors (LWE) or ring-learning-with-errors (RLWE) based encryption.

The main focus of the theoretical cryptographic research community is currently on LWE and RLWE based FHE (\citet{brakerski2014efficient}, ~\citet{gentry2012better}, ~\citet{gentry2012fully}). LWE based was introduced by~\citet{regev2009lattices}, and has been shown to be as hard as the worst case lattice problems. This problem has been extended to work over rings by~\citet{lyubashevsky2013ideal}, and this extension increases the efficiency of LWE.

Integer based schemes were introduced by~\citet{van2010fully} as a theoretically simpler alternative to lattice based schemes and have been further developed to offer similar performance to existing lattice based schemes by~\citet{coron2011fully}, ~\citet{coron2012public}.

Despite different math basis have different performance, none of them is efficient enough for applications with time constraints. For example, key generation in Gentry and Halevi’s lattice based scheme in ~\citet{gentry2011implementing} takes from 2.5 seconds to 2.2 hours. And for the evaluation step, a recent research by ~\citet{gentry2012homomorphic} shows a homomorphic evaluation of AES-128 requires 36 hours which is actually incredibly slow compared with the speed of hundreds of MB/s for AES-128 on modern PC’s CPU. Very few applications can stand such delays.

Another important limitation of FHE is with the memory usage. FHE generates very large cipher text and uses public key sizes to guarantee adequate security to prevent against possible lattice-based attacks. Gentry and Halevi’s FHE scheme~\cite{gentry2011implementing} uses public key sizes ranging from 17 MB to 2.25 GB.

Current research is aiming at improving performance of FHE either by focusing on new fundamental math to reduce computation complexity or by implementing the existing FHE algorithms on different hardware (GPU or nanotechnology). New algorithms are expected to provide with an actual breakthrough in term of performance; however, on another hand,  hardware progression is relatively limited with regards to the need for a vast deployment of FHE.

\subsection{Performance study}

In this section, we provide current research results about performance of the existing algorithms and their implementations. We are adding our own implementation for comparison. As we mentioned earlier, theoretical breakthrough of algorithm may bring a revolution in term of acceptance of FHE, this may need many years of work. In the meantime, it is interesting to search for possible optimized solutions including by using existing powerful hardware to determine whether FHE is ever usable. Although many research articles have claimed the performance of FHE are slow or far from application, it seems important to characterize how slow FHE really is. Performance of the underlying crypto-primitives such as modular reduction and large multiplication are required in many of the FHE schemes. Actually, they are critical these operations could be significantly improved through the use of GPGPU, FPGA, or ASIC technology.




\begin{table}[htbp!] 
\caption{Performance study of different based FHE and SWHE modified according to~\cite{doroz2015acceleratingoverview}}
\centering
\label{table:fhe}
\begin{tabular}{c c c c}
\hline
 Designs & Schemes & Platforms & Performance  \\
\hline
\multicolumn{4}{c}{CPU Implementations}  \\ 
\hline
AES~\cite{gentry2012homomorphic} & BGV-FHE & 2.0 GHz Intel Xeon  & 5 min/AES block  \\

AES~\cite{doroz2014homomorphic} & NTRU-FHE & 2.9 GHz Intel Xeon  & 55 sec/AES block \\

Full FHE~\cite{rohloff2014scalable} & NTRU-FHE & 2.1 GHz Intel Xeon & 275 sec/bootstrap \\

\textcolor{blue}{Full FHE (our test)} & BGV-FHE & 3.0 GHz Intel I7 & 3-5 min/bootstrap \\
\hline
\multicolumn{4}{c}{GPU Implementations}  \\ 
\hline
NTT mul/reduction~\cite{wang2012accelerating} & GH-FHE & Nvidia C 250 & 0.765 ms \\

NTT mul~\cite{wang2012accelerating} & GH-FHE & Nvidia GTX 690 & 0.583 ms \\

AES~\cite{dai2014accelerating} & NTRU-FHE & Nvidia GTX 680 & 7 sec/AES block \\

\textcolor{blue}{NTT mul (our test)} & GH-FHE & Nvidia GTX 780 &  0.81 ms \\
\hline
\multicolumn{4}{c}{FPGA Implementations}  \\ 
\hline
NTT transform~\cite{wang2013fpga} & GH-FHE & Stratix V FPGA & 0.125ms \\

NTT mod/enc~\cite{cao2013accelerating} & CMNT-FHE & Xilinx Vitrex-7 FPGA & 13 ms/enc \\

AES~\cite{doroz2015acceleratingoverview} & NTRU-FHE & Xilinx Virtex-7 FPGA  & 0.44 sec/block \\

\hline
\multicolumn{4}{c}{ASIC Implementations}  \\ 
\hline
NTT mod~\cite{doroz2013evaluating} & GH-FHE & 90 nm TSMC & 2.09 sec \\

Full FHE~\cite{doroz2015accelerating} & GH-FHE & 90 nm TSMC & 3.1 sec/recrypt \\
\hline

\end{tabular}
\end{table}

The first GPU implementation of a FHE scheme was presented by~\citet{wang2012accelerating} in 2012. The authors implemented the small parameter size version of Gentry and Halevi’s lattice-based FHE scheme in~\citet{gentry2011implementing} on an NVIDIA C2050 GPU using the FFT algorithm, achieving speed up factors of 7.68, 7.4 and 6.59 for encryption, decryption and the recryption operations, respectively. The Fast Fourier Transform (FFT) was used to target the bottleneck of this lattice-based scheme, namely the modular multiplication of very large numbers.

An overview of FHE implementations on different platforms is shown in Table 1 in~\citet{doroz2015acceleratingoverview}. Clearly, since the platforms vary greatly according to available memory, clock speed, area/price of the hardware a side-by-side comparison is not possible and therefore this information is only meant to give an idea of what is achievable on various platforms. 
 
Much of the development so far has focused on the Gentry-Halevi FHE~\citet{gentry2011implementing}, which intrinsically works with very large integers (million bit range). Therefore, a good number of works focused on developing FFT/NTT (Number Theoretic Transform) based large integer multipliers in~\citet{doroz2013evaluating}, ~\citet{doroz2015accelerating}, ~\citet{wang2012accelerating}. Currently, the only full-fledged (with bootstrapping) FHE hardware implementation is the one reported by~\citet{doroz2015accelerating}, which also implements the Gentry-Halevi FHE. At this time, there is a lack of hardware implementations of the more recently proposed FHE schemes, i.e. ~\cite{coron2011fully} and~\citet{coron2012public}, BGV-style FHE schemes~\citet{gentry2011implementing} and~\citet{yagisawa2015fully} and NTRU based FHE, e.g. ~\citet{lopez2012fly} and~\citet{stehle2011making}.

\subsection{Discussion}

Results for different FHE algorithms and for limited evaluation functions (AES-128 bit here) were presented in ~\tablename~\ref{table:fhe}. We can use this table to conclude as in the European H2020 project~\cite{Heat2015} that FHE is still far from real application. But here, we can quantify the issue. The AES block is processed in around 1-5 mins on an Intel Xeon CPU which is the type of CPU currently used in workstations. A good GPU (Nvidia GTX 690) could help reducing this processing to about 7 secs. However, considering the AES is processed at a hundreds MB/s on PC’s CPU~\cite{dai2007crypto++}, which equals almost 1 million blocks processed per second, the performance of FHE-AES is far too slow to get considered usable. Even if the hardware upgrades, even if the performance of FHE-AES is improved one thousand times faster, it is still too slow for general use.

~\tablename~\ref{table:fhe} shows that we still need to progress by 2 or 3 order of magnitude before deploying FHE. Our own code is on par with current publications for similar schemes and similar platforms. The only hope would be to use partial homomorphic encryption (PHE) or somewhat homomorphic encryption (SHE) but their usage will be very limited to niche applications.

\section{Traditional full encryption}

Cryptography is the science of writing in secret code and is an ancient art; the first documented use of cryptography in writing dates back to circa 1900 B.C. when an Egyptian scribe used non-standard hieroglyphs in an inscription. Some experts argue that cryptography appeared spontaneously sometime after writing was invented, with applications ranging from diplomatic missives to war-time battle plans. It is no surprise, then, that new forms of cryptography came soon after the widespread development of computer communications. In data and telecommunications, cryptography is necessary when communicating over any untrusted medium, which includes just about any network, particularly the Internet.

Within the context of any application-to-application communication, there are some specific security requirements, including:

\begin{itemize}
\item \textbf{Authentication:} The process of proving one's identity. (The primary forms of host-to-host authentication on the Internet today are name-based or address, both of which are notoriously weak.)

\item \textbf{Confidentiality:} Ensuring that no one can read the message except the intended receiver.

\item \textbf{Integrity:} Assuring the receiver that the received message has not been altered in any way from the original.

\item \textbf{Non-repudiation:} A mechanism to prove that the sender really sent this message.
\end{itemize}




Encryption is one of the principal means to guarantee \textbf{privacy and confidentiality} of information. Traditional encryption algorithms in the recent several decades, which is also widely used in information security in telecommunication fields, perform various substitutions and transformations on the plaintext (original message before encryption) and transforms it into ciphertext (scrambled messages after encryption). The goal of encryption is to make the plain information unreadable, invisible or unintelligible to keep it secure from any unauthorized attackers.

Encryption algorithms are traditionally split into two groups: Symmetric key encryption (also called secret key) and Asymmetric key encryption (also called public key). Symmetric key encryption is a form of cryptosystem in which encryption and decryption are performed using the same key like DES, AES, 3DES, IDEA, etc. It is also known as conventional encryption. The security of symmetric encryption algorithms relied on very large key space and normally faster than asymmetric encryption on modern communication devices.

Asymmetric encryption is a form of cryptosystem in which encryption and decryption are performed using different keys (like RSA) – one public key and one private key. It is also known as public-key encryption. This two-key crypto system makes two parties possible to securely communicate on a non-secure channel without the problem of sharing the single key like in symmetric encryption systems. The most famous asymmetric key algorithm is Rivest-Shamir Adelman (RSA by~\citet{rivest1978method}). The asymmetric encryption algorithms are much slower than the symmetric ones because they use much more complex math calculations rather than just bit-level operations.




\section{Selective encryption}

\subsection{Basic concept of selective encryption}

\begin{figure}[htbp!] 
\centering    
\includegraphics[width=0.8\textwidth]{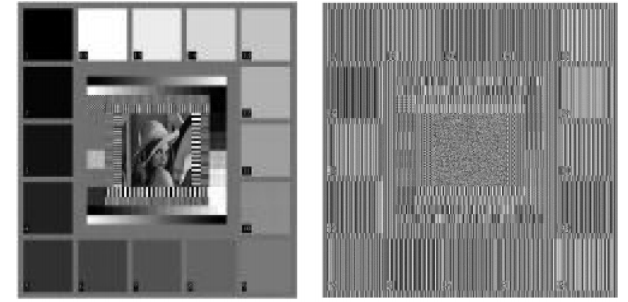}
\caption{One example in~\cite{grangetto2006multimedia}: Original image (left) compared with AES-encrypted image (right).}
\label{aesnotenough}
\end{figure}

Selective encryption (SE) used for protecting data especially multimedia data has been introduced more recently. The basic idea is to go as fast as possible to reduce the overhead involved by securing data. Although traditional data encryption techniques such as Advanced Encryption Standard (AES)~\cite{rijmen2001advanced} have become very popular, they have some clear limitations for multimedia applications. The main problem is that the majority of existing encryption standards such as DES and AES have been developed for i.i.d. (independent and identically distributed) data sources~\cite{clauset2011brief}; however, multimedia data are typically non i.i.d. which will lead to poor speed of encryption pointed out in~\figurename~\ref{aesnotenough} by~\citet{grangetto2006multimedia}. This is because the statistics for image and video data are strongly correlated and have strong spatial/temporal redundancy that makes them differ a lot from classical text data. And as pointed by Lookabaugh in~\cite{lookabaugh2004selective, lookabaugh2003security}, the relationship between plaintext statistics and ciphertext security is already highlighted by Shannon in~\cite{shannon1949communication}: a secure encryption scheme should remove all the redundancies in the plaintext; otherwise, the more redundant the souce code is, the less secure the ciphtertext is~\cite{massoudi2008overview}. Based on this viewpoint, the naïve full encryption algorithms are not suitable for protecting the multimedia contents and SE methods are designed to fit the need.

\begin{figure}[htbp!] 
\centering    
\includegraphics[width=0.8\textwidth]{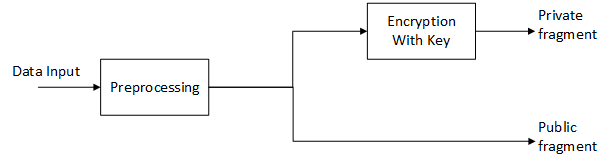}
\caption{Basic concept of selective encryption.}
\label{basicSE}
\end{figure}

SE consists in applying encryption to a subset of the original content with or without a preprocessing step like shown in~\figurename~\ref{basicSE}. The main goal of selective encryption is to reduce the amount of data to encrypt while achieving a required level of security. The general approach is to separate the content into two fragments. The first fragment is the \textit{public fragment}, it is left unencrypted in most SE cases and made accessible to all users. The second fragment is the \textit{private fragment} which is encrypted. Only authorized users have access to the protected private fragments. One important feature in selective encryption is to make the private fragment as small as possible.

The main question for SE is how to select the private fragment to encrypt while keeping the rest without an information leak. There is no general answer to this question because as shown in related works, the SE methods are most used for soft encryption purposes that make them have different protection standards. For example, in some applications (video on demand, database search, etc.), it could be important to encourage customers to pay for the entire content. To this purpose, only a soft visual degradation is achieved, so that everyone could still understand the content but have to pay to access the full-quality original content. In some other use cases like sensitive data (e.g., military images/videos, etc.), hard visual degradation could be desirable to completely disguise the visual content. And sometimes only a part of the image is recognized and protected~\cite{viola2001rapid}. Moreover, according to~\citet{massoudi2008overview}, many kinds of different methods are adapted to protect different multimedia formats (JPEG, MPEG, etc.) or different multimedia contents respectively, however, state of the art SE methods are designed to protect a given type and nature of data (e.g. bitmap image, jpeg image, mpeg video, etc.). Consequently, they can protect only the kind of data format which they were designed for.

In summary, different use cases and different formats of multimedia contents determine and restraint different purposes of SE designs. The most important trade-off is to make the private fragment as small as possible in order to reduce processing time while securing the whole data content according to a specific requirements.

\subsection{Related work of SE}



\begin{figure}[htbp!] 
\centering    
\includegraphics[width=1.0\textwidth]{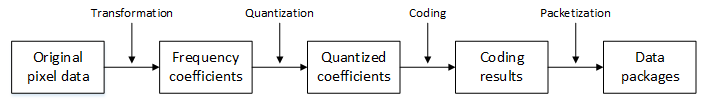}
\caption{The process of how most multimedia data is generated.}
\label{processMultimedia}
\end{figure}

SE methods have been described and discussed in many previous works (see an overview by~\citet{massoudi2008overview}). According to~\citet{massoudi2008overview}, SE methods can be classified by when the encryption is performed with respect to compression (there are very few multimedia formats that are uncompressed such as bitmap are not within this scope.). So three classes of SE methods are listed: (1) Precompression, (2) Incompression and (3) Postcompression. This classification is based on how most multimedia content is generated from initial pixel information to packets transmitted on Internet (see~\figurename~\ref{processMultimedia}). 

According to~\citet{massoudi2008overview}, in the process shown in~\figurename~\ref{processMultimedia}, the coding process is always seen as the compression step as the  widely used coding techniques especially entropy coding schemes~\cite{lei1991entropy} can efficiently reduce the multimedia data size before transmission. So if the selective encryption is performed at the frequency coefficients step, the SE methods are classified as precompression; if it is performed during the coding process, the SE methods are classified as Incompression and the other methods that perform SE after coding step are postcompression.  

\subsubsection{Precompression SE methods}

This category of SE methods are mainly protecting data at its frequency space. In~\figurename~\ref{processMultimedia}, the transformation methods like Discrete Cosine Transform (DCT) ~\cite{ahmed1974discrete} and Discrete Wavelet Transform (DWT) ~\cite{burrus1997introduction} are commonly used to generate frequency coefficients in the first step. As from a viewpoint of energy distribution in frequency domain, low frequency areas take less storage space while carrying most of the energy. Studies on the Human Visual System (HVS) have confirmed that humans are more sensitive to lower frequencies than to higher ones~\cite{puech2005crypto}. So the most important visual characteristics are to be found in the low frequencies, while details exist in the higher frequencies. And these considerations have had fundamental impacts on image or video compression techniques and also given the hint about the design of SE methods. In fact, most SE methods exploit this energy concentration in their designs.


The very initial SE method based on DCT is proposed by~\citet{tang1997methods} in 1996 to protect some of the DCT coefficients in the I-frame of a MPEG video~\cite{le1991mpeg}. The author used DES in CBC mode~\cite{coppersmith1994data} to protect the DC coefficients and randomly permutated the AC coefficients instead of the zigzag scans.

However permutation of the AC coefficients is not enough. As shown by~\citet{qiao1997mpeg} and~\citet{uehara2000chosen}, with setting DC coefficient to a fixed value, a chosen or known plaintext attack \cite{anderson2008security} can get a semantically good reconstruction. As long as the DC coefficient in the DCT represents the average intensity of the corresponding DCT block which is critical from an energy viewpoint, the rest AC coefficients still carries some information that can help to reconstruct and get an acceptable visual result. 

This situation is seen again in~\citet{puech2005crypto}, although protecting only the DC value can highly degrade the visual quality of image or even make an image totally unreadable, the DC coefficients can be recovered from the remaining coefficients which makes the reconstruction of the image possible as pointed by~\citet{uehara2006recovering}.

More recent works in~\citet{krikor2009image} and~\citet{yuen2011chaos} protect not only the DC values but also some AC values as well. These methods seems more promising as the coefficients protected (DC coefficient and first 5 AC coefficients in~\citet{krikor2009image}) carry more than 96\% of the whole energy in an image use case. However, this is still not enough as a protection method. Because in some cases there are sharp edges or many detail information contained in an image that makes the rest high frequency coefficients could show some hints about what the image is without any recovery of the protected coefficients. As shown in~\citet{qiu2015fast}, the reconstructed image by padding random number for the protected DC and first 5 AC coefficients is still able to be understood.

Indeed, protecting the low frequency coefficients of DCT can efficiently degrade the visual quality which fits some use cases. However, degrading the visual quality does not mean providing a good protection of the content. After all, as images are very different, it is difficult to generally determine how many low frequency coefficients should be protected to achieve a good level of protection.

Wavelet based SE methods are also shown in related works like in~\citet{chen2005optical},~\citet{taneja2011selective} and ~\citet{martin2005efficient}. The techniques include frequency selective encryption, block shuffling, encryption of wavelet packet tree structures, etc. Although there are no publications pointing that these techniques can be attacked, however, as pointed by~\citet{massoudi2008overview}, these SE methods mainly aim to degrade the visual quality and it necessarily is still difficult to evaluate its security level. That is to say, harder visual distortion does not imply more security.

\subsubsection{Incompression SE methods}

In 2003, ~\citet{pommer2003selective} proposed a SE method that encrypts only the head information of the wavelet packets which specifies the subband tree structure. This method can be attacked by chosen plaintext attack as the statistical properties of the wavelet coefficients remain unprotected which gives the possibility to reconstruct the approximation subband. Protecting only head information is far from enough to secure the content, however, their use case could justify this approach.

In 2001,~\cite{wu2001fast} and~\cite{wu2001efficient} gives a new viewpoint that SE methods can be done during the entropy coding stage. One method they proposed uses the multiple Huffman tables (MHTs) to protect audio and visual contents by generating millions of different Huffman tables using Huffman tree mutation~\cite{wu2001fast, wu2001efficient}. Indeed, decoding a Huffman coded stream without any knowledge about the Huffman coding tables is very difficult as shown in~\citet{gillman1996breaking}. However, the basic MHT could still suffer from known and plaintext attacks as shown in~\citet{zhou2007security}.

The other method proposed by~\citet{wu2001fast} and~\citet{wu2001efficient} is to protect during the process of QM arithmetic encoding~\cite{li1999embedded} (an enhancement of the Q coder~\cite{pennebaker1988overview}). As long as the QM coder is based on an initial state index as an entry, 4 published initial state indices is picked and used in a secret order according to the author. There are no known attack to this method but it can only be used for the multimedia format with a QM coding stage inside (e.g. JPEG standard~\cite{pennebaker1992jpeg}).

The similar technique shows up to protect JPEG2000~\cite{taubman2012jpeg2000} images when MQ coder (an enhancement of QM coder, see~\cite{taubman2012jpeg2000}) is used in JPEG2000 standard. In 2006, ~\citet{grangetto2006multimedia} used a randomized MQ coder that randomly the two alternative coding intervals that can achieve very good visual degradation. In 2014, ~\citet{xiang2014secure} gives another protection method for JPEG2000 images by replacing the initial lookup table during the MQ coding process. These methods can be efficiently used by embedding into the JPEG2000 coder and decoder but are also highly format reliance.

\subsubsection{Postcompression SE methods}





In 2000,~\citet{cheng2000partial} proposed a SE method at the output of quadtree compressor~\cite{markas1992quad}. The author takes the quadtree structure values as the private fragment to encrypt and leave the rest leaf values unencrypted. However, as pointed by~\citet{massoudi2008overview}, the brute force attack is practical for low information images and for high information images, the encrypted fragment can reach about 50\% of the original image size.

In 2008, ~\citet{massoudi2008secure} designed a SE method dedicated for JPEG2000 images on packets level that can degrade the visual quality of images by protecting only a small part of the original data. However, this method can be applied only on JPEG2000 format and the performance could be weak when high level protection is required.

These kind of methods are also seen in~\citet{wu2004compliant},~\citet{stutz2006format} and~\citet{engel2007format}. These methods did protect the code blcok contribution to packets (CCPs) which can achieve high level of visual degradation but may be weak against side channel attack.

\subsection{Our SE approach}

\begin{figure}[htbp!] 
\centering    
\includegraphics[width=1\textwidth]{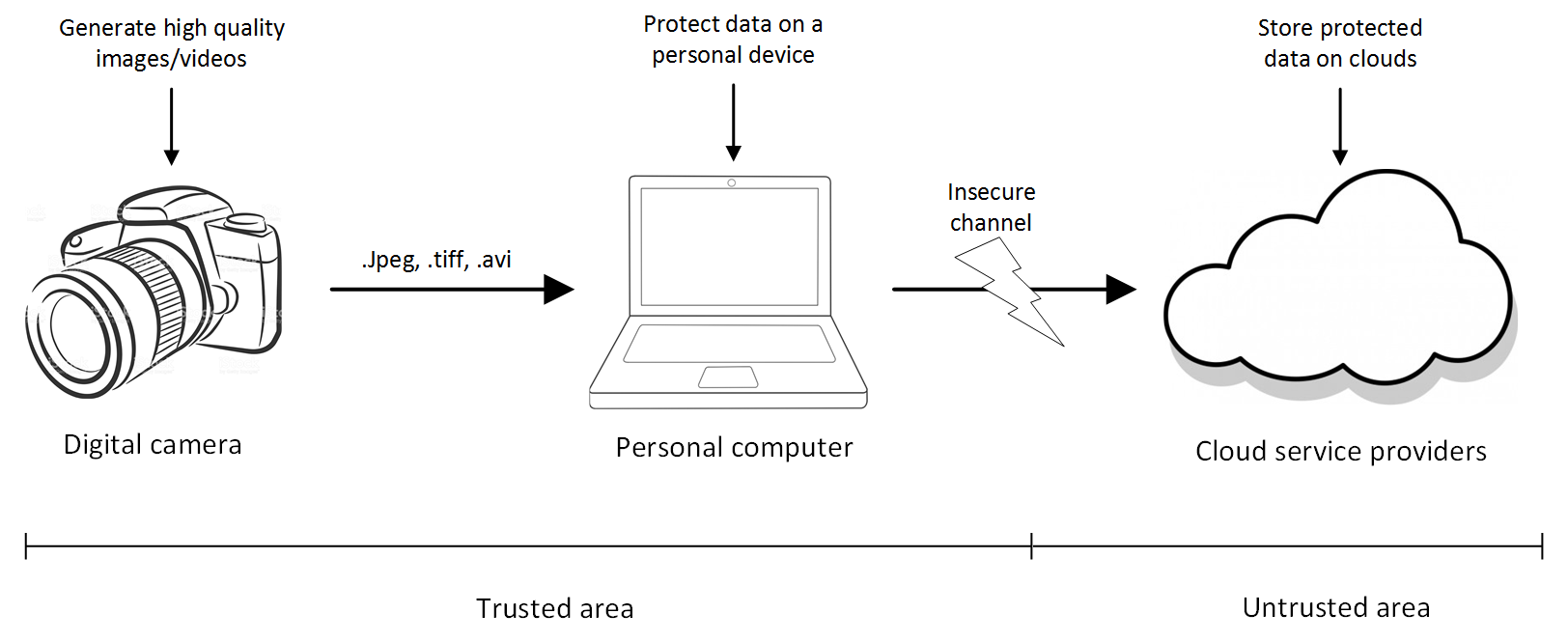}
\caption{A use case that multimedia data should be protected before sending to could storage.}
\label{cameratocloud}
\end{figure}

~\citet{massoudi2008overview} classified SE methods related works in three categories by when the encryption is done with regards to the compression process. However, SE designs are not practical for an end user. As shown in~\figurename~\ref{cameratocloud}, if we consider SE designs from an end user point of view, the digital devices that a normal end user have are normally a digital device that will generate multimedia content (also including multimedia contents downloaded from Internet) and a device that owns limited calculation capacity (a low-end laptop or a high-end desktop, etc.). In this case, if the end user wants to store the multimedia data (photos or videos) to a cloud server or to share these data through a cloud server, the data has to be protected before going to the insecure channel. However, as long as today's digital cameras are not equipped with hardware or software available for any security calculation, the very first data that an end user gets is the formatted package-level data like JPEG or MP4 files directly generated from the camera (see~\figurename~\ref{cameratocloud}).

Such a situation is not favorable to SE methods belonging to precompression or incompression are efficient because the only way to use these methods would consist in decoding the package-level data like JPEG on the laptop until transformation step (reverse the process of~\figurename~\ref{processMultimedia}) and applying the SE method to reformat everything again, which is of course very time costly and complex. Moreover, even if this kind of scheme is used, the data is still vulnerable as indicated in the previous section: many SE methods have not been published with enough security analysis and are proven either exposed to attack or can be reconstructed somehow leading to information loss.

In our work, we reconsider SE in an end user scenario. On the one hand, nowadays, data security is more important than a decade ago because we have security threats not only from insecure channels as usual but also from possible information leak from cloud providers (see~\citet{prism2013news}); on the other hand, however, the multimedia data have many kinds of formats with very different designs which makes using format reliance SE methods difficult. The use case we consider is based on an end user viewpoint that data of an end user should be protected from not only the insecure channel but also the cloud service providers (the whole untrusted area in~\figurename~\ref{cameratocloud}). Moreover, the SE design should be efficient enough compared with the full encryption methods.

\begin{figure}[htbp!] 
\centering
\includegraphics[width=1\textwidth]{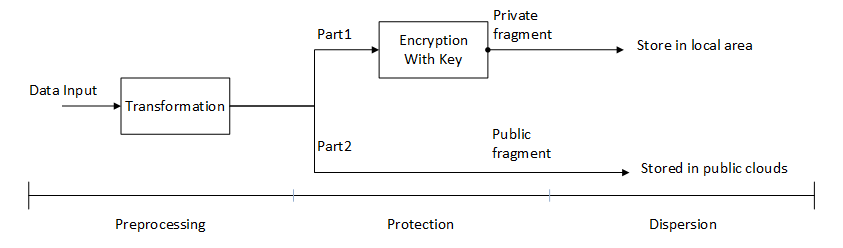}
\caption{SE conceptual design combines fragmentation with an example use case of efficient dispersion .}
\label{ourSE}
\end{figure}

The general concept of our view is shown in~\figurename~\ref{ourSE}. Three main steps are defined: Preprocessing, Protection and Dispersion. 

In this scenario, data first goes to a preprocessing step that will perform the transformation to help separating data into two fragments (sometimes more than two fragments) with different levels of importance. This is the concept of fragmentation introduced by our work in our SE design. In fact, fragmentation is not a new idea but a general concept used in computer science in many different applications and usages (by operating system to optimize disk space management, by database management or distributed systems to gain in performance particularly in latency, by routing algorithms in communication to increase reliability and support disaster recovery when combining replication and fragmentation together). Here the usage of fragmentation is done by the transformation like DCT or DWT in our design.

The second step is the protection for different fragments. Indeed, data fragments of different security levels should be protected with different encryption methods for efficient purpose. The encryption method used in our design for the most important data fragment is AES-128. Since 2001, Advanced Encryption Standard (AES) \cite{daemen1999aes}, is selected as a standard specification for the encryption of electronic data by the U.S. National Institute of Standards and Technology (NIST), it has become the most widely used symmetric encryption algorithm in the world. Although many proposals of side-channel attacks for AES were published in recent years (\citet{piret2003differential},~\citet{ors2004power},~\citet{schramm2004collision},~\citet{bertoni2005aes}), AES is still considered secure as long as no key abuse. Moreover, encryption algorithm for the private fragment can be easily replaced by another one if need be. In~\figurename~\ref{ourSE}, we fragment data into two parts: \textit{private fragment} and \textit{public fragment}. The \textit{public fragment} can be as large as needed but carry as little information as possible. And the protection method for the \textit{public fragment} should be light weighted or no protection at all with the target that no recovery should be possible only from the public fragment.

Then the dispersion step should be performed to store different fragments into different storage areas making for an additional hurdle for attackers. In~\figurename~\ref{ourSE}, we design to store the private fragment in a trusted area and the public fragment in a public and untrusted area. This design fits the real scenario shown in~\figurename~\ref{cameratocloud} which can let most of the data stored in public clouds without information leak and save storage space on the user's local device. For the transmission purpose, we have discussed in Chapter 5 that the private fragments can be encrypted and transmitted through different channel which allows our design fits the needs of both secure data storage and secure data transmission.

\subsection{Performance issue of SE}

Speed is a critical criterion and a key rationale for developing SE methods: encrypting a small part of the data ought to be faster than doing it in full. However, the gain in performance is not that obvious, as in some use cases, a complete SE approach adds a preprocessing step that could lead to overall worse performance than full encryption. After all, the proprecessing step also costs time and very few papers discuss and show performance of SE algorithms implementation (\citet{khashan2014performance}).

Therefore, we should benchmark any new proposed method against existing ones–in particular, the standard full encryption methods (today, AES)– using end to end comparison and similar hardware. The need for regular benchmarking is reinforced by the fast paced progression of hardware architecture and software implementations of full encryption methods, as a particular implementation could reach best in class performance on a well-adapted platform~\cite{dai2007crypto++}. Overall, when accounting for every step of the process, it is not so clear that full encryption is slower than SE, especially for some methods with the intensive steps.

One point worth consideration is that some random position permutation methods (\citet{li2008general},~\citet{li2011optimal}, ~\citet{zhang2013vulnerability}) and chaotic based cryptosystems (\citet{liu2010color},~\citet{bhatnagar2012selective}, ~\citet{zhang2013edge}) are used to encrypt entire or partial image data. These approaches do not have the performance issues we mentioned before. However, the security level of the random position permutation schemes is weak against the known plaintext attack. ~\citet{zhao2004decryption} proposed to recover the corresponding original image. Moreover, the main constraint of chaos based encryption schemes is that the finite accuracy of numerical calculations on modern computers can lead to an arbitrary change of major chaos properties such as the external parameters or initial conditions. In summary, although these methods have generally high performance, their security levels are not good enough as pointed by~\citet{amigo2007theory} and~\citet{kulkarni2009multimedia}.


Here we give a simple example to compare one DCT algorithm implementation (implemented based on~\cite{obukhov2008discrete}) and AES-128 bit (this simple example uses only two very common AES modes: CBC and CFB modes implemented based on~\cite{dai2007crypto++}) on two different PCs with Intel CPUs. The result in~\tablename~\ref{aesdctspeed} shows that DCT $8\times8$ is around $45\%$ slower than AES-128 bit. Moreover, AES has a counter mode (AES-CTR \cite{tran2011parallel}) which can be implemented in parallel on modern CPUs with multiple cores. The speed is normally three or four times faster (according to number of cores) than CBC mode on CPUs.

In summary, this brief comparison indicates the SE method using DCT $8\times8$ like~\citet{krikor2009image} has serious performance problems given that the preprocessing step (DCT $8 \times 8$) alone is much slower than standard encryption algorithms such as AES.

\begin{table}[htbp!] 
\caption{Benchmark of AES 128-bit and DCT $8\times8$ on current CPUs.}
\centering
\label{aesdctspeed}
\begin{tabular}{c c c c}
\toprule

Computer CPU & AES/CBC 128-bit & AES/CFB 128-bit & DCT $8\times8$  \\
\hline
Intel I7-3630QM & 374 MiB/s & 362 MiB/s & 203 MiB/s  \\
\hline
Intel I7-4770K & 494 MiB/s & 480 MiB/s & 267 MiB/s  \\

\bottomrule
\end{tabular}
\end{table}


In this work, we define a SE algorithm implementation as \textit{usable} if this algorithm meets both a suitable level of security with regard to the needs for the special use case and a level of performance comparable or better than a full encryption algorithm (in this work, we use AES 128-bit as the standard encryption algorithm to compare). Based on this definition, many of the SE algorithm implementation using DCT $8\times8$ in the literature are actually ‘unusable’ as DCT $8\times8$ implementation is not faster than AES running on the same CPU. 

This issue could be solved by introducing additional calculation resource available, such as the common GPUs on today's PCs.


\chapter{Hardware acceleration}

\ifpdf
    \graphicspath{{Chapter3/Figs/Raster/}{Chapter3/Figs/PDF/}{Chapter3/Figs/}}
\else
    \graphicspath{{Chapter3/Figs/Vector/}{Chapter3/Figs/}}
\fi

In this chapter, the development of parallel computing especially General Purpose Graphic Processing Unit is presented. Both the hardware and software development are given to illustrate the huge improvement of GPGPU in the last decade. Then, the GPGPU of Nvidia is chosen as the platform used for this thesis and the detail information is given.

\section{Background of parallel computing}

Moore’s law~\cite{schaller1997moore}, in the form of doubling the number of transistors in a dense Integrated Circuit (IC) every two years was proven to be met from the 60s to late 90s. In the meantime, clock speed, which determines the main frequency of the chip and is a key criteria to measure the commodity computer CPU's performance, also doubled about every 18 months until 2000~\cite{brodtkorb2013gpu}. In this period, \citet{bixby2002solving} pointed that from 1987 to 2000, performance of commercial Linear Programming solvers were increased one million times faster: 1000 times coming from better methods and the other 1000 times benefited from general improvement in performance in computers technology.

From the 1970s, when the first generation of CPU was created, to the year of 2004, most of computer CPUs used a serial model of execution for calculation tasks. The main improvements were more transistors, higher clock speed, and better memory technology. 

Among these factors, the clock speed, linked to the IC technology, determines the minimal time one CPU round needs, always increased at every new generation of computer CPUs until 2004. As pointed by~\citet{brodtkorb2013gpu}, the main frequency of computer CPUs seem to reach some physical limit in early 2000s. It is also reported by~\citet{owens2007gpu} that CPU main frequency increased from 0.5 GHz in 1991 (HP PA-RISC) to 3.6 GHz in 2005 (Intel Xeon). But nowadays, clock frequencies seem stabilized: on Intel CPUs we see even less than 3.6 GHz (commonly between 2.0 GHz and 3.5 GHz without boost). At the same time, however, we see parallelism keeping growing in CPU architecture from two cores inside one CPU to dozens of cores integrated within one CPU. From then on, parallel implementation for calculation tasks became so important that solutions for complex algorithms need to be optimized to fully exploit the multi-core architecture of modern CPUs.

Parallel computing is not a new idea. Since the 1960s, as the first computers with multiple processors were built up and deployed, parallel computation became a wide spread programming technique. Indeed, according to~\cite{brodtkorb2013gpu}, there are different types of parallel computing in different levels and formats: e.g. parallelism at IC instruction level is common today~\cite{wall1991limits}; parallelism for tasks and for data are the main optimization used by modern IC designs. The task parallelism consists on processing a large  number of input elements in to a pipeline that feeds output of each successive task into the input of the next task. This is commonly seen on a computer CPU's working way that divides this pipeline by time and calculate each pipeline stage in turn. However, data parallelism has a different approach that divides the calculation of the pipeline by space instead of time. This model makes it possible that different parts of hardware can be customized with dedicated-purpose for different task calculation to achieve a generally greater computation efficiency over a general-purpose solution.

The different parallel designs for computing are according to the application needs. In this recent decade, huge number of applications for digital contents especially multimedia contents show up with a different feature for the needs of computing. An important feature of these applications need is that the data can be processed independently and in any order on different processing elements for similar operations which is called throughput computing~\cite{livny1997mechanisms}. The throughput computing applications are also seen as the most important classes of future applications~\cite{asanovic2006landscape,KKwsn2017}. 

In such a situation, the traditional philosophies of designing CPUs which is to provide calculation capacity for different applications and fast response time for a single task were not suit for these application needs now. Moreover, due to the cost of technology complexity and power consumptions, the main stream CPUs in recent years are integrating only a small number of processing general-purpose cores on one die like Intel-I7 series CPUs~\cite{casazza2009intel}.

At the same time when we see the parallelism keeps growing in CPU architecture, Graphic Processing Unit (GPU), built on different initial philosophies, as an alternative parallelism model, showed up to fit the needs of these application calculation. In the beginning, designed as subordinate processors, GPUs are built specially for rendering and other graphics applications for multimedia data. This category of applications determined Single-Instruction-Multiple-Data (SIMD) as the basic execution model of GPU. This is borrowed from vector computers~\cite{bailey1987vector} built in 1970s.

In this recent decade, driven by the needs of multimedia applications especially gaming industry and needs for accelerating some general-purpose applications that fits more data parallelism, GPU was well developed with both hardware upgrades and software adaption that gives rise to a wider General-Purpose-Graphic-Processing-Unit (GPGPU) field~\cite{owens2008gpu}. And until today, not only three of the world's five fastest supercomputers use GPU acceleration~\cite{top500}, but also almost every personal computer is equipped with a high performance GPGPU to accelerate special applications.

\section{Development of modern GPGPU}

The initial role that the GPU play was just a normal component in common PCs. Nowadays, high performance GPUs are common on not only on professional workstations, servers, or super-computers but also personal computers with different capability. Initially, GPU cards are dedicated to video memories and special calculation units. Today, the need for speed of dedicated memories and calculation units are still the main requirement for the GPU performance. 

In this section, the development in hardware and software of GPUs for personal computers is presented to elaborate on how GPUs become so efficient for calculation tasks. The Nvidia GeForce series GPUs (for PC users) will be used as examples as we will be comparing their evolution. However, the development of dedicated GPGPUs for workstations or super-computers will also be briefly mentioned but they are not utilized for the use cases we discuss and evaluate.

\subsection{Hardware development}

\begin{figure}[htbp!] 
\centering    
\includegraphics[width=0.6\textwidth]{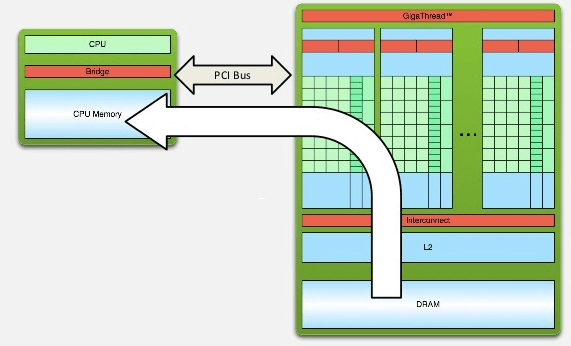}
\caption{Memory architecture of CPU and GPU on modern PCs \cite{nvidia2007compute}.}
\label{cpugpum}
\end{figure}

In this section, the hardware evolvement of modern PCs is introduced. As shown in \figurename~\ref{cpugpum}, the host memory (CPU memory) is controlled by CPU and communicates with GPU through PCI Bus. And there is a specific memory (DRAM) for GPU.

In the last two decades, the hardware and industrial process for making GPUs have improved so much that now owning a high performance GPU on a personal computer is common. However, unlike the development of CPUs in the past 40 years, the most performance gain of GPU does not come from the increase of main frequency but from the increasing number of calculation cores and architecture of their dedicated memory.

\subsubsection{Memory}

GPU memory, also called as video RAM, is an independent memory card integrated on GPU board communicating with the host memory on motherboard through bus. The GPU memory we mentioned in this section is only about the memory on GPU board (called 'global memory' in Nvidia CUDA) not in GPU chip (caches, called 'shared memory' or 'texture', etc in Nvidia CUDA). 

The speed of GPU memory is measured by the memory bandwidth, which is basically the speed of the read and write operations of the dedicated video memory by the calculation cores. Normally, it's measured in gigabytes per second (GB/s). The reason why there is an independent memory for GPUs is that the GPU cores are calculating much faster than the bus transfer speed and in recent decades, even the speed of the host memory cannot meet the calculation needs. If the memory is not fast enough, GPU cores will wait for data transfer after each operation and the memory can become a series bottleneck. As a result, since a decade ago, dedicated memory became widely used in GPU with size ranging from 512 MB to today's 2-12 GB.

The memory bandwidth today is mainly determined by two factors: memory clock and memory width. The memory clock means the clock rate of the memory chips and memory width is the width of the interface bus. They are all determined by the standard processing at each generation~\cite{jedec2012jedec}: DDR (Double Data Rate), DDR2, DDR3/GDDR3, DDR4/GDDR4, DDR5/GDDR5 and GDDR5X. If we consider Nvidia GPUs as examples, in 2006, when DDR2 memory is still used for host memory by PCs, Nvidia GeForce 8800 Ultra has the DDR3 memory clock rates at more than 1000 MHz. Today, as DDR3 memory is used commonly by CPU memory, Nvidia GPUs are equipped with GDDR5 or GDDR5X that provides more than 5000 MHz clock rates~\cite{nvidiawhite1080}. Moreover, the DDR5 generation memory supports 256-bits or 384-bits for bus width which produces the theoretical maximum bandwidth by multiplying memory width and memory clock. 




\begin{figure}[htbp!] 
\centering    
\includegraphics[width=0.85\textwidth]{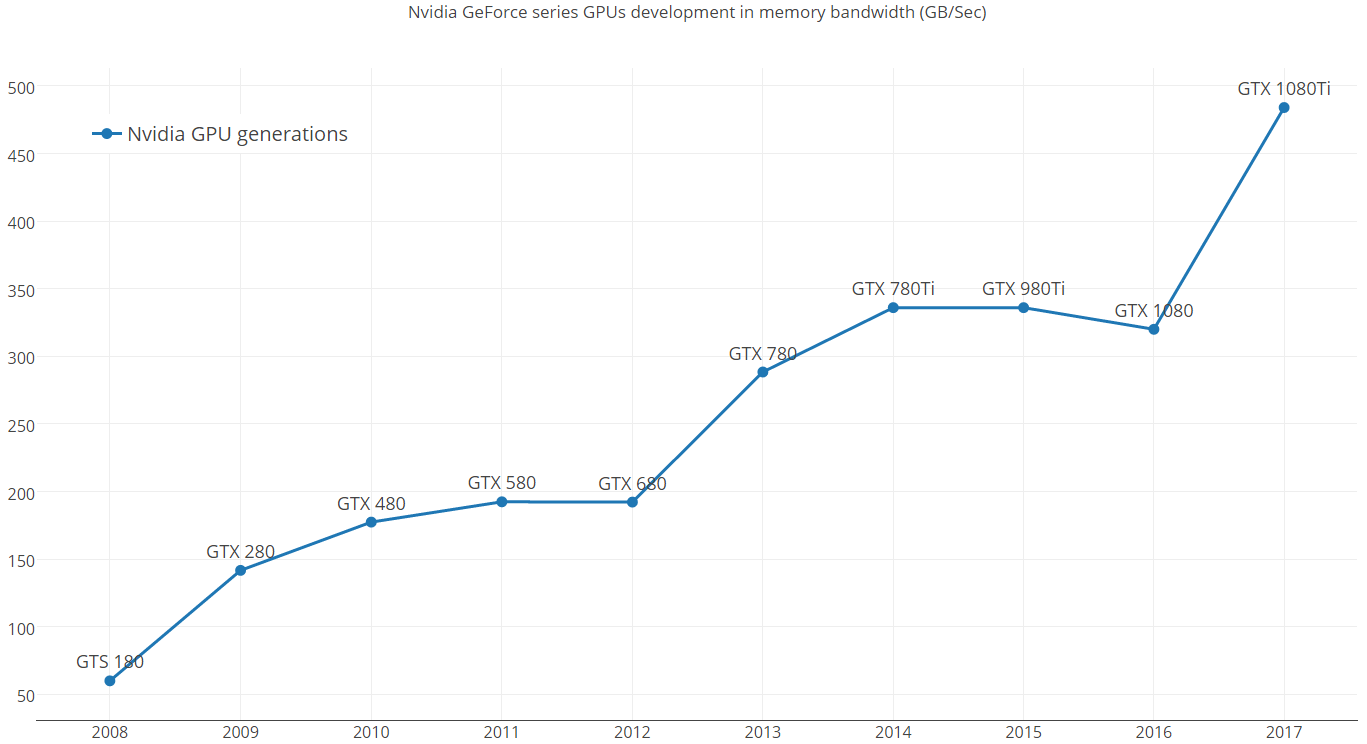}
\caption{Increase of memory bandwidth (CUDA cores on Nvidia GeForce series GPUs since 2008).}
\label{nvmems}
\end{figure}

The memory bandwidth of the best Nvidia GeForce series GPUs (high-end GPUs for PCs) are shown in \figurename~\ref{nvmems} in each year from 2008 to 2017 (data collected from Nvidia website). The memory bandwidth increases about 10 times faster in the past 9 years from about 50 GB/Sec to 484 GB/Sec. One point should be noticed is the GTX 780Ti is the enhanced version of GTX 780 but with a difference that Ti means a very enhanced version. The GTX 780 is one of the GPU card used in this thesis.


\subsubsection{Calculation cores}

Today, all GPU manufacturers including both AMD and NVIDIA are building architectures with unified, massively parallel programmable units at their cores. However, as pointed by~\citet{owens2008gpu}, a decade ago, the GPU was just a fixed-function processor, building around the graphics pipeline, it could excel at three-dimensional (3-D) graphics but little else.


In fact, the initial design of GPUs was to treat computer graphics primitives such as vertices and pixels inputting as a stream model. For one piece of data input, there is a vertex processor calculating points (seen as multiple component vectors) and another processor calculating pixel color and so on. Inside one GPU chip, many processing units with this simple architecture are integrated and connected via data flows to perform this simple operation and use the spatial parallelism of graphic applications (e.g. for one frame, which pixel is calculated first is not important). In 2003, the GPU ATI R300 had eight-pixel pipelines handling single-instruction, multiple-computing processing~\cite{macedonia2003gpu}.

\begin{figure}[htbp!] 
\centering    
\includegraphics[width=0.85\textwidth]{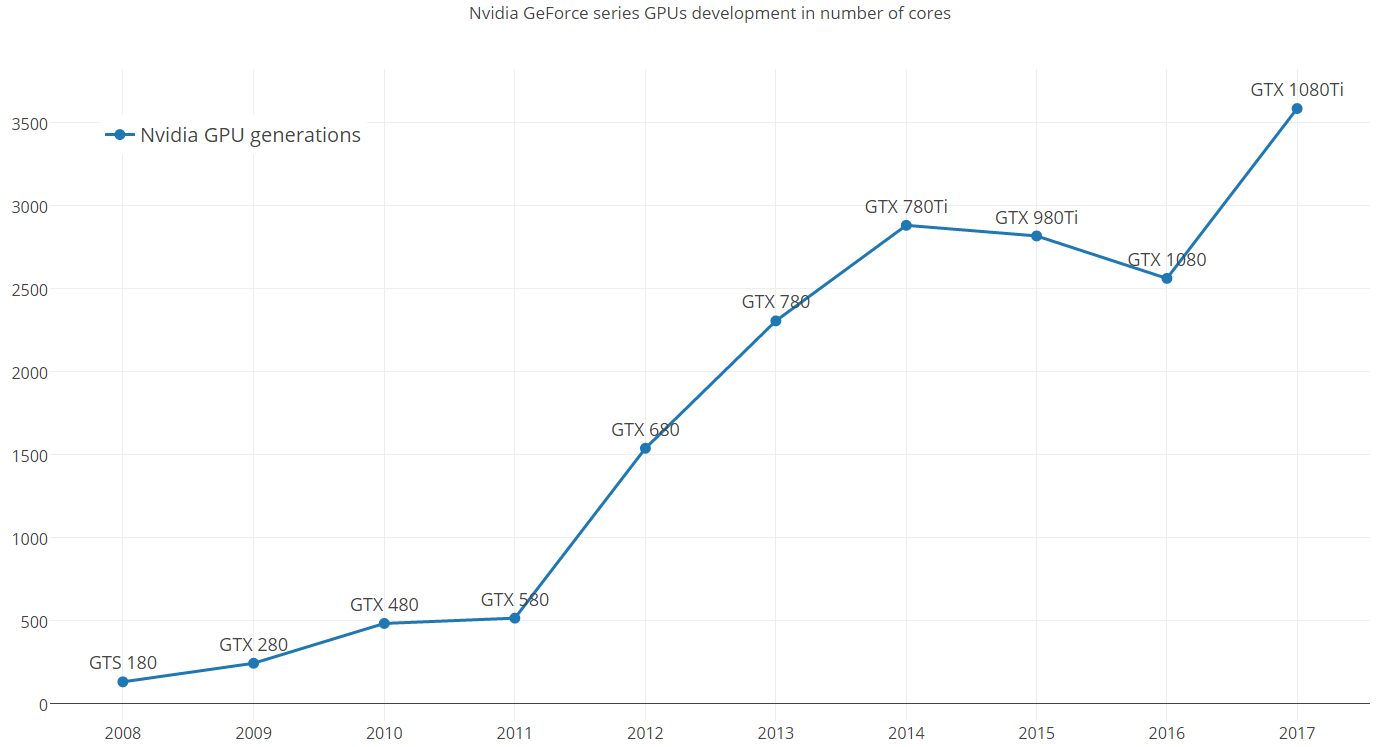}
\caption{Increase of calculation cores (CUDA cores on Nvidia GeForce series GPUs since 2008).}
\label{nvcores}
\end{figure}

This simplicity of GPU architecture made it possible to use large areas of chip real estate for computation engines. In 2003, the ATI R300 chip has more than 110 million transistors which is almost same as Intel’s Xeon microprocessor with 108 million in the same year. However, more than 60\% transistors of the Xeon are devoted to cache~\cite{macedonia2003gpu}. Nowadays, in 2016, the Intel CPU for PCs contains 1-2 billion transistors in die~\cite{intel4770k} and Nvidia GPU has 5.2 billion transistors on GTX 980 and 7.2 billion transistors on GTX 1080~\cite{nvidiawhite1080} with a much larger improvement than CPU's progression compared with one decade ago.

Another direct comparison of hardware improvement is the number of CUDA cores in Nvidia series GPUs in \figurename~\ref{nvcores}. The CUDA cores counted here is the special single-precision calculation cores of Nvidia GPUs in GeForce series (designed for PCs, mainly for gaming purpose) and more details of CUDA will be explained in following sections. From 2008 to 2017, the Nvidia GPUs produced in each year have evolved from less than 200 CUDA cores to more than 3500 CUDA cores.


\subsection{From GPU to GPGPU}

Initially driven by specific needs for gaming applications, the computation capacity of GPUs are mainly fixed-function. Since 2006, as pointed by~\citet{owens2008gpu}, the GPU has evolved into a powerful programmable processor and GPU evolution has been focusing on the programmable aspects of the GPU. This is due to the development of calculation capacity, it became more and more biverse application utilizing GPUs as accelerators for computing bound tasks in general-purpose computing.

In the early days of programming, graphics specific APIs such as OpenGL~\cite{woo1999opengl} or DirectX~\cite{gray2003microsoft} were be used to perform computations. And the shader programming \cite{engel2004shaderx2} is the most common method to execute user defined computation on GPU. For example, the operation of adding two matrices on GPU is one in following steps: creating a window with each pixel corresponding to one output element; rendering one quadrilateral to cover this window; then the texture unit will render this quadrilateral with two textures as every color value inside each texture means the value of the input matrices; finally the color value will be added to get a new texture which can get the output result based on the output quadrilateral. During this process, as long as there was no API for matrix addition, the operation had to be written to fit the existing API. This can be a really cumbersome process when dealing with more complex general-purpose operations like matrix multiplication or DCT transform like in~\citet{fang2005techniques}. In fact, ~\citet{fang2005techniques} achieve 50\% more performance gain with shader programming compared to CPU with SSE implementation which is not a huge improvement.

\begin{figure}[htbp!] 
\centering    
\includegraphics[width=0.85\textwidth]{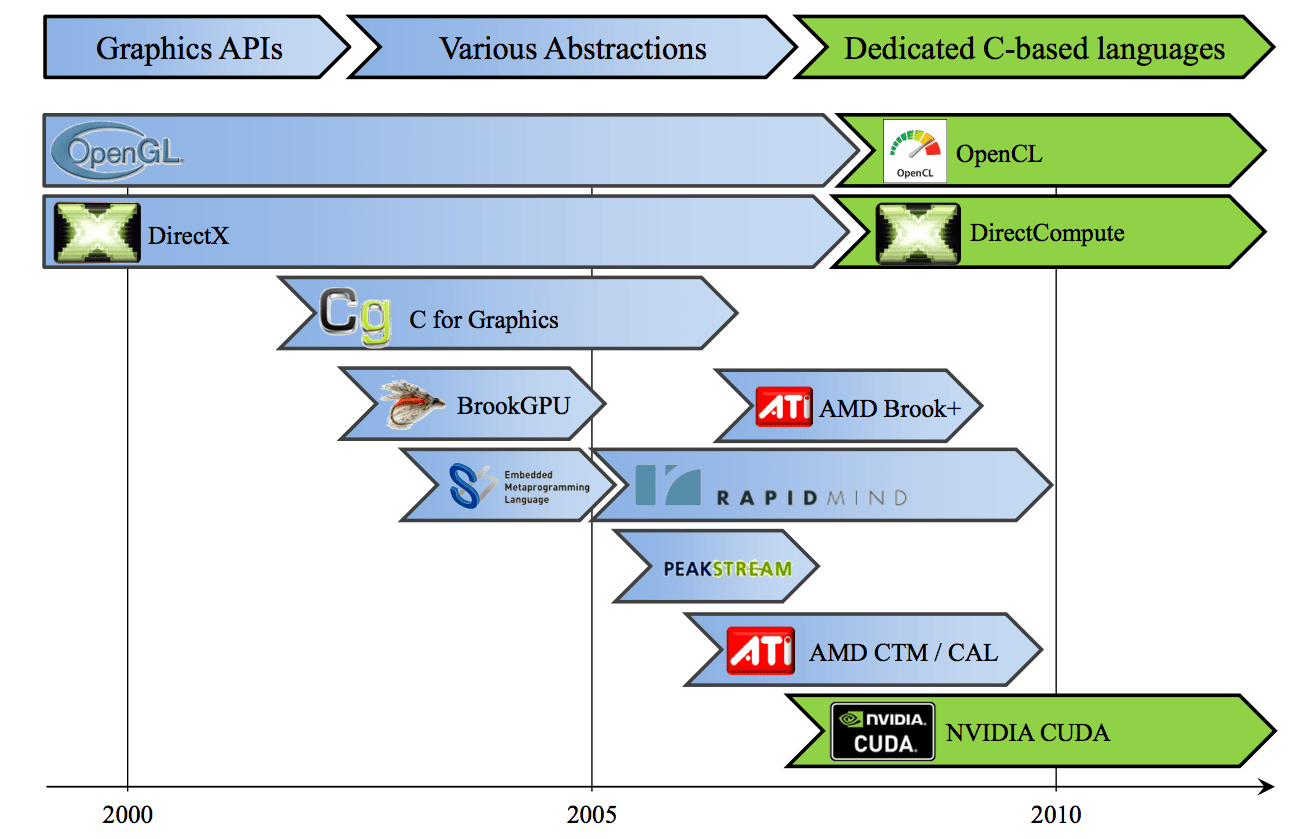}
\caption{Evolution of GPU programming languages. Initially: since 2007 general purpose
languages such as CUDA, DirectCompute, and OpenCL have appeared.~\cite{brodtkorb2013gpu}}
\label{gpulanguagedevelop}
\end{figure}

In 2003 parts of GPUs' fixed-function pipeline became programmable with the release of the NVIDIA GeForce 256 GPU and C for Graphics language~\cite{fernando2003cg} (see~\figurename~\ref{gpulanguagedevelop}).

In 2006, GPUs started to support the unified Shader Model 4.0 on both vertex and fragment shaders~\cite{Blythe2006ds10}. The instruction set specially started to support both 32-bit integers and 32-bit floating-point numbers and the hardware allowed an arbitrary number of both direct and indirect operations from global memory (texture) which makes the single-precision calculation much easier to accelerate. Since then, the design of GPUs are increasingly focusing on the programmable units in the graphics cores and instead of being seen as a a fixed-function pipeline, GPUs started to be described as a programmable engine supported by large number of high efficient fixed-function units.

In 2007, NVIDIA released the first general-purpose language for
programming GPUs, Compute Unified Device Architecture (CUDA~\cite{nvidia2007compute}). Also, as shown in~\figurename~\ref{gpulanguagedevelop}, two other alternative tools to CUDA have emerged: OpenCL (successor of OpenGL) and DirectCompute (successor of DirectX).

Since then, the GPGPU developing entered a new era that the designing of GPU hardware are for different purpose computing with parallel model and platform gives programmers direct access to the GPU's virtual instruction set, parallel computational elements and arbitrary memory operations. It is possible to implement and optimize complex computation tasks at very low level on GPU and the recent researches show the performance gain compared with CPU are increasing very fast~\cite{brodtkorb2013gpu}. In the following sections, CUDA platform and corresponding Nvidia series GPUs are chosen to be elaborated on implementation and architecture details.




\section{CUDA platform}

As indicated by~\cite{nvidia2007compute}, the Nvidia Compute Unified Device Architecture (CUDA) programming model was created as an inexpensive (since it is present as a graphic card in every computer), highly parallel hardware and software architecture available to a continuously larger community of more and more various application developers. 

The main purpose of this CUDA platform is to manage computations on the GPU as a data-parallel computing task without mapping them to a graphics APIs. Also, not only the software level is designed to give a new general-purpose C-like programming language, but also the hardware is adapted to support multi-threads at a hardware level.

It is available for the Nvidia GeForce series GPUs (for PCs), Nvidia Quadro series GPUs (for professional rendering), Nvidia Tesla series GPUs (for science especially math calculation) and Nvidia Tegra GPUs (for mobile platforms). Although different Nvidia series GPUs have the same CUDA architecture, the design purpose and hardware configurations are very different. In this subsection ,we elaborate the three most important factors in CUDA paltform and explain the details for implementing calculation on CUDA-enabled GPUs. More details of different GPU series will be mentioned later in this chapter.

\subsection{CUDA cores}

\begin{figure}[htbp!] 
\centering    
\includegraphics[width=0.7\textwidth]{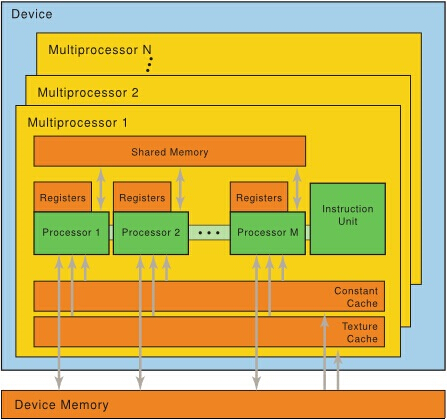}
\caption{CUDA-enabled Nvidia GPGPU device architecture.~\cite{kirk2007nvidia}}
\label{CUDAcoresmemorys}
\end{figure}

The most common way to measure a Nvidia GPU calculation capacity is counting the CUDA cores. High-end PC GPUs today have more than 2000 CUDA cores. However, it is not fair to compare GPU and CPU by them. Basically a core in a CPU means an independent core with large cache that can handle each single operation a computer does including calculation, memory fetching, I/O, interrupts with a highly complex instruction set. In CUDA, the corresponding concept should be Multiprocessors (namely Streaming Multiprocessor, shortly as SM) instead of CUDA cores as shown in~\figurename~\ref{CUDAcoresmemorys}. 

As shown in~\figurename~\ref{CUDAcoresmemorys}, each SM (Multiprocessor in Figure) has its own instruction unit, registers, different caches and from 8 to 128 CUDA cores based on different version of architectures (also called stream processors, SP in previous version of Nvidia docs). This SM can perform the complex functions that similar to what a CPU core can do. The actual CUDA cores inside each SM are just weak calculation cores that can only perform simpler calculations but in a real hardware-level parallel (all CUDA threads are mapping to different CUDA cores and executed in different hardware calculation unit in parallel). This is the SIMD model that lets the CUDA cores within the same SM execute same instruction on different pieces of data. Besides, CUDA cores on the same generation of GPUs are the same, and different generations of CUDA cores are similar except different technology process and power consumptions, etc.



\subsection{CUDA threads model}

Normally in a CPU scenario, thread, a component of a process, means the smallest sequence of programmed instructions that is handled by the operating systems. Also, definition of Multi-threads on a computer architecture normally corresponds to how many physical cores the CPU has. For instance, systems with a single processor generally implement multi-threads program by slicing the time which is to make the CPU switch between different software threads. This switch between processes make the user cannot tell that the threads are not physically parallelized. Modern CPUs equipped with several physical cores like Intel PC CPUs can execute multiple threads in physical parallel with every processor or core executing a separate thread simultaneously.

Threads used in GPU case have a different definition with the one used for CPUs. Based on a totally different design method, the GPUs are designed mainly for calculation instead of managing tasks or logic operations. So the usage of physical parallel threads is much more important than the number in CPU case (normally more than tens of thousands). These threads are more likely to be only simple calculation tasks. Not like the CPU threads, CUDA threads have to be in a very regular fashion with no branches and inter-thread communication to maintain the efficiency. For implementation, CUDA threads normally are patched into warps and sent down to the pipeline together. As a result, the irregular and branch operations are difficult for GPU threads.

\begin{figure}[htbp!] 
\centering    
\includegraphics[width=0.85\textwidth]{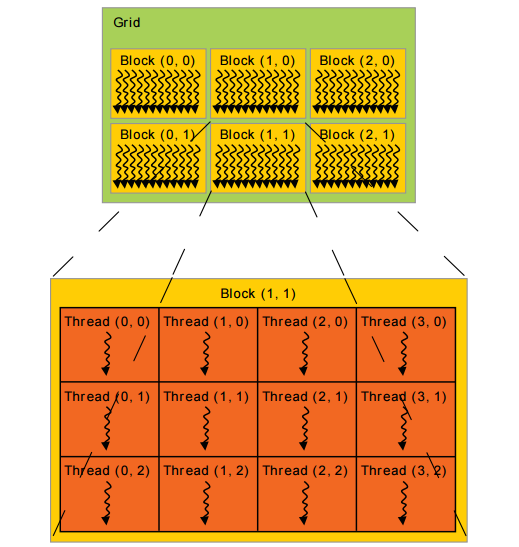}
\caption{CUDA threads model in layers built of grid and block~\cite{nvidia2007compute}.}
\label{gputhreads}
\end{figure}

For Nvidia GPUs, CUDA platform extends C language by allowing the programmer to define C functions, called 'kernels', that, when called, are executed N times in parallel by N different CUDA threads, as opposed to only once like regular C functions. As shown in \figurename \ref{gputhreads}, the thread hierarchy architecture is that threads are grouped into blocks and blocks are grouped into a grid. In the end, a kernel is executed as a grid of blocks of threads. The key factors of the threads hierarchy architecture are:

• Each thread is executed by one core

• Each block is executed by one SM and does not migrate

• Several concurrent blocks can reside on one SM depending on the blocks’ memory requirements and the SM’s memory resources

• Each kernel is executed on one device

• Multiple kernels can execute on a device at one time


Thread blocks are required to be executed independently which is to say they can be executed in parallel or in series. This independence requirement allows thread blocks to be scheduled in any order across any number of SMs which enables programming to adapt with the number of cores. This feature provides adaptability to CUDA programs that allow to fit different GPUs and to get efficiency even the configuration of these GPUs are different (e.g. GPUs equipped with different numbers of CUDA cores or SMs).

\subsection{CUDA memory access management}

The memory bandwidth increase shown in~\figurename~\ref{nvmems} are the largest piece of memory on GPU board also called as global memory which is also the largest and the slowest memory on GPU board. Other types of memory are mainly many different kinds of caches and registers inside the GPU chip as shown in \figurename~\ref{CUDAcoresmemorys}.

There are many memory technologies both in software level and hardware level for the computer to accelerate the memory operations. The most common way for hardware level is the usage of faster memory and wider bus. As shown in \figurename~\ref{nvmems}, GPUs are always using the state-of-the-art memory hardware which can provide the best memory bandwidth. On the other hand, for the software level of design, the most common way to accelerate is to exploit memory by not only using faster memory hardware integrated on chip but also optimizing to avoid cache miss. However, although the CPU has multiple levels of cache with high performance, the operations with these caches are managed only by the operating system and are not accessible to programmer.

Unlike CPUs, CUDA-enabled GPUs allows user to access different kinds of memory during execution. As shown in \figurename~\ref{CUDAcoresmemorys}, each thread corresponding to a 'Processor' has a private local memory which is the registers; each thread block  corresponding to a SM has shared memory visible to all threads of this block and with the same lifetime as this block; and all threads can access to the global memory. There are also two additional read-only memory spaces accessible by all threads: the constant and texture memory spaces. These two kinds of memory are designed for some specific data formats. The constant memory is used to stored the constants which can reduce the required memory bandwidth. Because the constant memory space is cached, a read from constant memory costs one memory read from device memory only on a cache miss; otherwise, it just costs one read from the constant cache. For the threads, reading from constant memory is as fast as reading from registers as the constant memory is cached on chip. The other special memory, texture memory, is designed for threads likely to read from an address “near” the address that nearby threads read. Normally it is used when the program ought to read the data often but update the data rarely and the reading access fits the pattern of spatial locality. For example, in matrix multiplication, the nearby threads access nearby locations of memory (neighbour matrix elements) which can profit the texture memory. Moreover, texture memory can provide additional speedups if we utilize some of the conversions that texture samplers can perform automatically, such as unpacking packed data into separate variables or converting 8-bit and 16-bit integers to normalized floating-point numbers.

In summary, the core design of GPU memory access management is to not only provide fast memory hardware as in \figurename~\ref{CUDAcoresmemorys} but also optimize the software level to let the user have the access to different elements of a complex memory hardware architecture. This could accelerate calculation tasks quite noticeably. However it makes virtualization like in cloud computing difficult for GPUs.

\section{Different hardware platforms}

Different GPU hardware architecture may have huge difference in all aspects. As limited by factors like cost design purpose, or power supply, GPU configurations can be categorized as low-end PC GPUs (for laptops or some low-end desktop), high-end PC GPUs (normally for high-end desktops, but also on some gaming laptops, professional GPUs (for professional math calculations) and mobile GPUs (with totally different designs). In this section, several main GPU platforms will be mentioned and compared. Also, details of the two very different GPU platforms used in our evaluation will be given.

\subsection{PC GPU platform }

In this subsection, a brief overview of the GPU hardware specifications is given by introducing PC GPUs including low-end laptop GPUs, high-end desktop GPUs, state-of-the-art PC GPUs. The main terms that determines a GPU's performance is the number of CUDA cores, memory configuration, and hardware version (also named "compute capacity version" in Nvidia official documents). 


We can notice that in 2011, the calculation speed of GPU varied a lot (can up to more than 50 times faster) according to different GPU types because of their different hardware configuration. This huge difference is rarely seen between PC CPUs. According to \citet{gregg2011data}, the Geforce GTX 480 card runs sort algorithm more than 10 times faster than 330M card. The reason for this situation is because of different design purposes and limit of cooling system or power supply in different PC computers. In recent years, the gap between high-end and low-end GPUs increased even more rapidly.

We categorize PC GPUs into three main categories according to which type of computer they equip: laptop GPUs, desktop GPUs and cutting edge ones. Professional gaming laptops equipped with very powerful GPUs and low configured desktops equipped less powerful GPUs should be considered. 

However, this category fits most of the use cases that high-end GPUs are more widely used on desktop for gaming experience and low-end GPUs are normally designed for laptop to reduce power consumption and physical space usage. The most advanced ones in recent two years are listed as well to compare the main hardware configuration.

In \tablename~\ref{gpuconfigs}, we compare six Nvidia GPUs along three product lines. The performance of Nvs 330M GPU and GeForce gtx 480 GPU (manufactured ain the same year 2010) used in \citet{gregg2011data} have very different CUDA cores inside which leads to different performance for the same algorithm. The next laptop GPU (Nvs 5200M) and desktop GPU (GeForce gtx 780) released in 2012-2013 both increased in all aspects including CUDA cores and memory space. However, the huge difference between laptop GPU and desktop GPU still exists due to the initial design purpose. By the year of 2016-2017, the newest generation of Nvidia GPUs are using the newer generation technology in the micro-architecture manufacturing. As pointed out in \cite{nvidiawhite1080}, the GeFroce gtx 1080 GPU combines benefits of the new Pascal architecture and implementation which is 16 nm FinFET manufacturing process, and the latest GDDR5X memory technology. These benefits allows it to be 3 times more efficient than the GeForce gtx 780 with the similar CUDA core numbers and even less memory width.

\begin{table*}[!htbp]
\centering
\caption{Main characteristics of a laptop GPU and a desktop GPU.}
\label{gpuconfigs}
 \vspace{-1ex}
\begin{tabular}{c c c c c c c c}
\hline
 & Nvidia card & Year  & \begin{tabular}[c]{@{}l@{}}Hardware\\version\end{tabular} & \begin{tabular}[c]{@{}l@{}}CUDA \\ cores\end{tabular} &  \begin{tabular}[c]{@{}l@{}}Memory \\ (MB) \end{tabular} & \begin{tabular}[c]{@{}l@{}}Clock \\ (MHz) \end{tabular} &\begin{tabular}[c]{@{}l@{}}Memory \\ Width\end{tabular}   \\ \hline

\multirow{ 2}{*}{Laptop}& Nvs 330M &2010 &\centering{1.2}  & 48 & 256 & 1265 & 128 bit \\
& Nvs 5200M  &2012 &\centering{2.1}       & 96         & 1024        & 1344       & 64 bit       \\ \hline

\multirow{ 2}{*}{Desktop}& GeForce gtx 480  &2010 &\centering{2.0}       & 480         & 1024        & 1401    & 320 bit  \\
& GeForce gtx 780 &2013 &\centering{3.5}       & 2304        & 3072       & 941      & 384 bit  \\ \hline

Most & GeForce gtx 1080 &2016 &\centering{6.1}       & 2560         & 10k        & 1733  & 256 bit    \\
advanced& GeForce gtx 1080Ti &2017  &\centering{6.1}       & 3584         & 11k        & 1582  & 352 bit    \\ \hline
\end{tabular}

\end{table*}

The Nvs 5200M and GeForce gtx 780 are the GPUs used to implement our design in this thesis. In fact, the two different GPUs are used to present the results for two different use cases: a laptop with a limited calculation power GPU and a desktop with a much more powerful GPU. 

In the Chapters 4 and 5, benchmark evaluation for different calculation tasks on two different GPU platforms shows that implementation details can be varied a lot due to the different performance of GPUs. As long as CPU are also involved in calculation, it is important to consider at least two very different GPU platforms since the difference in performance of the GPU may modify the level of involvement of the CPU.

\subsection{Mobile GPU platform}

During the past decade, mobile phones especially smart phones have changed from just handling dull text-based menu systems to a device equipped with powerful calculation cores and being able to render high-quality graphics at high frame rates. In recent years, due to the development of the battery technology and circuit design, the CPU and GPU in today's smart phones have more capabilities than just being used for rendering graphics.

According to \citet{akenine2008graphics}, a mobile device (mobile phone) is by definition powered with batteries and also has to be small in size in order to be portable. As a result, most limitations stem from constraints of battery-driven and small size. To provide long use-time on the battery, the system of the mobile phones are designed to save energy which limits the calculation power. 

For the CPU case, the main difference between the mobile phone and personal computer is that the mobile phone CPU normally has limited CPU instruction set and a lower clock frequency (e.g., sometimes the division instruction is missing and often floating-point support is not available). Moreover, due to lack of fans or other cooling devices,even if batteries would suddenly become much more powerful, CPU calculation power could not be increased rapidly just like in PC scenarios. 

\begin{figure}[htbp!] 
\centering    
\includegraphics[width=0.7\textwidth]{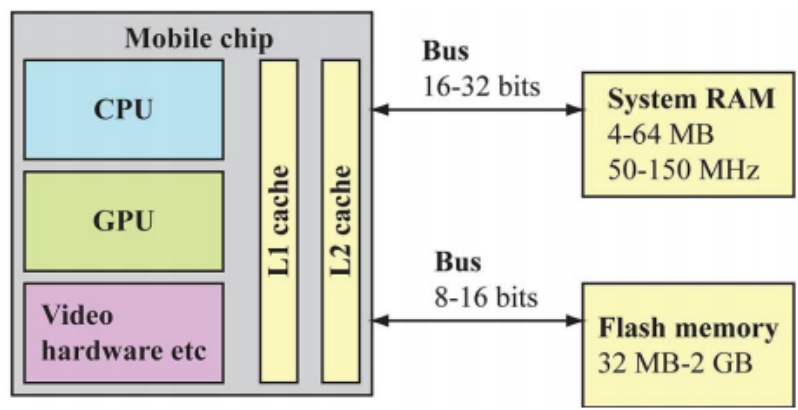}
\caption{One example for mobile phone CPU and GPU system shown in~\cite{akenine2008graphics}.}
\label{phonegpu}
\end{figure}

For the memory design, the memory architecture is quite different from that of PC systems as in \figurename~\ref{phonegpu}. A flash memory often plays the role of 'hard disk' which can keep the data even when the power is off. And there is a small system RAM that is located off chip. As the feature of flash memory is the reading operation is faster than writing, some data that are often used like videos or photos are stored in flash and loaded into RAM when needed. In many cases, there is no
dedicated graphics memory and no separate bus for graphics-related memory designs.

In summary, because of all the factors listed above, the mobile phone GPU today cannot be such power calculation chip as for PC GPGPUs. However, there is still a large development both for hardware and software level. For instance, as pointed by \citet{akenine2008graphics}, in 2008, the L2 cache shown in \figurename~\ref{phonegpu} is still seldom seen. However, today, the A8 processor used by iPhone 6 has a per-core L1 cache of 64 KB for data and 64 KB for instructions, a L2 cache of 1 MB shared by both CPU cores, and a 4 MB L3 cache that services the entire chip \cite{iphone6}. 

More importantly, in recent years, many researchers have already been exploiting to make the dedicated graphics APIs on phone GPUs fit to general-purpose calculation. In \cite{chou2014implementation} the computational speed of the FAST corner detection algorithm is increased 24 times by using GPU parallel computing on an iPhone 4. In \cite{tveit2016deeplearningkit}, the Metal \cite{metal1} and Swift based Deep Learning library for Apple devices like iPhone or Apple TV is introduced and the authors aim to make the iPhone GPU support using deep learning models trained with popular Deep Learning frameworks.

Although the most popular tools used to develop the general-purpose computing task on a mobile phone GPU are still graphics dedicated, it is possible to predict a general-purpose platform like CUDA will emerge and the potential calculation resources of mobile phones will be exploited.



\section{Discussion}

Over the last several decades, parallel computing has evolved significantly both on software aspects and hardware aspects. The computation process has moved from dedicated algorithm on costly equipped super computers to general programming model on almost every personal computers and even many smart phones. Today, the field of parallel computing is having one of its best moments in history of computing and its importance will only grow as long as computer architectures keep evolving to a higher number of processors.

And as shown in our comparison of the Nvidia series GPGPUs released recently, the Nvidia GPGPUs have multiple branches of products with very different hardware configurations which will lead to different software designs. On the other hand, the rapid development pace in hardware manufacturing will continually influence the algorithms implementation.


It seems like almost all problems of calculation could be accelerated by the usage of GPGPU. However, there are still some reasons that limits the GPGPU usage. One of the most common problems is the bottleneck of memory transfer between the host memory and GPGPU's memory. As pointed out by \citet{gregg2011data}, due to the reason that the memory transfer speed is not so fast as the calculation speed, the GPU could be idle while the PCI bus are busy which leads to the decrease of whole performance. This fact rises two open challenges which are how to update the hardware to accelerate the memory transfer and how to design the algorithm to make the execution time of GPU overlap the memory transfer time.

\clearpage

%
%
%
%
%
%
%

\chapter{DCT based selective encryption for bitmaps}

\ifpdf
    \graphicspath{{Chapter4/Figs/Raster/}{Chapter4/Figs/PDF/}{Chapter4/Figs/}}
\else
    \graphicspath{{Chapter4/Figs/Vector/}{Chapter4/Figs/}}
\fi


%

In this chapter, the improved selective encryption methods are shown based on DCT and GPGPU acceleration. There are two levels of design with different purpose. First, the DCT related bitmap protection is introduced. It is pointed that high frequency coefficients of DCT could be used to recover some of the contents. Then, the performance issue mentioned in chapter 2 is solved by employing GPGPU as a hardware accelerator. The two levels of our improved designs are followed with many details to guarantee the minor loss of image quality. Then, the security analysis is presented. At last, the calculation allocation for accelerating process speed is presented to make the designs fit into two typical different hardware platforms.

\section{DCT transformation and selective image protection}

The Discrete Cosine Transform (DCT) is a Fourier-like transform, which was first proposed by~\citet{ahmed1974discrete}. The purpose of DCT is to perform decorrelation of the input signal and to present the output in the frequency domain just like other transformation algorithms. Compared with Fourier Transform, which represents a signal as a combination of sines and cosines, DCT performs only the cosine-series expansion. 

DCT is widely used in many selective encryption algorithms (~\citet{shi1999mpeg}, ~\citet{chiaraluce2002new}, ~\citet{yen1999new}, ~\citet{tang1997methods}, ~\citet{tosun2000efficient}, ~\citet{qiao1997mpeg}, ~\citet{shi1998fast}, ~\citet{zeng2003efficient}, ~\citet{wang2003dct}, ~\citet{wu2001fast}, ~\citet{wu2001efficient}, ~\citet{kankanhalli2002compressed}). The reason that DCT is used in many SE methods is because DCT itself is widely used in multimedia content compression algorithms. And DCT only has the cosines coefficients which makes it map real numbers to real numbers. Compared with DCT, the FFT algorithm always has complex numbers that is difficult to store and process. The second reason is that DCT is known for its property of very high 'energy compaction', meaning that the transformed low frequency coefficients are very large and high frequency coefficients are relatively very small. As a result, this transformed results can be easily compressed by using quantization to keep only a few low frequency components (see JPEG standard~\cite{wallace1992jpeg}). 

DCT has different types shown in~\citet{kresch1999fast}. The most popular DCT algorithm is two-dimensional symmetric variation of the transform that operates on $8 \times 8$ blocks (DCT $8 \times 8$) and its inverse (iDCT $8 \times 8$). This DCT $8 \times 8$ is utilized in JPEG compression routines~\cite{wallace1992jpeg} and has become an important standard in image and video transformation algorithms and many other areas. The two-dimensional input signal is divided into the set of non-overlapping $8 \times 8$ blocks and calculation for one DCT $8 \times 8$ two-dimensional block is defined as follows:

\begin{equation}
C(u,v) = \alpha (u)\alpha (v)\sum_{x=0}^{7}\sum_{y=0}^{7}f(x,y)\cos \left [ \frac{\pi(2x+1)u}{16} \right ]\cos \left [ \frac{\pi(2y+1)v}{16} \right ]
\end{equation}

The inverse of two-dimensional DCT  $8 \times 8$ is defined as:

\begin{equation}
f(x,y) = \sum_{u=0}^{7}\sum_{v=0}^{7}\alpha (u)\alpha (v) C(u,v)\cos \left [ \frac{\pi(2x+1)u}{16} \right ]\cos \left [ \frac{\pi(2y+1)v}{16} \right ]
\end{equation}

where

\begin{equation}
\alpha (u) = \begin{cases}
 & \sqrt{\frac{1}{8}} \text{ , } u=0 \\ 
 &\hspace{4.2mm} \frac{1}{2} \text{\hspace{1.2mm}, } u \neq 0
\end{cases}
\end{equation}

As can be seen from equation 4.1, especially, in the forward DCT $8 \times 8$, the substitution of $u,v = 0$ yields:

\begin{equation}C(0,0) = \alpha (0)\alpha (0)\sum_{0}^{7}\sum_{0}^{7}f(x,y)\end{equation} 

which is eight times of the mean of $8 \times 8$ sample. In fact, this value is called the DC coefficient of the transform results and the others are called the AC coefficients which are independent of the average value. Normally in the image compression case, the DC coefficient is relatively large in magnitude while the AC terms become lower in magnitude as they move farther from the DC coefficient. This means that by performing the DCT $8 \times 8$ on the input raw image, the representation of the image (the main elements carried by an image) is concentrated in the upper left coefficients of each of the output $8 \times 8$ matrix (i.e. low frequency area), while the lower right coefficients of the output matrix contains less important information like details (high frequency area).

From the energy viewpoint, the DC coefficient takes most of the signal energy of the input matrix. In most DCT-based compression algorithms like JPEG standard, there is a quantization step (see~\figurename~\ref{processMultimedia}) to rounding mainly the high frequency coefficients. However, for protecting the bitmap image, compression is not an option.

\begin{figure}[htbp!] 
\centering    
\includegraphics[width=1\textwidth]{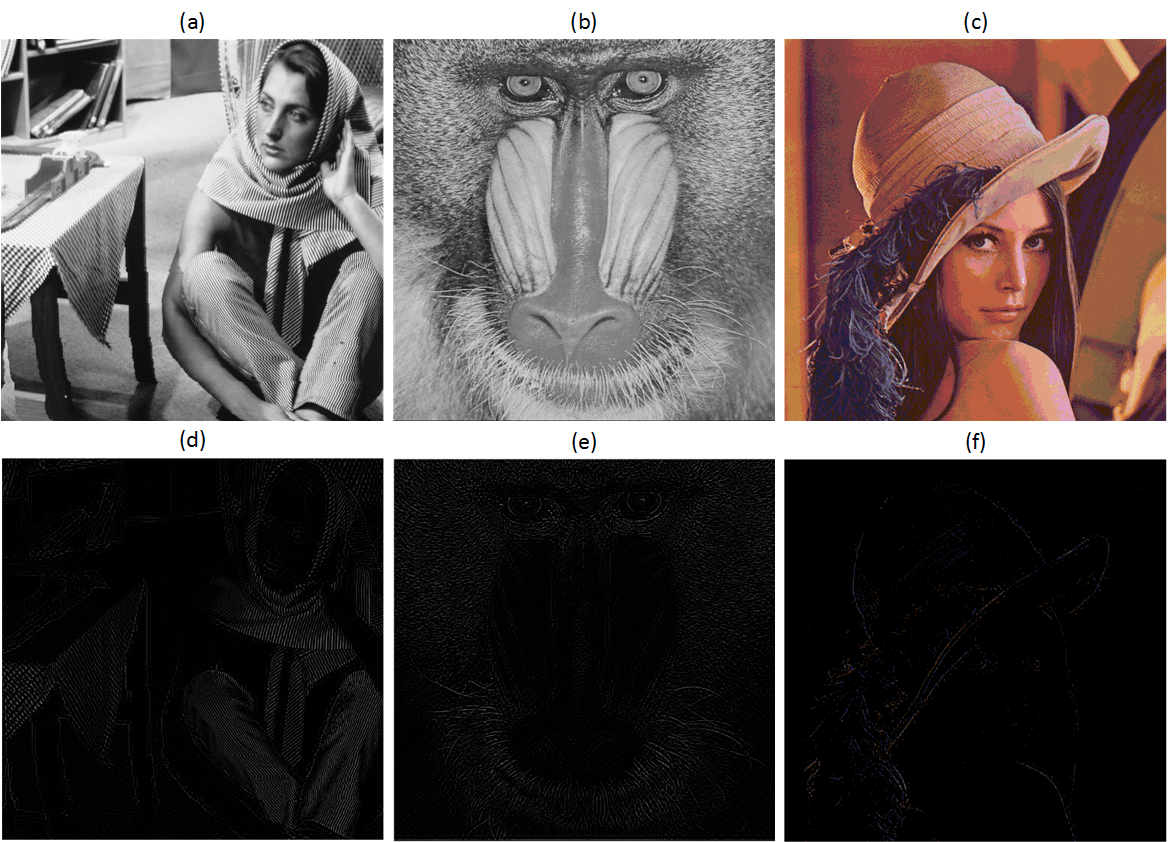}
\caption{Theoretical protection results in~\cite{krikor2009image, qiu2014fast}: original images (a-c) and images that the low frequency area is padded with zeros (d-f).}
\label{c4_1stlvl_protection}
\end{figure}

As pointed out in Chapter 2.4, protecting only the DC coefficient of each $8 \times 8$ block for an input bitmap image is far from enough to guarantee security. Other researches explored in protecting some of the important AC coefficient as well~\cite{krikor2009image}, as long as in most image use case, the DC coefficient and first 5 AC coefficients take more than $96\%$ of the signal energy. If the DC and first 5 AC coefficients (chosen as \cite{krikor2009image} but this is a changable parameter) and padded with zeros and iDCT with the rest 58 AC coefficients, the visual content that can be seen is really limited as shown in~\figurename~\ref{c4_1stlvl_protection}(d-f). In fact, selecting more AC coefficients could help the visual degradation but does not help providing security.

However, SE methods like~\cite{krikor2009image} (protecting the first 6 coefficients of each $8\times 8$ sub-matrix of a bitmap's DCT $8\times 8$ results) is still not enough if the purpose is to protect the whole image content rather than to only disguise the image visual quality. As the remaining 58 AC coefficients carries very little energy, the iDCT $8 \times 8$ result for these 58 coefficients seems almost clean. However, in some cases when the bitmap has many sharp edges, just padding zeros to the first 6 coefficients can show some critical contour to help attackers guess the original content by just visual like shown in \figurename \ref{c4_1stlvl_protection} (d-f).

Moreover, there are possibilities to guess the protected DC values from the know high frequency values in each of the blocks. In fact, in the JPEG standard, original image data will subtract 128 from each pixel intensity in each block to form the range [-128, +127]. And in some SE methods this subtract is not performed. We calculate means of absolute values of the rest 58 AC coefficients for each $8 \times 8$ block and renormalize them to a interval [0, 1]. The guessing DC value is given by multiplying these renormalized means and 2048 (upper limit of DC coefficients if no subtraction) or 1024 if there is subtract.

\begin{figure}[htbp!] 
\centering    
\includegraphics[width=1\textwidth]{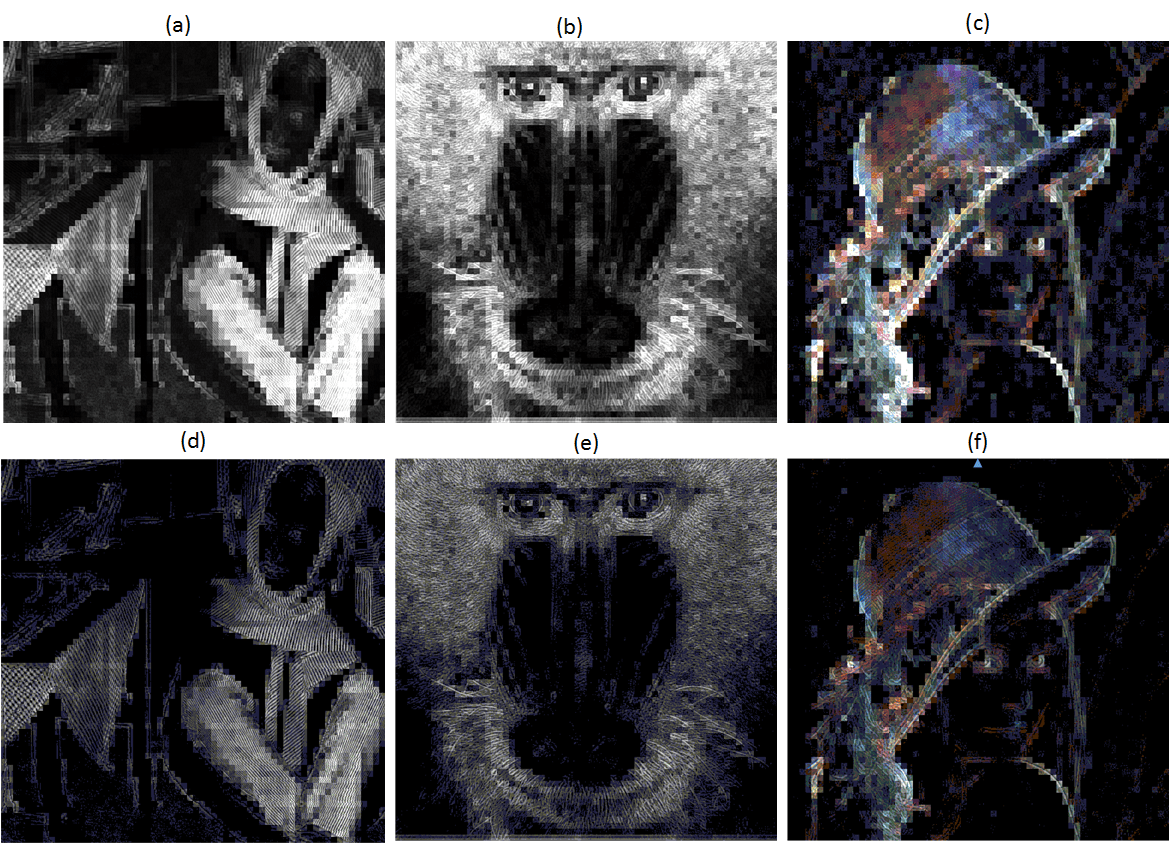}
\caption{Three examples to recover original images by guessing DC coefficient.}
\label{c4_1stlvl_problem}
\end{figure}

\clearpage

 ~\figurename~\ref{c4_1stlvl_problem} (d-f) and (g-i) give the recovery results for no subtract and with subtract respectively which can clearly show the original image contents. This is because a smaller range and more accurate of the DC coefficient of a block can be estimated from the remaining 58 AC coefficients of the same block. Moreover, as pointed by~\citet{li2011recovering}, recovering an arbitrary set of missing DCT coefficients (except for the case when all DCT coefficients are missing) at an acceptable level is possible.




In summary, although all DCT coefficients of one block can be seen as separate layers with different importance according to energy distribution, SE methods based on protecting only few low frequency area can just fit the use cases of disguising the image quality. When 
protecting the image content is the purpose of SE design instead of degrading the visual quality, protecting only the low frequency coefficients and leaving the rest coefficients as plain is far from enough.


\section{DCT acceleration on GPGPU}


\subsection{DCT implementation on CPU}

A lot of effort has been put into optimizing DCT routines on existing hardware. Most of the implementations of DCT $8\times8$ on computer CPUs are well-optimized, which includes the transform separability utilization on high-level and fixed point arithmetic, cache-targeted optimization on low-level~\cite{frigo2005design}. 

However, very few papers discuss and show performance of SE methods based on DCT $8\times8$ by giving implementation. It is important to benchmark DCT $8\times8$ to watch over performance as long as hardware evolves at a fast pace which makes DCT implementations outperforming against a full encryption. E.g. the AES 128-bit can reach about 200 MB/s on a PC’s CPU~\cite{dai2007crypto++} in 2009 but the same code can run almost twice faster today. In this section, we compare the DCT implementation on CPU and GPU today and elaborate on how DCT $8\times8$ is accelerated by GPU.

\subsection{DCT implementation on GPU}

GPU acceleration of DCT $8\times8$ has been possible since creation of shader languages long time ago. However, it requires a specific setup to use common graphics API such as OpenGL~\cite{woo1999opengl} or Direct3D~\cite{engel2002direct3d} for general-purpose computing which is difficult to implement. Since the appearance of CUDA~\cite{nvidia2007compute}, it is easy to have a transparent implementation of GPU accelerated DCT $8\times8$ on a recent Nvidia GPU with a natural extension of C language programming. 

We should notice that most of the traditional low-level optimizations commonly used in CPU implementations are unnecessary in GPU scenario. The floating point calculation is native to GPU and the MUL, ADD and MAD operations on chip are executed with the same speed. The optimization of DCT on GPU is mainly based upon two aspects: first to make the calculation process fit the GPU calculation model; then to optimize the implementation at hardware level (memory usage, data transfer, avoid bank conflict, replace multiplications by reciprocals or arithmetic shifts, etc).

Accelerating DCT by CUDA is discussed in many previous works \cite{obukhov2008discrete}. Normally, the DCT $8\times8$ on two dimensions is actually a separable transform according to equation (4.1). By definition, DCT is firstly applied to the columns of the input $8\times8$ block (on one direction of the 2D block), and then DCT is calculated along the rows of results in last step (on the other direction of the block). Each time the DCT on one direction is applied, it is actually a matrix multiplication of the value matrix and cosine value matrix (or its transpose). This can be seen as twice matrix multiplication as shown in the following equation:

\begin{equation}
  DCT_{2D} = C \times Input \times C^{T} 
\end{equation}

In this matrix multiplication implementation of DCT, the cosine value matrix (presented as $C$) is never calculated on the fly but pre-calculated and stored as a $8\times8$ constant matrix. The value of $C$ is given in following equation

\begin{equation}\makeatletter\def\f@size{10}\check@mathfonts
C = \begin{bmatrix}

0.35355&  0.49039&  0.46194&  0.41573&  0.35355&  0.27779&  0.19134&  0.09755& \\
0.35355&  0.41573&  0.19134& -0.09755& -0.35355& -0.49039& -0.46194& -0.27779& \\
0.35355&  0.27779& -0.19134& -0.49039& -0.35355&  0.09755&  0.46194&  0.41573& \\
0.35355&  0.09755& -0.46194& -0.27779&  0.35355&  0.41573& -0.19134& -0.49039& \\
0.35355& -0.09755& -0.46194&  0.27779&  0.35355& -0.41573& -0.19134&  0.49039& \\
0.35355& -0.27779& -0.19134&  0.49039& -0.35355& -0.09755&  0.46194& -0.41573& \\
0.35355& -0.41573&  0.19134&  0.09755& -0.35355&  0.49039& -0.46194&  0.27779& \\
0.35355& -0.49039&  0.46194& -0.41573&  0.35355& -0.27779&  0.19134& -0.09755 
\end{bmatrix}
\end{equation}

As we mentioned in Chapter 3, GPU can accelerate calculations like matrix multiplications. As shown in ~\citet{patel2009jpeg}, the authors use this twice matrix multiplication methods to accelerate DCT/iDCT $8\times8$ process on GPU and get a performance gain of 10 to 20.



However, as pointed by~\citet{obukhov2008discrete}, the elements in matrix $C$ still contains many clear symmetric which can be used to accelerate the calculation again. Here, we use the DCT optimized algorithm from Nvidia technical report~\cite{obukhov2008discrete}. In~\tablename~\ref{dcton2gpus}, we compare the acceleration for DCT $8 \times 8$ on two different computers, a laptop with an Nvidia 5200M GPU and an Intel I-7 3630QM 2.4GHz CPU and a desktop with an Nvidia GTX 780 GPU and an Intel I7-4770K 3.5GHz CPU.

From this evaluation, we first see that CPUs on a laptop and desktop computer are not that different as performance of CPUs mainly account on factors like main frequency and caches. For the same generation of Intel CPUs, performance are actually quite similar compared with the vast difference of GPU performance of similar generation which mainly relies on the number of the CUDA cores (which could vary from 100 to 2300). Secondly, as performance gain for the laptop (low-end GPU use case) reaches a factor 10 which is similar to the gain found in \cite{patel2009jpeg}, the acceleration obtained for the desktop  (high-end GPU use case) reaches a factor over 70.

\begin{table}[htbp!]
\caption{DCT $8\times8$ accelerated by GPU for laptop and desktop GPUs.}
\centering
\label{dcton2gpus}
\begin{tabular}{c c c c c}
\toprule

Image size&	$1024\times 768$ &	$1600\times1200$ &	$3240\times2592$ &	$4800\times4800$   \\
\hline
Laptop CPU time  &	3.78ms &	9.24ms &	39.4ms &	108.4ms \\
Laptop GPU time	 &  0.41ms &	0.79ms &	3.67ms &	9.98ms \\
Performance gain &	9.2	   &    11.7   &	10.7   &	10.8 \\
\hline
Desktop CPU time &	2.88ms &	7.0ms  &	29.9ms &	82.4ms \\
Desktop GPU time &	0.04ms &	0.09ms &	0.41ms &	1.12ms \\
Performance gain &	72	   &    77.8   &	72.3   &	73.6 \\
\bottomrule
\end{tabular}
\end{table}

As we pointed out in Chapter 3, according to~\citet{gregg2011data}, the calculation ability of GPU varies a lot (can up to more than 50 times faster) in different GPU types because of their different hardware configuration. ~\citet{gregg2011data} show that the Geforce GTX 480 card (high-end desktop GPU) runs sorting algorithm more than 10 times faster than 330M card (low-end laptop GPU). We see the similar difference between our two GPUs in~\tablename~\ref{dcton2gpus}. The huge difference of hardware configurations results in corresponding huge difference of GPU performance. This is an important difference which explains why a SE architecture dedicated to a laptop can benefit from using CPU and GPU while the corresponding desktop solution will look at using only the GPU.



\section{Design of SE for bitmaps based on DCT}

As mentioned in Chapter 2, different SE methods are designed for different protecting purpose and use cases. Here, in this section, we introduce two designs of SE based on DCT $8\times 8$ for bitmap images: a first level of protection when speed is of the essence and disguise image visual quality fit with the use case requirements, and a more complex second level of protection when a more global protection of the image is required. One the one hand, our second level of design proves that by encrypting DC and a few AC coefficients while still protecting the rest AC coefficients in DCT $8\times 8$ blocks can achieve a good level of protection; on the other hand, our work shows that the data-parallel execution model of GPU fits nicely with the preprocess step, DCT $8\times 8$. This fitness will make GPU provide a critical performance gain for selective encryption based on DCT $8\times 8$ since it is only through a GPU implementation that SE is more efficient than a full encryption. Also we point the allocation for arranging calculation tasks would change depending on hardware configuration by providing evaluations on two typical different computers. The following results have been published in \cite{qiu2014fast} and \cite{qiu2015fast}. We provide here additional test cases and a few more details for instance or accuracy.


\subsection{First level protection}

\begin{figure}[htbp!] 
\centering    
\includegraphics[width=1\textwidth]{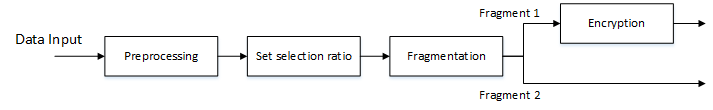}
\caption{General design method for first level protection where Fragment 2 is let to be plain.}
\label{c4_1st_lvl_design}
\end{figure}

In Chapter 2, we defined the fragmentation step to label the preprocessed data with different levels of importance. Here, firstly, the input data will be preprocessed using DCT $8\times 8$. Then the results of the DCT $8\times 8 $ which are the frequency coefficients will be fragmented into two parts according to the selection ratio with respect to the required visual disguise level. The encryption system will be used for the private fragment (Fragment 1 in~\figurename~\ref{c4_1st_lvl_design}) and the public one (Fragment 2 in~\figurename~\ref{c4_1st_lvl_design}) is let to be plain.

Although the visual information cannot be totally protected and having the risk the coefficients could be somewhat recovered, this method significantly reduces the need for the data to be fully encrypted and improve the output performance. The fragmentation step possesses selection ratio done in the frequency domain to increase or decrease the private fragment to be encrypted allowing the user to increase or decrease the desired level of protection. The selection ratio can be set by user, however, the coefficients [0,0], [0,1], [1,0], [2,0], [1,1], [0,2] of the DCT $8\times 8$ block cited from~\citet{krikor2009image} constitute the default selection and is recommended from experience.


This protection method will erase most important visual characters from an image (like people’s face) as shown in~\figurename~\ref{c4_1stlvl_protection} (d-f). It is recommended if for instance, the user’s target is to protect against a mild level of attack from knowing who is in the image. More generally, this first level of protection is good for soft encryption when high performance is required at the same time. In the following sections we will elaborate how GPU is used to accelerate the whole process and how to achieve the best performance on different hardware platforms.

\subsection{Second level protection}

\begin{figure}[htbp!] 
\centering    
\includegraphics[width=1\textwidth]{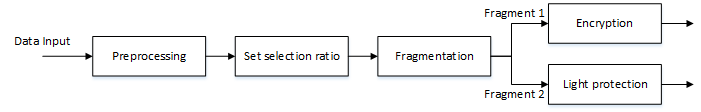}
\caption{General design method for second level protection where Fragment 2 is also protected.}
\label{c4_2ndlvl_design_general}
\end{figure}

As pointed out in Chapter 2.4 and elaborated in Section 4.1, Coefficients in high frequency area of DCT $8\times 8$ sometimes can be used to reveal some information about the original image especially some sharp edges or clear details are contained. For some use cases like protecting whole image content without any information leak, a second level protection is needed to protect ont only low frequency coefficients but also high frequency coefficients. In order to guarantee that performance for the whole SE process will still be better than the full encryption, a lightweight protection method is used to protect the high frequency coefficients (see~\figurename~\ref{c4_2ndlvl_design_general}, generally, encryption for Part1 and light protection for Part2).

A dispersion step is added to separate the storage of the two parts of protected data. This fragmentation method is using Fragment 1 to build up a lightweight protection for the high frequency coefficients (Fragment 2).

\begin{figure}[htbp!] 
\centering    
\includegraphics[width=1\textwidth]{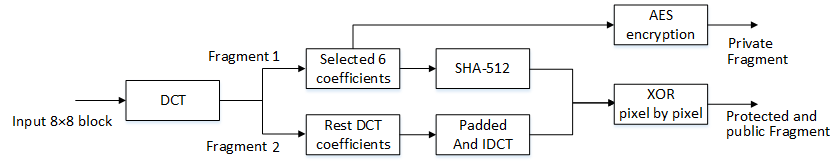}
\caption{Design to enhance the protection level.}
\label{c4_2ndlvl_design}
\end{figure}

We go in more details in~\figurename~\ref{c4_2ndlvl_design}: the lightweight protection step uses the SHA-512 function~\cite{toolkit2006secure} to get a unique fixed-length string (512-bit long) from the 6 selected coefficients (Fragment 1 in~\figurename~\ref{c4_2ndlvl_design_general}). The SHA-512 function has a feature that can generate two totally different and unpredicted fixed length strings even if only one bit of the input string is different. Moreover, according to~\cite{toolkit2006secure}, it is not possible to recover the input data if only the output 512-bit string is known. This feature will guarantee that the 512-bit string cannot be used to do prediction, recovery or guessing even adjacent $8\times 8 $ blocks of a bitmap image are very similar. Because the block we processed is $8\times 8 $, the reverse DCT (iDCT in~\figurename~\ref{c4_2ndlvl_design}) result of the rest DCT coefficients (DC position padded with 1024 and rest AC coefficients padded with zeros) contains 64 pixels. If we store these 64 pixels in 8-bit integers (Byte), the total length is exactly 512 bits. Then the XOR step can protect every bit pixel by pixel within each block.

\section{Storage space usage and numeric precision}


There is a classic trade-off between the memory occupation (both footprint and storage space) and numeric precision when it comes to handling floating point numbers. This problem is rarely considered in traditional DCT based SE methods. The possible information loss caused by not designing the numeric format transform between integers and floating-point numbers seems to be ignored by many related works (~\citet{krikor2009image}  ~\citet{pareek2006image} ~\citet{puech2005crypto} ~\citet{guan2005chaos}). This is because most of the DCT based SE methods are protecting contents with the compression step which is the quantization step that rounded most of the high frequency coefficients, so no need of the design to store the floating point numbers.

However, in bitmap protection cases, if we take ~\cite{krikor2009image} as an example, each pixel in bitmap files is usually stored as an 8-bit integer (two more bytes are used for ‘Highcolor’ and two additional bytes are used for ‘Truecolor’). During the forward DCT computation, these integers are transformed into floating point (32 bits) numbers increasing the footprint by a factor 4. At the end of the computation, numbers are turned back to integer and this is repeated during the computation of iDCT. These value type changes involves truncation as the range of encrypted coefficients are very different and some rounded values have to be ignored in order to fit the storage space. This could lead to image distortion or  information loss when decrypting and rebuilding the original image. Moreover, recursive rounding error cannot be avoided in this case if one bitmap image is encrypted and decrypted multiple times by this scheme.

In our implementation, we firstly calculate the possible value range to determine how to store the results. Then we measure the possible distortion or information loss in image processing by using PSNR (Peak Signal-to-Noise Ratio)~\cite{huynh2008scope} to show how much signal is lost for different images. Moreover, more specific comparison between original images and multiple encrypted and decrypted images are compared to show the possible information loss.


\subsection{Storage space design}

For the bitmap case, we consider the gray scale image as an example (for 3 layer color images, each color layer can use the same scheme for one gray scale image).

For each pixel of a gray scale bitmap, the input range is between 0 and 255. Since DC coefficients are bounded by 1024 (subtract 128 from each pixel to reduce DC value range from [-2048, +2048] to [-1024, +1024]), we used an 11-bit storage space to store only the integer part (1 bit for sign and 10 bits for absolute value). For AC coefficient, it is easy to calculate that the range is within -1023 to +1023 by using the definition and equations (4.5), (4.6) (no matter the input range is from 0 to 255 or from -128 to 127). Our design is to use 11-bit storage space for each selected AC coefficients: first bit for the sign and the other 10 bits for the value (10-bit can store integers from 0 to 1023). For color images, the protection is done for the different layers of pixels respectively. We call this design as 11-bit store method and use it as default in this design.

After storing the selected important values, the remaining AC coefficients will need an iDCT to re-transform to a $8\times 8$ matrix with each element as an 8-bit integer. Here the initial DC value is padded with 1024 and the 5 selected AC values are padded with zeros. The reason why not just pad every selected element as zero is because as long as the DC values and first 5 AC coefficients are zeros, the iDCT results will contain a lot of negative values or positive values very near to zero. It is difficult to store all these values within 8-bit storage space. However, if the DC value is set to be 1024, the average value is set to be 128 in the iDCT result which is easier for keeping the coefficients. Then the value range is to be set as between 0 and 255 to round all iDCT coefficients to 8-bit unsigned integers. 

\begin{figure} [htbp]
\centering    
\includegraphics[width=0.9\textwidth]{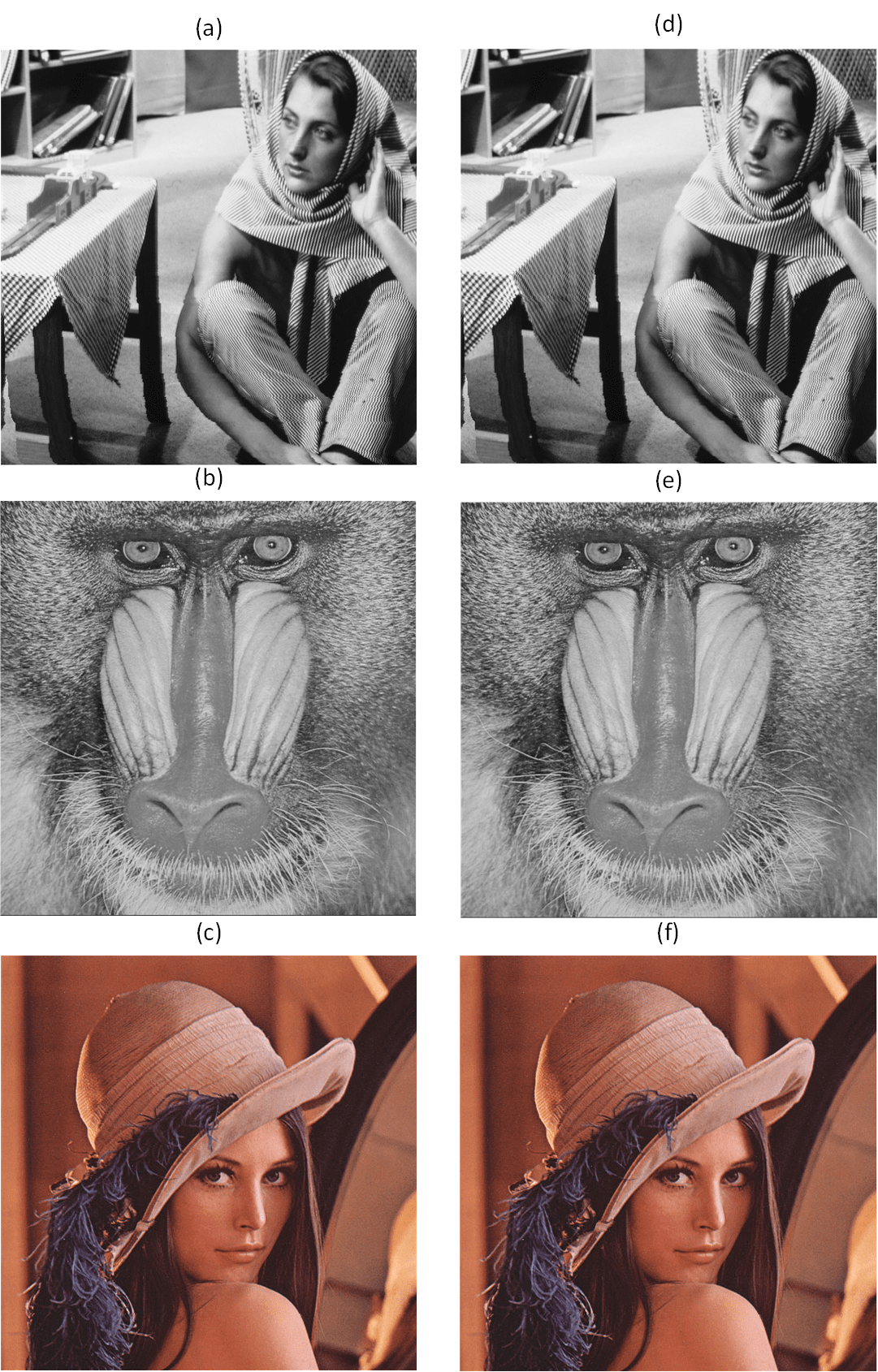}
\caption{Visual effect of 11-bit store method (a-c): original images and (d-f): decrypted and rebuilt images with PSNR values are 63.1, 62.7 and 62.9 respectively.}
\label{c4_store_psnr}
\end{figure}

The extra storage space of 11-bit store method is just the extra bits used to store the selected DC coefficient (11 bits) and the 5 AC coefficients (55 bits in total) which is 66 bits more per $8\times8$ block. In summary, the total extra storage usage is $66/512 = 12.9\%$ of the original space usage. Considering the fragmentation step, the private fragment only takes 66 bits and the protected and public fragment takes the same storage space with the original image. The visual results for the designs is shown in \figurename~\ref{c4_store_psnr}. 

\subsection{Numeric precision analysis}

In~\figurename~\ref{c4_store_psnr}, we gives the example of visual results of the decrypted images (d-f) compared with the original images (a-c). And the value of PSNR after encryption and decryption is more than 62 dB as shown in \tablename~\ref{c4_store_psnr_table}. PSNRs of 50 dB for 8bpp (bit-per-pixel) images usually results in almost identical images according to \cite{huynh2008scope}.

\begin{table}[htbp!] 
\caption{PSNR for the selective encryption of different images.}
\centering
\label{c4_store_psnr_table}
\begin{tabular}{c c c}
\toprule

Image size & PSNR (enc and dec)  \\
\hline
$256\times 256$ & 62.78 dB  \\

$512\times 512$ & 63.10 dB \\

$1024\times 768$ &  62.76 dB \\

$1600\times 1200$ &  62.82 dB  \\

$3240\times 2592$ &  62.85 dB \\ 

$4800\times 4800$ &  62.89 dB \\ 

\bottomrule
\end{tabular}
\end{table}

In~\tablename~\ref{c4_store_psnr_table}, we tested multiple bitmap images as input to show the PSNR of the images before and after protection keeps about 62 dB which means the loss caused by rounding in the design is really tiny. In fact, there are two places where the information details are missing: first one is the rounding step of the DC coefficient and selected 5 AC coefficients; second one is to store the remaining coefficients after rounding each of the iDCT results into 8-bit unsigned integers. second one is to store the remaining coefficients after rounding each of the iDCT results into 8-bit unsigned integers.



\begin{figure}[htbp!] 
\centering    
\includegraphics[width=1\textwidth]{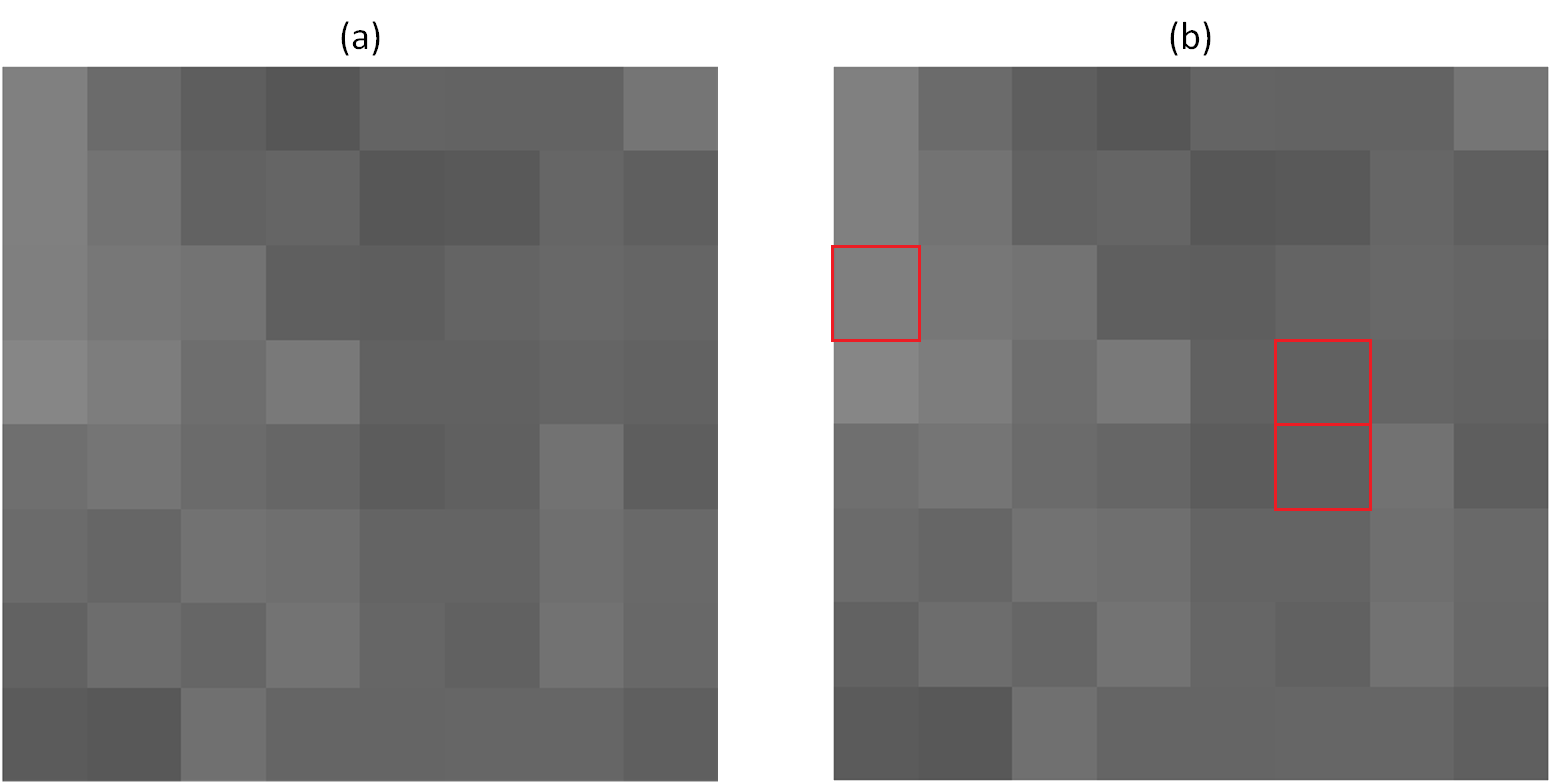}
\caption{Visual effect of one $8\times8$ block with only 3 pixel value different (original block and reconstructed block).}
\label{showdiff}
\end{figure}

In order to give a complete idea of the bit value loss due to the integer and float conversion, a more straight forward way is used by comparing the different values in the gray pixel value. Firstly, we take one image 'fofo' (image (b) in \figurename~\ref{c4_store_psnr}) as the plain image. And each pixel value of the reconstructed image is compared with the initial plain image. There are only about $3\%$ of the pixel values are different due to the integer and float conversion. Then we select randomly one block to show the visual difference with this kind of minor pixel value difference on visual effect. The following two blocks are the pixel values and only three values are slightly different and are located at $(3,1)$, $(4,6)$ and $(5,6)$. In \figurename~\ref{showdiff}, the minor difference in visual is shown by comparing the original block with the reconstructed block. The red small blocks in the right image are the ones different with original ones and all pixel values corresponds to the following matrix.

\[
\begin{bmatrix}
128	& 107 & 94	& 86 &	100 &	99 &	99 &	117\\
128	& 115 & 98	& 101 &	87 &	89 &	102 &	95\\
127	& 119 & 115	& 95 &	94 &	100 &	104 &	101\\
134	& 125 & 110	& 121 &	97 &	97 &	101 &	98\\
111	& 117 & 107	& 102 &	92 &	96 &	114 &	94\\
107	& 102 & 114	& 111 &	100 &	100 &	111 &	105\\
98	& 109 & 102	& 115 &	102 &	97 &	114 &	104\\
91	& 88 & 112	& 101 &	101 &	102 &	102 &	95\\
\end{bmatrix}
,
\begin{bmatrix}
128	& 107 & 94	& 86 &	100 &	99 &	99 &	117\\
128	& 115 & 98	& 101 &	87 &	89 &	102 &	95\\
(128) & 119 & 115 & 95 & 94     & 100 &	104 & 101\\
134	& 125 & 110	& 121 &	97 &	(98) &	101 &	98\\
111	& 117 & 107	& 102 &	92 &	(97) &	114 &	94\\
107	& 102 & 114	& 111 &	100 &	100 &	111 &	105\\
98	& 109 & 102	& 115 &	102 &	97 &	114 &	104\\
91	& 88 & 112	& 101 &	101 &	102 &	102 &	95\\
\end{bmatrix}
\]

\begin{figure}[htbp!] 
\centering    
\includegraphics[width=1\textwidth]{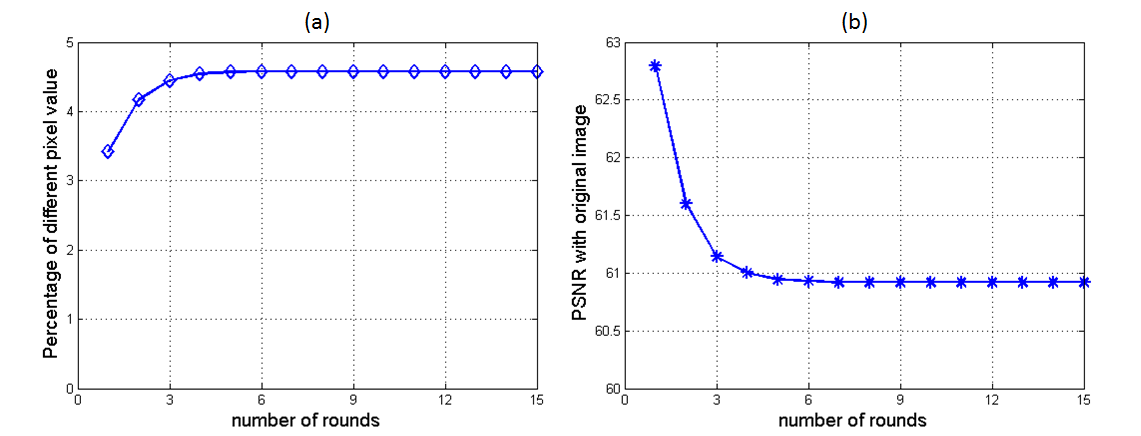}
\vspace{-2em}
\caption{Percentage of different pixel values (a) and PSNR (b) stop changing after several rounds for fofo image.}
\label{diffpsnrfofo}
\end{figure}

\begin{figure}[htbp!] 
\centering    
\includegraphics[width=1\textwidth]{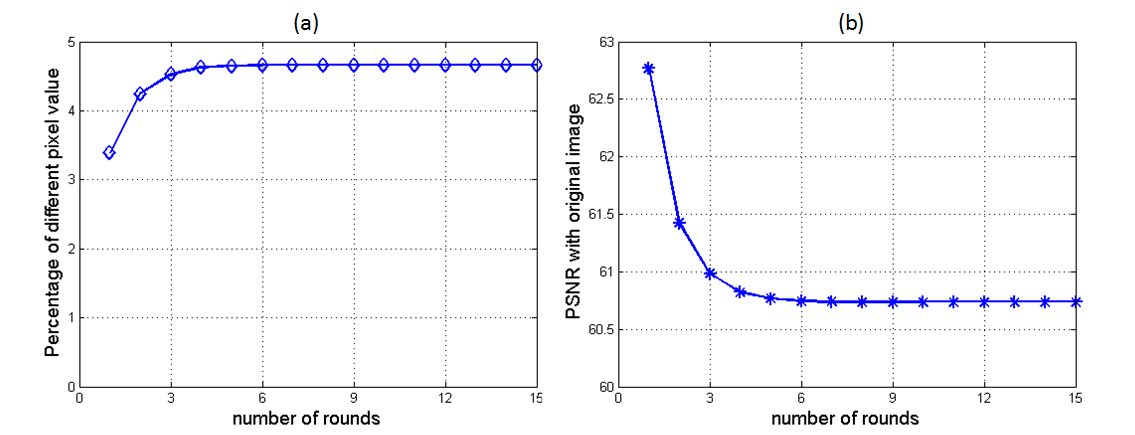}
\vspace{-2em}
\caption{Percentage of different pixel values (a) and PSNR (b) stop changing after several rounds for barbara image.}
\label{diffpsnrbar}
\end{figure}

\begin{figure}[htbp!] 
\centering    
\includegraphics[width=1\textwidth]{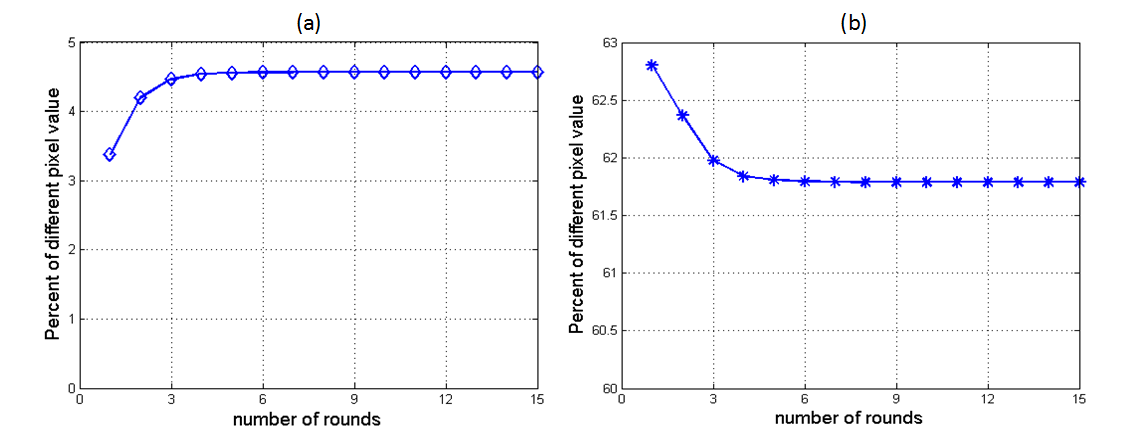}
\vspace{-2em}
\caption{Percentage of different pixel values (a) and PSNR (b) stop changing after several rounds for lena image.}
\label{diffpsnrlena}
\end{figure}

As long as every time the protection and rebuild process would introduce the truncation and rounding, recursive rounding error \cite{jpgrec} is also introduced. In \figurename \ref{diffpsnrfofo}, \figurename \ref{diffpsnrbar}, and \figurename \ref{diffpsnrlena}, 15 rounds of protected and rebuild process for three bitmaps (images (a): fofo, (b): barbara, and (c) lena in \figurename \ref{c4_store_psnr}) are done, and in each round, two results are compared between the rebuilt image in current round with the original image: PSNR and the percentage of changed pixel values. From the experimentation, the PSNR and percentage of changed pixel values keep unchanged after several rounds which means the storage design avoids the recursive rounding error after some rounds loss. In the end, PSNR is 60.919 for fofo case, 60.735 for barbara case, and 61.788 for lena case. And totally less than 5\% (around 4.6\%) pixel values are slightly changed like in \figurename \ref{showdiff}.

\section{Result analysis}

In this section, probability density function (PDF) and correlation coefficient computation are used to evaluate the protection quality. As pointed out before, our method will fragment the image into two (a confidential fragment to be stored locally or in a high-level security place, a public fragment to be stored in a public server). Also, the first level protection method is only for disguising the image quality instead of protection, only the second level protection of our designs will be analyzed here. And the analysis is only for the protected public fragment as long as the private fragment is seen as secure by employing AES (It is easy to replace AES to any other kind of encryption algorithms).
 
As pointed by \citet{pareek2006image}, it is known that many encryption algorithms have been successfully analyzed by statistical analysis and several statistical attacks. In most cases, visual degradation is used to evaluate the security property of SE methods for images. To test the robustness of our encryption method, we will perform statistical analysis by giving the PDF and the correlations for two adjacent pixels in the protected public part.

\subsection{Probability Density Function analysis}

\begin{figure}[htbp!] 
\centering    
\includegraphics[width=1\textwidth]{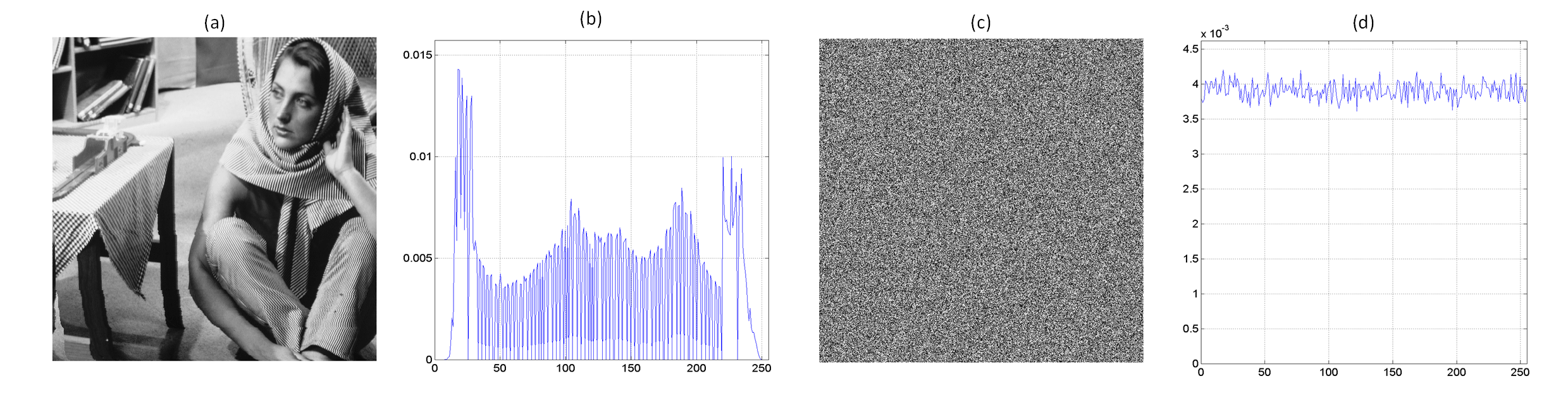}
\caption{Plain gray-scale image (a) and its PDF (b) compared with protected and public fragment (c) and its PDF (d).}
\label{c4_histogram}
\end{figure}

A probability density function (PDF) of the image byte representation illustrates how pixels in an image are distributed by graphing the number of pixels intensity level. For a gray scale image case, \figurename~\ref{c4_histogram} gives an example that the PDF of the protected and public fragment of the image are fairly uniform and significantly different from the respective PDF of the original image. 

\begin{figure}[htbp!] 
\centering    
\includegraphics[width=1\textwidth]{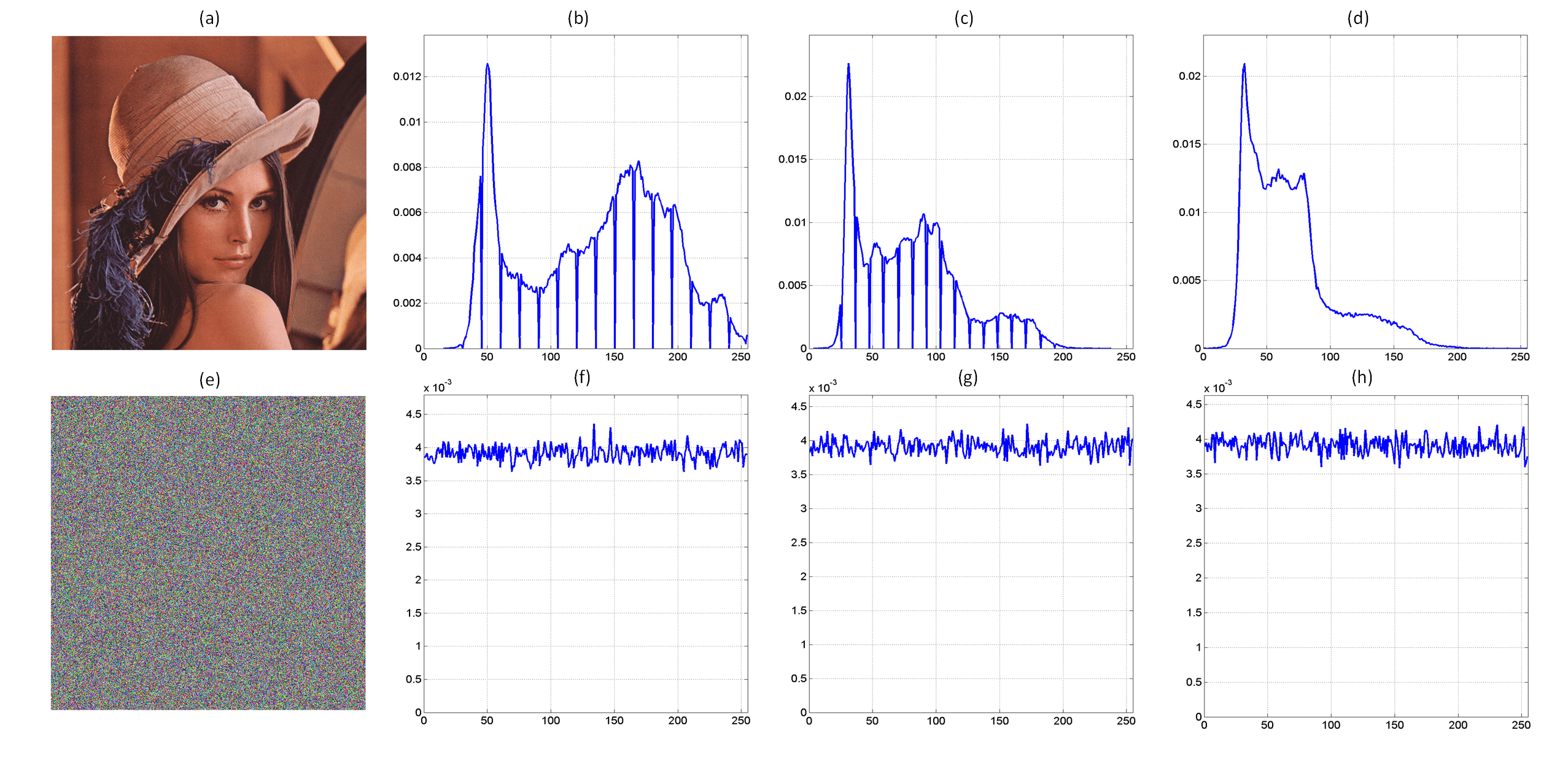}
\caption{Plain RGB image (a) and its PDF on RGB layers ((b),(c),(d) correspond to red, green, blue layers) compared with protected and public fragment (e) and its PDF on RGB layers ((f),(g),(h) correspond to red, green, blue layers).}
\label{lena_histogram}
\end{figure}

Respectively, ~\figurename~\ref{lena_histogram} gives an example that for a truecolor image (RGB image) case, the distribution for three color layers are all uniformly distributed after protection. This property guarantees that the protected and public fragment of the image will not provide any clue to employ any statistical attack~\cite{pareek2006image}.

\vspace{-1em}
\subsection{Coefficients analysis}

Lower correlation between original and encrypted data is an important factor that permits to validate the independence between them. Having a correlation coefficient uniformly distributed means that a high degree of randomness is obtained. According to~\citet{wang2011new}, to test the correlation between two adjacent pixels, the following procedures are carried out. First, randomly select 10,000 pairs of two adjacent pixels in horizontal, vertical and diagonal direction, then compute the correlation coefficient $r_{xy}$ of each pair using:
 
\begin{equation}
r_{xy}={}{}\frac{cov (x,y)} {\sqrt{D(x)\times{D(y)}}}
\end{equation}
where 
\begin{eqnarray}
&& \; E(x)=\frac{1}{N}\times \sum_{i=1}^N x_i \nonumber\\
&&{} \;D(x)=\frac{1}{N}\times \sum_{i=1}^N (x_i-E(x))^2\nonumber \\
&& \; cov (x,y)=\frac{1}{N}\times \sum_{i=1}^N (x_i-E(x))(y_i-E(y))  \nonumber 
  \end{eqnarray}
 
 $x$ and $y$ are values of the two adjacent pixels in the image for the gray scale image case.

\begin{figure}[htbp!] 
\centering    
\includegraphics[width=1\textwidth]{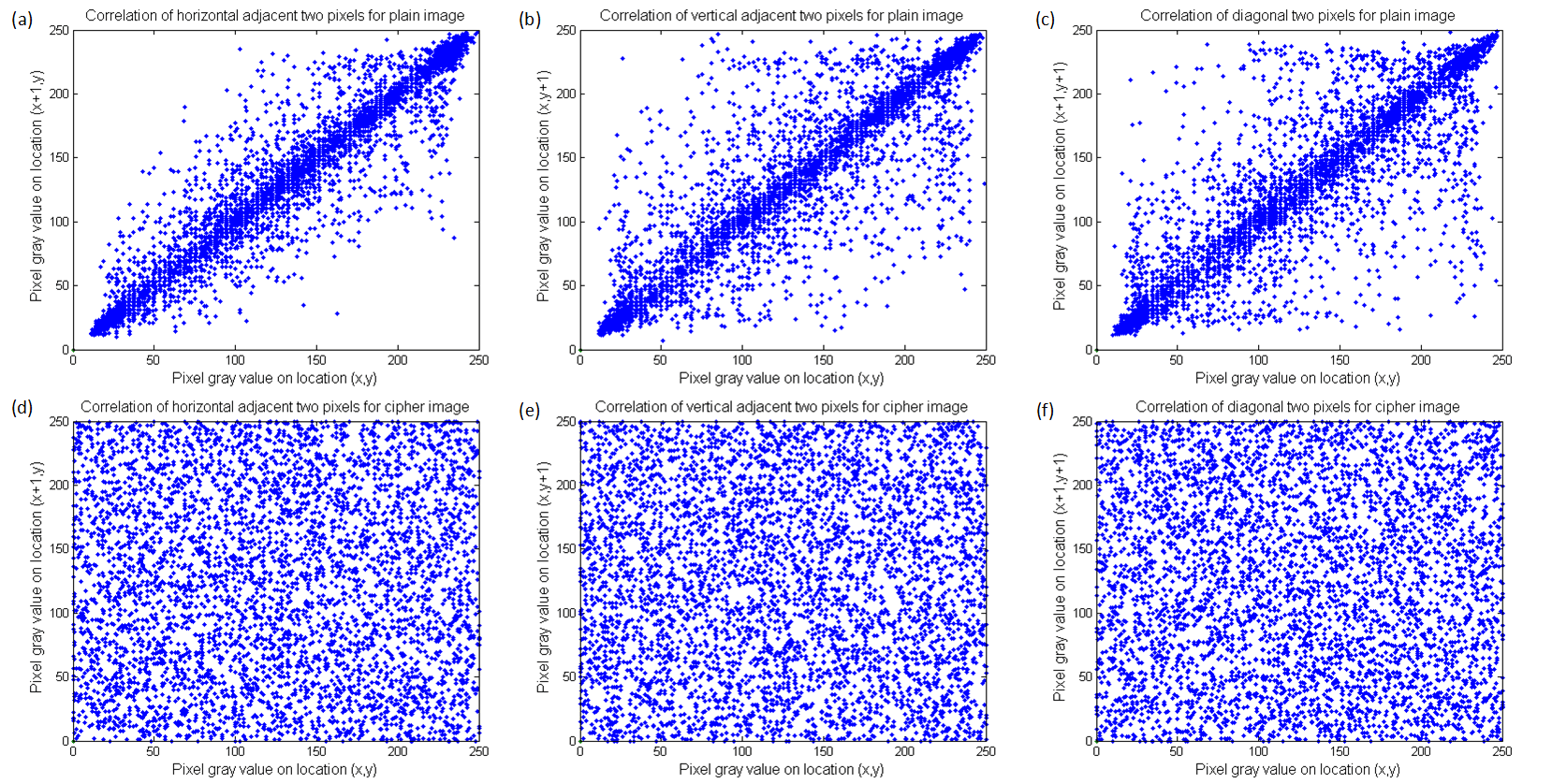}
\caption{Correlation of adjacent pixels in horizontal, vertical and diagonal direction for a gray scale bitmap image.}
\label{correlationsgray}
\end{figure}

\begin{figure}[htbp!] 
\centering    
\includegraphics[width=0.8\textwidth]{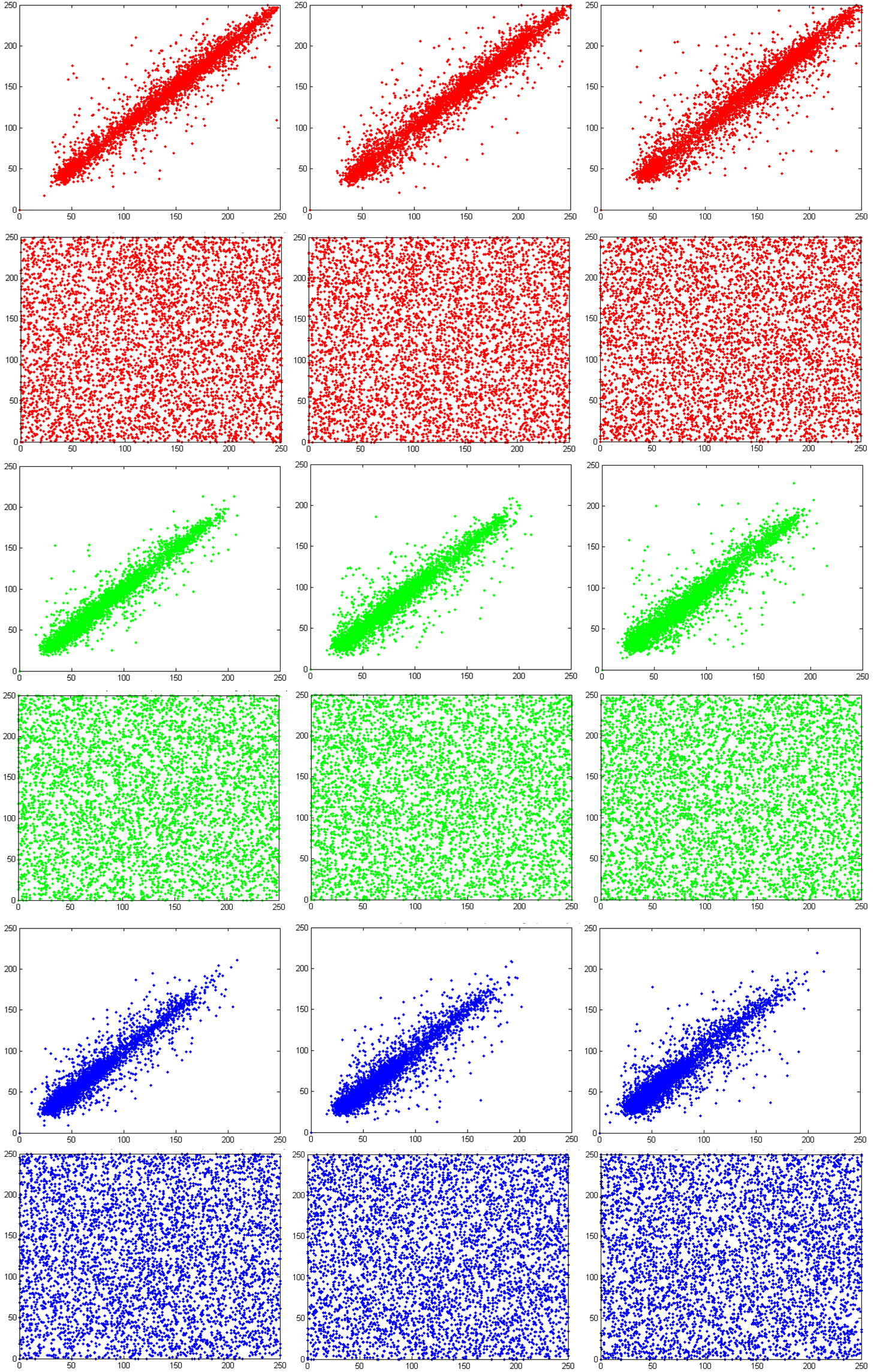}
\caption{Correlation of adjacent pixels in horizontal, vertical and diagonal direction for Red, Green and Blue layers.}
\label{correlationscolor}
\end{figure}

 Then, the same operations are performed along the vertical and the diagonal directions. As shown in~\figurename~\ref{correlationsgray}, the correlation coefficient distributions of the cipher images seem uniform compared with the original plain image. In~\figurename~\ref{correlationscolor}, the analysis for the protection results of a three layers of a RGB format bitmap image is given. In this format, every pixel is stored using 24 bits with every 8 bits for one color layer. The protection will mix the correlation coefficient distributions for each of the layer.

\section{Evaluations with different computer architecture}

In this section, we mainly discuss the implementation of allocating calculation tasks to GPU and CPU and evaluate their performance. As we pointed out in Chapter 3, the huge difference between low-end GPU and high-end GPU makes the calculation and design for program very different. Here a common low-end laptop GPU (Nvidia Nvs 5200M equipped with a CPU of Intel I7-3630QM) is used to test allocation. Then using a high-end desktop gaming GPU (Nvidia GTX 780 equipped with a CPU of Intel I7-4770K) leads to change and improve the design.

One circumstance of the hardware worth pointing out is that the other devices equipping by the two computers are similar (CPU, bus, motherboard, host memory, etc). Therefore the difference in performance allows comparing with the GPUs and the respective designs.

\subsection{Allocation of calculation tasks for a moderately powerful GPU (laptop)}

\begin{figure}[htbp!] 
\centering    
\includegraphics[width=0.6\textwidth]{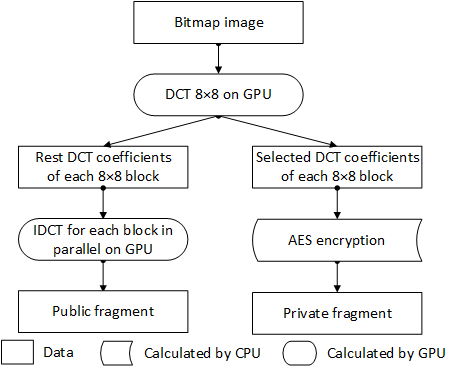}
\caption{Process steps for first level protection.}
\label{design1lvl}
\end{figure}

~\figurename~\ref{design1lvl} shows the design steps to process a single bitmap image. The image content will be copied into GPU memory and fragmented by GPU after DCT $8\times 8 $ preprocess. Then the selected coefficients which are considered as the private fragment will be transferred to host memory and encrypted using AES 128-bit by CPU. In parallel, the remaining DCT coefficients will be padded and transformed by iDCT $8\times 8 $ to build the public part. Then the public fragment will be transferred to host memory for further dispersion.

This design is aiming at fully utilizing both the CPU and GPU resources on a laptop by accelerating DCT processing using GPU. The total execution time depends on a race between CPU and GPU. As we evaluate separately the execution times of on CPU and GPU: the time spent by DCT on GPU is greater than the AES time spent by AES on CPU as shown in~\tablename~\ref{c4_1lvl_dct_aes}. This is because although GPU is able to accelerate DCT, CPU just need to encrypt a small part of the original data. The overlay and parallel design is very simple: it uses the GPU to calculate DCT and then when GPU is calculated, the iDCT for the public part, CPU is only calculating the AES for the private part.

\begin{figure}[htbp!] 
\centering    
\includegraphics[width=1\textwidth]{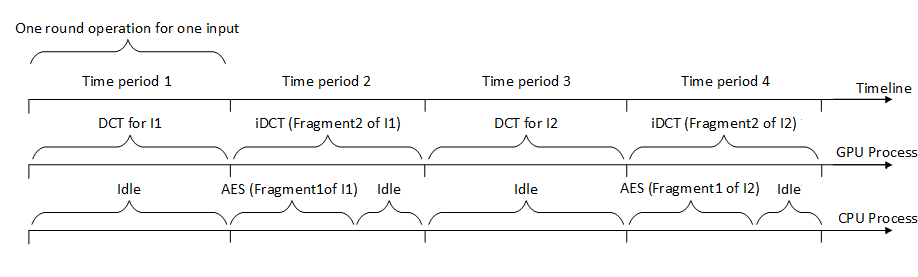}
\caption{Time overlay design of first level protection for multiple bitmap images as series input.}
\label{c4_overlay2}
\end{figure}

This design based on laptop hardware configuration works well for a series of images in the same format (bitmap) as input because of the overlay design of the GPU and CPU. As GPU are calculating the DCT $8\times 8 $ and iDCT $8\times 8 $ of each input image and CPU are encrypting the selected parts (data amount about 10\% of the original image) in parallel. The total run time depends on which processor is slower (on laptop use case, the GPU execution is always slower due to its limited calculation capacity). The working flow of operations is shown in~\figurename~\ref{c4_overlay2}.

\begin{table}[htbp!] 
\caption{DCT time on GPU and AES time on CPU of the laptop use case.}%
\centering
\label{c4_1lvl_dct_aes}
\begin{tabular}{c c c c c}
\toprule

Image size & $1024\times 768$ & $1600\times 1200$ & $3240\times 2592$ & $4800\times 4800$  \\
\hline
DCT on laptop GPU & 0.41 ms & 0.79 ms & 3.67 ms & 9.98 ms  \\

AES on laptop CPU & 0.19 ms & 0.47 ms & 2.05 ms & 5.87 ms \\ 

\bottomrule
\end{tabular}
\end{table}

\begin{figure}[htbp!] 
\centering    
\includegraphics[width=0.6\textwidth]{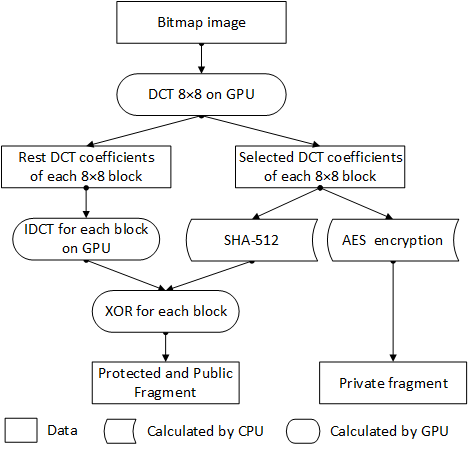}
\caption{Process steps for the second level protection.}
\label{design2lvl}
\end{figure}

In fact, as long as the encryption run time on CPU for Fragment 1 of I1 is less than total time of period 2 and period 3 in~\figurename~\ref{c4_overlay2}, this design of overlay makes it adapted for the second level protection method. The main difference is the hash function which becomes a new calculation task. The initial plan for second level protection is to use the idle time space of CPU to calculate the hash value of Fragment 1 of the image to be protected along the scheme (\figurename~\ref{design2lvl}). This is the right option as long as the laptop GPU is limited and cannot calculate hash function fast enough. However, \tablename \ref{c4_1lvl_dct_aes_sha} shows that
the SHA-512 calculation becomes the key element and our time overlay design in~\figurename~\ref{c4_overlay2} will not suit anymore as we would love to let GPU hold at time Period 2 to wait for the hash calculation. So the overlay design in~\figurename~\ref{c4_overlay2} should be modified.

It is important to evaluate time cost for all calculation tasks for second level protection in laptop scenario. As shown in~\tablename~\ref{c4_1lvl_dct_aes_sha}, the execution time of SHA-512 algorithm and AES on CPU is still possible but it has to be covered by the execution time of DCT and iDCT on GPU which makes a new overlay design still possible in the case of a series of images as input.

\newpage

\begin{table}[htbp!] 
\caption{DCT time for one input image on GPU; AES and SHA-512 time for selected DCT coefficients on CPU for the laptop use case.}%
\centering
\label{c4_1lvl_dct_aes_sha}
\begin{tabular}{c c c c c}
\toprule

Image size & $1024\times 768$ & $1600\times 1200$ & $3240\times 2592$ & $4800\times 4800$  \\
\hline
DCT on laptop GPU & 0.41 ms & 0.79 ms & 3.67 ms & 9.98 ms  \\

AES on laptop CPU & 0.19 ms & 0.47 ms & 2.05 ms & 5.87 ms \\ 

SHA-512 on laptop CPU & 0.29 ms & 0.73 ms & 2.8 ms & 7.69 ms \\ 

\bottomrule
\end{tabular}
\end{table}

\begin{figure}[htbp!] 
\centering    
\includegraphics[width=1\textwidth]{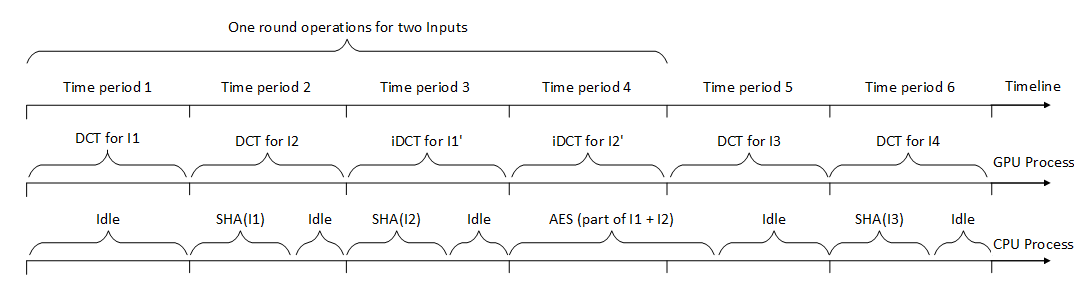}
\caption{Time overlay design for the second level protection.}
\label{c4_2lvl_overlay}
\end{figure}

If we consider the two same size bitmaps (e.g. two $512 \times 512$ bitmap images) as the input together, while the SHA function for the first image is calculated by the CPU (one hash per block), the GPU will calculate the DCT for the second input image in parallel. For every two DCT operation on GPU, the GPU turns to calculate the iDCT for the first image (XORed with the hash results) while CPU continues the SHA calculation for the second input image. And AES will be performed for the selected coefficients of two images together while iDCT is performed for the second image. The time flow is shown in~\figurename~\ref{c4_2lvl_overlay} to elaborate the whole design. The 'Idle' period in the CPU timeline can be considered as the redundancy prepared for the possible delay caused by the memory transfer between host memory and GPU memory (this memory transfer is always controlled by the CPU instructions).

The evaluation of performance is shown in~\tablename~\ref{c4_1lvl_aes_gpu_laptop}. The execution time for one input image is not exactly twice of the DCT time on GPU because the DCT and iDCT are not exactly the same. The evaluation shows that the total run time is almost the half of the DCT speed on GPU which is over 1.1GB/s (faster than full AES on CPU in~\cite{dai2007crypto++}).

\begin{table}[htbp!] 
\caption{Speed of full AES for the input image on CPU, our SE design, and AES for the input image on GPU for laptop scenario.}%
\centering
\label{c4_1lvl_aes_gpu_laptop}
\begin{tabular}{c c c c c}
\toprule

Image size & $1024\times 768$ & $1600\times 1200$ & $3240\times 2592$ & $4800\times 4800$  \\
\hline
AES on laptop GPU & 5.5 ms & 13.5 ms & 59.2 ms & 162.3 ms  \\

AES on laptop CPU & 2.1 ms & 5.0 ms & 21.9 ms & 60.2 ms  \\

SE on laptop CPU + GPU & 0.89 ms & 1.94 ms & 8.38 ms & 20.91 ms  \\

\bottomrule
\end{tabular}
\end{table}

However, as shown by~\citet{li2012implementation} and~\citet{gervasi2010aes}, GPUs can also accelerate AES computation. However, as shown in~\tablename~\ref{dcton2gpus}, GPU run times vary widely according to their architecture. In~\tablename~\ref{c4_1lvl_aes_gpu_laptop}, we compare the performance of full encryption using AES on CPU and GPU with our SE method. Due to the limitation of the GPU compute capability on laptop, the AES on GPU is even slower than on CPU. This means the design of calculation task arrangement is highly based on the hardware configuration. And for a low-end GPU laptop use case, using GPU and CPU in parallel for both first and second level SE methods is the best implementation option.

\subsection{Allocation of calculation tasks for a powerful GPU (desktop)}

We saw in Chapter 3 that the evolving of the GPUs are so fast that the computation allocation can vary over time according to GPU configurations. In our work, this fact makes it possible to do all SE steps including DCT $8\times 8 $ and AES on the GPU in some special situations. The situation worth dealing with the situation on most desktops today where GPUs are so powerful that can calculate DCT for all piece of data still faster than AES for only a small part of data on CPU. This leads to a serious question: the overlay design of~\figurename~\ref{c4_overlay2} and~\figurename~\ref{c4_2lvl_overlay} are not working anymore. Evaluations in~\tablename~\ref{overlay_problem} shows that, on Nvidia GTX 780, the GPU time period for computing iDCT is so short that it cannot cover for the calculation of AES on CPU.

\begin{table}[htbp!] 
\caption{Run time in period 2 on desktop GPU and CPU.}%
\centering
\label{overlay_problem}
\begin{tabular}{c c c c c}
\toprule

Image size & $1024\times 768$ & $1600\times 1200$ & $3240\times 2592$ & $4800\times 4800$  \\
\hline
iDCT on desktop GPU & 0.04 ms & 0.09 ms & 0.41 ms & 1.12 ms  \\

AES on desktop CPU & 0.16 ms & 0.38 ms & 1.72 ms & 4.67 ms  \\

\bottomrule
\end{tabular}
\end{table}

According to~\citet{li2012implementation}, AES speed can reach more than 50 Gbps on an Nvidia GPU of a desktop machine with CUDA implementation (In our work, it can reach almost 40 Gbps on our desktop as shown in~\tablename~\ref{aes_cpu_gpu_1lvl}). In such a situation, the parallel design in~\figurename~\ref{c4_overlay2} is definitely not suitable anymore as we are getting a scheme of GPU always idle and CPU always fully used. Although the performance is still better than the full AES on CPU, the hardware resource in GPU is not fully exploited. 

In fact, according to these evaluations, we move all SE calculations including DCT and AES to GPU for the first level of SE. This design uses GPU to work three steps in sequential for each input image: DCT for original image, AES for selected coefficients and iDCT for the rest coefficients. In~\tablename~\ref{aes_cpu_gpu_1lvl}, we list the evaluation for full encryption on CPU and GPU compared with SE on GPU.




\begin{table}[htbp!] 
\caption{Speed of AES on CPU and GPU, our SE (first level protection) on GPU.}%
\centering
\label{aes_cpu_gpu_1lvl}
\begin{tabular}{c c c c c}
\toprule

Image size & $1024\times 768$ & $1600\times 1200$ & $3240\times 2592$ & $4800\times 4800$  \\
\hline
AES on desktop GPU & 0.19 ms & 0.46 ms & 1.91 ms & 5.46 ms  \\

AES on desktop CPU & 1.56 ms & 3.8 ms & 16.6 ms & 45.5 ms  \\

SE on desktop GPU & 0.10 ms & 0.21 ms & 1.01 ms & 2.76 ms  \\

\bottomrule
\end{tabular}
\end{table}

We can see that the SE we use is still faster than naïve AES on either CPU or even on GPU. This results benefits from the idea that all the calculations of SE are moved to the GPU. Based on these observations, we can see that using a GPU as an accelerator for our SE algorithm is always a better choice compared with naïve AES. The main reason for this situation is because although the AES can be accelerated a lot by GPU, the DCT $8\times 8$ calculation itself suits better than AES to the GPU design. A deeper reason is that DCT $8\times 8$ is optimized by many previous works that the calculation is adapted to fit Nvidia GPU architecture; in the mean time, the design of AES algorithm utilized logic operations at the bit level which is not as easy as DCT to optimize for GPU. This main difference makes AES always slower than DCT on the same GPU platform. Moreover, the SE method of both the first and the second level protection generates only about 13\% of original data to do the AES operation which indeed does not add much burden for calculation.





For the strong level of protection method on desktop, the only difference is how to allocate the hash calculation task. As pointed out in Chapter 3, according to tests based on programs in~\citet{steube2013oclhashcat}, the SHA-512 performance on the desktop GPU, Nvidia GeForce GTX 780 is around 136 MH/s (means 136 million hash calculation per second). We should notice that for each $8\times 8 $ block, there will be one hash calculation, so we can evaluate the run time by SHA-512 as in~\tablename~\ref{hash_gpu}.

\begin{table}[htbp!] 
\caption{Speed estimation of SHA-512 of once per $8 \times8 $ block on desktop GPU.}%
\centering
\label{hash_gpu}
\begin{tabular}{c c c c c}
\toprule

Image size & $1024\times 768$ & $1600\times 1200$ & $3240\times 2592$ & $4800\times 4800$  \\
\hline
SHA-512 once per block  & 0.09 ms & 0.22 ms & 0.96 ms & 2.65 ms  \\

\bottomrule
\end{tabular}
\end{table}

The speed is much faster than SHA-512 on CPU based implementation from~\cite{dai2007crypto++}. Also, SHA-512 implementation on CPU is too slow to fit the overlay design shown in~\figurename~\ref{c4_2lvl_overlay} in our desktop scenario. In the evaluation of second level protection, we allocate this hash task to the GPU also. In the end, we compare the strong level of protection method on GPU with AES on GPU in~\tablename~\ref{c4_2lvl_se_final}.

\begin{table}[htbp!] 
\caption{Evaluation of AES on GPU, our SE (strong level of protection) on GPU.}%
\centering
\label{c4_2lvl_se_final}
\begin{tabular}{c c c c c}
\toprule

Image size & $1024\times 768$ & $1600\times 1200$ & $3240\times 2592$ & $4800\times 4800$  \\
\hline

AES on desktop CPU & 1.56 ms & 3.8 ms & 16.6 ms & 45.5 ms  \\
AES on desktop GPU & 0.19 ms & 0.46 ms & 1.91 ms & 5.46 ms  \\

SE level 2 on GPU & 0.19 ms & 0.43 ms & 1.97 ms & 5.41 ms  \\

\bottomrule
\end{tabular}
\end{table}

~\tablename~\ref{c4_2lvl_se_final} shows that the SE in second level protection mode on desktop GPU has the same performance as AES-128 on GPU. In summary, the allocation of calculation of HASH and AES tasks always depends on the computation capacity of the GPU while the DCT task always can be allocated to the GPU.

\section{Discussions}

In this chapter, we implemented two levels of selective encryption methods both using DCT $8\times 8$ preprocessing based on GPU acceleration. We defined a first level of protection which is lightweight and is designed to disguise image quality. Then, we defined a strong second level of protection that can provide a good level of security.

The first level of protection design combines CPU and GPU resources available on most PCs, tablets, or even smartphones today. It provides a very fast speed to perform selective encryption in the frequency domain for bitmap images. The second level of protection method addresses the issue with better protecting the public fragment which is left plain in the first level of protection. The idea is to use a small number of high frequencies to rapidly protect the low frequencies of the public fragment; indeed, the second level of protection implementation also uses the acceleration offered by the GPGPU. Evaluations show that it is about twice faster than AES on a laptop; as \tablename \ref{c4_2lvl_se_final} shows that SE performance are comparable to AES with a high-end GPU as the ones equipped on a desktop. By two different statistical analyses, it shows that the second level protection method offers a good level of protection to resist recovery.

The separation of an image data into a private fragment and a public and protected fragment can be used to address the issue with efficiently protecting large amount of bitmap images using but not completely trusting remote storage servers like a cloud storage provider. We separate the original data as putting the important private fragment to be stored locally and putting the remaining fragment protected to a remote server. For instance, in a cloud with the additional protection offered by the cloud provider. Doing so, we make the best usage of the local memory where we store only about 13\% of the data depending of a tunable number of coefficients selected to constitute the private fragment. To perform one or the other of the two methods, we refined the implementation architecture using both the GPU and the CPU available on a PC and reach a level of performance that much faster than CPU based AES and comparable with GPU based AES and never slower.

Indeed, one have to realize that GPGPU architectures as well as encryption algorithms are progressing at a fast pace. For instance, late in 2014, a new generation Nvidia Geforce series GPUs (http://www.nvidia.com/object/geforce\_family.html) was released with more CUDA cores, higher clock frequency and wider memory bandwidth, improving effective speed by 40\% compared to GPU for desktop we used (manufactured in 2013). And this increase keeps showing up in 2015 and 2016 with different generations of calculation core architecture introduced, faster memory equipped. We are convinced that performance for computing the DCT $8\times 8$ and other algorithms benefiting from GPU like SHA-512 or even AES will still progress. As pointed by~\citet{gregg2011data}, the memory transfer between host and GPU memory could be a bottleneck due to the limitation of the PCI Express bus connecting them (normally several GB/s). During this work, we have seen that unfortunately, this can influence the load to assign to the CPU vs. the GPU in order to obtain the best performance. This would suggest developing a software adaptor to smartly allocate the computation task according to the hardware architecture available. As pointed in Chapter 3, the mobile platform that lets CPU and GPU use the same memory which maybe a solution to this problem. However, it is still difficult to see solutions for replacing PCIE bus showing up in today's PC manufactures.

Nonetheless, our work is clearly showing that selective encryption can potentially become widely used for bitmap image protection since it provides excellent processing time, a minimal visual content loss, a good level of protection by fragmenting bitmap into two separate storage space, and a moderate increase of its total memory storage.



\chapter{DWT for general purpose protection}

\ifpdf
    \graphicspath{{Chapter5/Figs/Raster/}{Chapter5/Figs/PDF/}{Chapter5/Figs/}}
\else
    \graphicspath{{Chapter5/Figs/Vector/}{Chapter5/Figs/}}
\fi

In this chapter, first we present the GPGPU based acceleration of DWT-2D. Then, the design is given with both general architectures and details. The security analysis followed, for different types of files, are presented to prove the good effect of our design. Then, the benchmark section gives a general comparison of our experimentation on two different computer platforms with other encryption algorithms. At last, we present the use case with the transmission and secure sharing architecture. 

\section{Discrete wavelet transform and GPU acceleration}

In previous sections, DCT (Discrete Cosine Transform) was used to support fragmentation decision before performing encryption for bitmap image protection. However, DCT cannot guarantee the total losslessness due to conversions between integers and floating point numbers which will result in rounding errors (sometimes even recursively). These rounding errors can be reduced by using more storage space with more detailed designs but cannot be totally avoided. This is the reason why DCT cannot provide the integrity required for dealing with any kind of data type.  

Discrete Wavelet Transform (DWT) \cite{burrus1997introduction} is sometimes used in selective encryption (see previous work \cite{gonccalves2015survey} \cite{sadourny2003proposal} \cite{pommer2003selective}), but most of the time it is used as a standard compression step for formatting rather than as a preprocessing step for selecting in multimedia use cases. In our design, DWT is used as a preprocessing step before fragmentation with a special filter Le Gall 5/3 \cite{burrus1997introduction} which has an important lossless property by mapping integers to integers. The DWT-2D based on Le Gall 5/3 filter fits best for our design for it can provide data integrity and also be efficient both in performance and storage space usage.

Performance against full encryption is constantly required. The transform used in the preprocessing step of SE can legitimately be removed from the benchmark when SE and compression are integrated and that transform is used by both applications. In these cases, SE performs a light weight protection within the compression or coding process for a specific format like MPEG4 \cite{richardson2004h} or JPEG2000 \cite{christopoulos2000jpeg2000}. However, our use case aiming at dealing with any kind of data will have to take into account the entire process when it comes to performance evaluation since it should be able to deal with agnostic data type and even formatted data. This will lead us to implement DWT on a GPGPU to benefit from its acceleration \cite{franco2009parallel}. 

%


\subsection{DWT}

\begin{figure}[htbp!] 
\centering    
\includegraphics[width=1\textwidth]{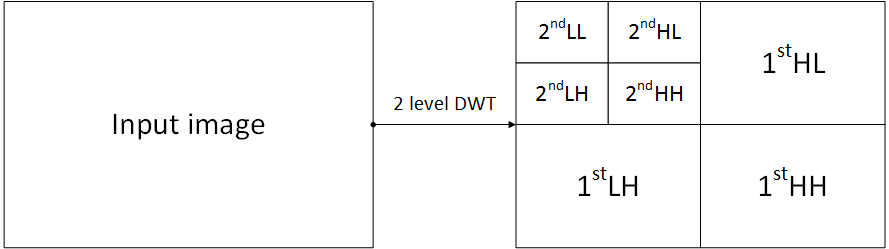}
\caption{Two level Discrete Wavelet Transform 2D result \cite{christopoulos2000jpeg2000}.}
\label{c5_dwtlvl2_max}
\end{figure}

DWT is a signal processing technique for extracting information mostly used in compression standard such as JPEG2000~\cite{christopoulos2000jpeg2000}. It can represent data by a set of coarse and detail values in different scales. Naturally, it is a one-dimensional transform. But, it also can be used as a two-dimensional transform as applied in the horizontal and vertical directions. For the image case, this DWT-2D transform will generate four small images which each one is one quarter size of the original image with one level transform: one with low resolution (LL), one with high vertical resolution and low horizontal resolution (HL), one with low vertical resolution and high horizontal resolution (LH), and one with all high resolution (HH). Then the second level transform will only be performed for the first quarter ('LL' part) of the first level’s result which is called dyadic decomposition as shown in ~\figurename~\ref{c5_dwtlvl2_max}.

To perform the forward DWT, a one-dimensional sub-band is decomposed into a set of low-pass samples and a set of high-pass samples. In our design, the \lq\lq{}Le Gall 5/3 filter\rq\rq{} by~\citet{burrus1997introduction} is used so no data will be lost due to numerical rounding. And the lifting-based filtering scheme \cite{acharya2006survey} is used which updates odd sample values with a weighted sum of even sample values, and updating even sample with a weighted sum of odd sample values. The lifting-based filtering for the 5/3 analysis filter is achieved by using equations (5.1) and (5.2):

\begin{equation}
y(2n+1) = x_{ext}(2n+1) -  \lfloor{\frac{x_{ext}(2n)+x_{ext}(2n+2)}{2}}\rfloor \qquad
\end{equation}
\begin{equation}
y(2n) = x_{ext}(2n) -  \lfloor{\frac{y(2n-1)+y(2n+1) +2}{4}}\rfloor \qquad
\end{equation}

where $x_{ext}$ is the extended input signal, $y$ is the output signal and $\lfloor{a}\rfloor$  indicates the largest integer not exceeding $a$.

DWT can be performed at different levels, we chose a two-level DWT as illustrated in~\figurename~\ref{c5_dwtlvl2_max}. In ~\figurename~\ref{SE1block}, the selected coefficients to build the private fragment are the $2^{nd}$ LL which takes about 1/16 of the storage space and carries the basic elements (coarse information) of the original image. The reason for using two-level DWT is that the one-level DWT still has a large low frequency part (1/4 of the whole DWT-2D result) to be protected. Three or more DWT levels make the value range of the high frequency coefficients too large to waste more storage space (more details about value range given later in this section).



\subsection{DWT acceleration based on GPGPU}

There are two main categories of DWT implementations on hardware: by convolution operations \cite{mallat1989theory} and by the lifting scheme \cite{sweldens1998lifting}. In early GPU-based DWT implementations (\citet{hopf2000hardware}, \citet{garcia2005gpu}), the convolution operations were preferred as the early developing tools for GPU such as OpenGL or Cg. The performance gain compared with CPU implementations was limited due to not only the limited GPU calculation capacity but also the lack of a general purpose GPU development platform to fully exploit the GPU parallel computing resources. In fact, the lifting scheme is more suitable for GPU computation as each coefficients in this scheme is computed using the coefficient that in the even or odd position and its two neighbours. As a consequence, all coefficients can be calculated without dependencies therefore can be performed in parallel. This important feature is not fully used until CUDA is released~\cite{kirk2007nvidia}. Not only it provided operation level parallelism, but it also gave access to arbitrary memory operations. CUDA architecture along with the Nvidia GPGPUs accelerates DWT 10 to 20 times faster than an optimized CPU implementation (multi-core CPU based on OpenMP) in 2009 \cite{franco2009parallel}.

The reason why GPGPU brought such a huge performance acceleration is because unlike modern CPUs with only a few powerful physical cores (4 or 8 on a Intel CPU for PC) that allows only limited number of actually parallel threads, a GPGPU could contain hundreds even thousands of threads at the same time. This can fit the scheme based feature to allow each of the output coefficients calculated separately with a hardware level parallelism. The CUDA platform allows to realize the first generation of DWT-2D implementation by simply using the parallel computing cores in CUDA enabled GPGPU since 2009. 

More optimization came out after 2009 like optimizing the memory usage to avoid the time consuming operations like matrix transpose. For example, the calculation rounds in \cite{franco2009parallel} are to load the data from global memory to shared memory and calculate the horizontal direction of DWT; then the result matrices are loaded back to global memory and transposed to become the input data for the next round. In the next round, the same calculation operations are performed as the data are transposed so the vertical DWT can be easily done. This method loads data between fast shared memory and slow global memory twice and transposes the matrix once which is not efficient as pointed out by \citet{enfedaque2015implementation}. 

Further improved methods \cite{matela2009gpu} explored minimizing memory transfers between the global memory and the shared memory by computing both the horizontal and the vertical filtering in one step. The improvement is based on carefully arranging the input matrix into rectangular blocks before computing and loading them to the shared memory by a thread block. Then both the horizontal and the vertical DWT can be calculated in these blocks which successfully avoided the matrix transpose or further memory transfers. However, there is a drawback of such an approach: adjacent blocks have data dependencies that can only be avoided by extending all blocks with some rows and columns that overlap with adjacent blocks.

More recent research \cite{van2011accelerating} shows better performance with more optimization in handling the problem of data block dependencies by more memory transfer steps. And the result shows that a 10 to 14 times speedup can be reached compared with a CPU implementation using instruction level accelerations (MMX and SSE extensions). In 2015, the fastest implementation of the DWT found in the literature is proposed by \citet{enfedaque2015implementation}. In this chapter, an optimized scheme is implemented with the state-of-the-art hardware (Nvidia GTX Titan GPU). The DWT-2D with “Le Gall 5/3” filter for a $4096\times4096$ image can be done in 0.467 ms.

In summary, as pointed out in Chapter 3, the development for CUDA enabled GPGPU is highly dependent on both the software and the hardware architecture. In recent years, there is continuing optimization for DWT as different CUDA versions and different GPGPU are released. 

In our implementation, the lifting scheme is used and the main method according to \citet{van2011accelerating}. As we are not focusing on the best optimization of implementing DWT-2D on GPGPU, the performance evaluations are just based on the hardware we have instead of the most powerful GPGPUs.

\section{Design of DWT based SE}

\vspace{-1ex}

\subsection{Designs}

\vspace{-1ex}

\begin{figure}[htbp!] 
\centering
\includegraphics[width=1\textwidth]{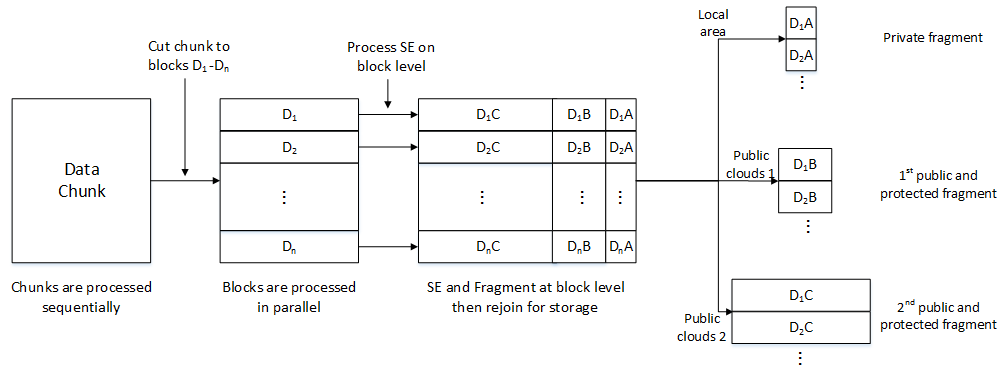}

\caption{SE General method for processing large amount of data.}
\label{fig:SEgeneral}
\end{figure}

In our design \cite{qiu2017efficient}, as shown in  \figurename~\ref{fig:SEgeneral}, in order to deal with sizable input data it is being proposed to cut it into several chunks of the same given size 2D matrix (e.g. seen as a $512\times512$ or $1024\times1024$ byte block which is chosen to accommodate further  transformation or the hardware platform architecture). Then every chunk ($D_{i}$) goes to the SE process to generate three fragments which are the private fragment $D_{i}A$, the first public and protected fragment $D_{i}B$, and the second public and protected fragment $D_{i}C$. Then the $D_{i}A$ fragments go to the trusted area like a local machine under the user’s control and the $D_{i}B$, $D_{i}C$ fragments may be transmitted to the public area like a public cloud with little fear of an attack since $D_{i}B$, $D_{i}C$ are supposed to carry little information and also to be protected. This will be shown later in Section 5.3 where a number of security analysis will be performed.

\vspace{-2ex}

\vspace{-2ex}

The main idea is to consider every block $D_{i}$ as a matrix and be treated as such by the SE process. That is to say any kind of data formats can be seen as a matrix by considering every byte of data as a pixel to form a bitmap gray scale image. Then every "image" $D_{i}$ is simply processed using the SE method block by block with block size $8\times8$ shown in \figurename~\ref{SE1block}. The block size chosen can be changed according to the size of the original data. This tiling step is used to achieve a nice fitting with the GPGPU architecture (will be mentioned later).

\begin{figure}[htbp!] 
\centering
\includegraphics[width=1\textwidth]{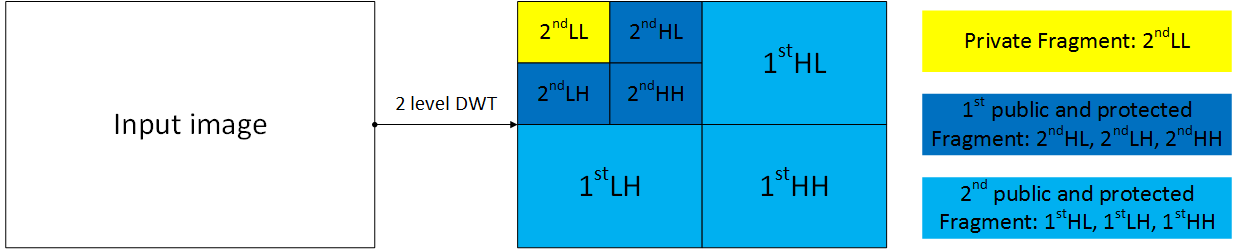}
\caption{DWT-2D and fragmentation process for the single $8\times8$ block.}
\label{DWT2D_fragments}
\end{figure}

\begin{figure}[htbp!] 
\centering
\includegraphics[width=1\textwidth]{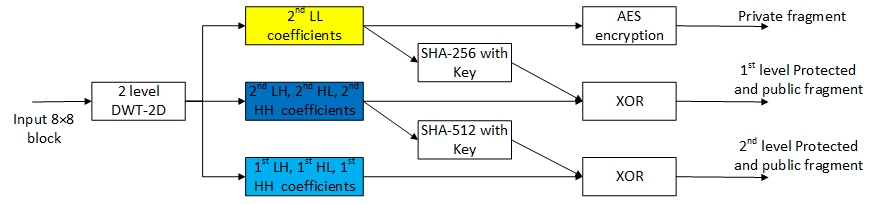}
\caption{SE process for the single $8\times8$ block.}
\label{SE1block}
\end{figure}

The first step for the $8\times8$ block is to do the Discrete Wavelet Transform (DWT). In our work we perform two successive levels of the DWT with the Le Gall 5/3 filter so the low frequency coefficients which are considered as the private fragment (shown in \figurename~\ref{DWT2D_fragments}). This fragment takes only $4$ out of $64$ coefficients (with $k=4$ in our implementation) but carries most of the original frequency feature. The AES-128 bit~\cite{giraud2004dfa} will be used (In our design, however, the implementation is structured such that another encryption algorithm can easily replace AES-128 if need be) to protect this fragment if further transmission is needed. 

\vspace{-1ex}

\vspace{-2ex}

Then the private fragment of each $8\times8$ block will be used to generate a 256-bit sequence by using SHA-256~\cite{toolkit2006secure} which can guarantee very different bit sequence generated even when the corresponding coefficients of the private fragments in neighbor blocks are very similar (encryption key is used to guarantee the key sensitivity in \figurename~\ref{SE1block}). This bit sequence is used to protect the $1^{st}$ public and protected fragment (the rest coefficients of $2^{nd}$ level DWT shown in \figurename~\ref{DWT2D_fragments}) by performing an XOR operation. This fragment is defined as the $1^{st}$ public and protected fragment as shown in \figurename~\ref{DWT2D_fragments}. For the rest DWT coefficients which forms the $2^{nd}$ public and protected fragment, a bit sequence generated from SHA-512~\cite{toolkit2006secure} results of $1^{st}$ public and protected fragment and encryption key is used to do protection.


\vspace{-1ex}

\vspace{-2ex}

The protection for the 'protected and public fragments' provided by a XOR operation is based on the randomness guaranteed by the HASH algorithms. For example, in bitmap case, as long as there are redundancies, the $L3$ coefficients could be very similar especially between neighbor blocks. However, the SHA-256 and SHA-512 will generate totally different bit sequences even when there is only one different bit in inputs. This randomness will be added by XORing the to the 'protected and public fragment' which is the next level of DWT coefficients and random hash value of the current fragment. For other kinds of files as input, this design also has good effect. More security analysis for the protection will be shown later.

\subsection{Evaluation of the storage necessary for DWT}

\subsubsection{Evaluation for the first level of DWT-2D transform}

\begin{figure}[htbp!] 
\centering    
\includegraphics[width=1\textwidth]{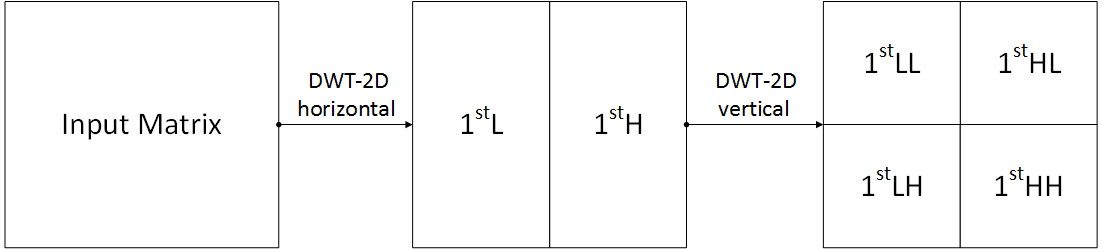}
\caption{Discrete Wavelet Transform 2D is calculated in two steps for the $1^{st}$ level.}
\label{c5_dwt2d_lvl1_max}
\end{figure}

Input matrix is a $8\times8$ matrix with all elements are 8-bit integers (consequently within a range 0 to 255 or -128 to +128). As shown in \figurename~\ref{c5_dwt2d_lvl1_max}, the Discrete Wavelet Transform has two steps: first, in horizontal direction. This will generate all coefficients in all lines: the first 4 are low frequency and last 4 are high frequency. It is easy to find out that the range for high frequency is -255 to +255 (double the range of input). And the range for low frequency is -192 to +192 (1.5 times of the input range). Then, the transform in vertical direction transforms the $1^{st} LL$ and $1^{st} H$ blocks respectively. This step will generate four blocks: $1^{st} LL$ (low frequency of the $1^{st} L$ block), $1^{st} LH$ (high frequency of the $1^{st} L$ block), $1^{st} HL$ (low frequency of the $1^{st} H$ block), $1^{st} HH$ (high frequency of the $1^{st} H$ block).

\vspace{-2ex}

It is easy to calculate the range for the $1^{st} HH$ is largest: $-511$ to $+511$ and the range for  $1^{st} LL$ is $1.5\times1.5 = 2.25$ times of the input range ($-128$ to $+127$). 

In summary, the possible max or min values of the $1^{st}$ level DWT-2D results can be stored with 10-bit long integers (-512 to +512).

\vspace{-2ex}

\subsubsection{Evaluation for the second level of DWT-2D transform}

\vspace{-2ex}

The $2^{nd}$ level transform only calculates the $1^{st} LL$ block (shown in \figurename \ref{c5_dwt2d_lvl1_max}) which is a $4\times4$ block. The calculation will also be performed in two directions. Finally, the transform generates three $4\times4$ blocks and four $2\times2$ blocks (\figurename~\ref{c5_dwtlvl2_max}).

In order to calculate the possible max and min value in the $2^{nd}$ level four blocks, we need to get the equations for the coefficients presented by the input values. In order to get the max and min values for the output, the input values are either -128 or +128 (reach the max abstract values). So the rounding process can be ignored as we just want to estimate the output values in this case. The equations (5.3) and (5.4) can be simplified as follows:

\begin{equation}
y(2n+1) = x_{ext}(2n+1) -  {\frac{x_{ext}(2n)+x_{ext}(2n+2)}{2}} \qquad
\end{equation}
\begin{equation}
y(2n) = x_{ext}(2n) -  {\frac{x_{ext}(2n-1)+x_{ext}(2n+1)}{4}} \qquad
\end{equation}

In this case, the transform can be calculated by using the input matrix to multiply the coefficients matrix. And for the $1^{st}$ level DWT, the coefficients matrix A is:

\[
\begin{bmatrix}   
    0.75 &    -0.125 &  0 &          0 &         -0.5 &  0 &      0 &      0\\
    0.5   &    0.25  &     0&           0 &          1 &      0 &      0 &      0\\
    -0.25  & 0.75    &  -0.125&   0 &          -0.5 & -0.5 &  0  &     0 \\
    0   &       0.25  &     0.25 &      0 &          0 &      1 &      0 &    0\\
    0   &       -0.125&   0.75 &     -0.125 & 0 &      -0.5 &  -0.5 &  0 \\
    0   &       0   &         0.25 &     0.25 &     0 &      0 &      1 &      0\\
    0   &       0   &        -0.125 &   0.625 &  0 &      0 &      -0.5 &  -1\\
    0   &       0   &         0 &           0.25 &      0 &      0 &      0    &   1\\

\end{bmatrix}
\]

If we define the input matrix as $IN$, the first step of wavelet for the horizontal direction can be presented as: $Output1 = IN
\cdot A$.
Then the vertical direction is calculated: In fact, it can be represented as $Output2 = Output1T\cdot A$ (use the transpose of $Output1$ to multiply matrix $A$ again) which will get the transpose of the result we want.
Then the coefficients of the $2^{nd}$ level DWT can be calculated:

\[
\begin{bmatrix}
0.75  & -0.125 & -0.5 & 0\\
       0.5 &     0.25 &     1   & 0\\
      -0.25 &   0.75 &  -0.5 & -1\\
       0    &    0.125 &   0 &   1\\
\end{bmatrix}
\]

It is the same to calculate the $2^{nd}$ level DWT but only on the $1^{st} LL$ block. If we define the final output matrix as $F$, we list the $F(4,4)$ as follows:

\begin{equation}
\begin{aligned}
F(4,4) = & 0.015625\times IN[2, 2] - 0.03125\times IN[2, 3] - 0.109375\times IN[2, 4] + 0.09375\times IN[2, 6] \\ 
&+ 0.03125\times IN[2, 7] - 0.03125\times IN[3, 2] + 0.0625\times IN[3, 3] + 0.21875\times IN[3, 4] \\
&- 0.1875\times IN[3, 6] - 0.0625\times IN[3, 7] - 0.109375\times IN[4, 2] + 0.21875\times IN[4, 3] \\
&+ 0.765625\times IN[4, 4] - 0.65625\times IN[4, 6] - 0.21875\times IN[4, 7] + 0.09375\times IN[6, 2] \\
&- 0.1875\times IN[6, 3] - 0.65625\times IN[6, 4] + 0.5625\times IN[6, 6] + 0.1875\times IN[6, 7] \\
&+ 0.03125\times IN[7, 2] - 0.0625\times IN[7, 3] - 0.21875\times IN[7, 4] + 0.1875\times IN[7, 6] \\
&+ 0.0625\times IN[7, 7]
\end{aligned}
\end{equation}

It is easy to calculate the max and min values if input of IN is within $-128$ to $+128$. The max value is $648$ and min value is $-648$. Then we calculate all the max values for all 16 values of $2^{nd}$ level DWT output and put them in the same matrix (min values are the same absolute value but negative):

\[
\begin{bmatrix}
338& 267 & 468 & 468\\
       267 & 211 & 369 & 369\\
      468 & 369 & 648 & 648\\
      468 & 369 & 648 & 648\\
\end{bmatrix}
\]

According to this calculation, there are only four values that could possibly exceed the range of $(-512, +512)$. The storage method for the four values should be designed as 11-bit long $(-1024, +1024)$. Then the $2^{nd} LL$ block (the private fragment) storage space is $40$ bits. The $2^{nd}HL, 2^{nd}LH, 2^{nd}HH$ blocks ($1^{st}$ public and protected fragment) take $40+40+44=124$ bits in total. The $1^{st}LL, 1^{st}HL, 1^{st}LH, 1^{st}HH$ blocks ($2^{nd}$ public and protected fragment) take $480$ bits.

\subsection{Storage space usage and numeric precision}

Because of the transformation step used for the SE, the footprint of the data before and after the transformation could be different. This would lead to the difference of the storage space usage or rounding errors caused by conversions between integers and floating point numbers. In~\citet{guan2005chaos}, the authors claim all variables are declared as type $double$ with a bit-length of 64 bits. This is unnecessary in our case as the input data are stored as integers especially $int$ type with a bit-length of 8 bits as the storage of the results will require 8 times more storage space compared with original data. In~\citet{qiu2015fast} we already designed how to optimize integer representation but still could not avoid possible rounding errors caused by the calculation of DCT.

In this chapter, the preprocessing step is the DWT based on \lq\lq{}Le Gall 5/3\rq\rq{} filter which is designed to be an invertible integer-to-integer map, such that the DWT Le Gall 5/3 is lossless. As a result, on one hand, any rounding error is avoided; on the other hand, the extra storage space usage caused by the $int$ to $float$ conversion does not exist. The only possible extra storage usage could be caused by the different value range of the input 8-bit $int$ and the output $int$ coefficients. And the output value range can be calculated as long as the input values are always stored Byte by Byte, the input value range (seen as unsigned value) is from $0$ to $255$ which can be considered as from $-128$ to $+127$ (the range is seen as from $-128$ to $+128$ during the following calculation). Then the storage methods can be optimized according to the value range distribution.

The first level DWT-2D transform is actually calculated by twice DWT-1D transforms (equation $(1)$ and $(2)$) on the $8 \times 8$ block in horizontal and vertical directions sequentially. The first horizontal transform generates two sub-matrices which are $1^{st}L$ and $1^{st}H$ that take each half of the result matrix horizontally. The vertical transform is done on each of the two sub-matrices which generates four sub-matrices like in Fig. 4 ($1^{st}LL, 1^{st}HL, 1^{st}LH, 1^{st}HH$). 

In the first horizontal transform, the range for $1^{st}H$ is -255 to +255 (double the range of input) and the range for the $1^{st}L$ is -192 to +192 (1.5 times of the input range). Then the transform in vertical direction, which is transform of the $1^{st}L$ and $1^{st}H$ blocks respectively, gets the following results: $1^{st}HH$ is from $-511$ to $+511$ and the range for $1^{st}LH$ is from $-384$ to $+384$. All the coefficients in the three sub-matrices of first level DWT-2D transform can be stored using 10-bits storage space.

The value range of second level DWT-2D coefficients are generated by the same two direction DWT-1D transform of the $1^{st}LL$ sub-matrices coefficients. Range of the second level DWT coefficients can be estimated by simplifying the equations (5.1) and (5.2) and then directly get results from calculating final formula of each elements in the four sub-matrices in ~\figurename~\ref{c5_dwtlvl2_max} ($2^{nd}LL, 2^{nd}HL, 2^{nd}LH, 2^{nd}HH$). The max and min values for each of the value estimated are shown in the following matrices. And the storage method for the second level DWT-2D coefficients is: 11-bit long for each of the lower left corner four coefficients ($2^{nd}HH$) and 10-bit long for rest of the coefficients.
 \vspace{-1ex}

\[
\begin{bmatrix}
    338       & 267 & 468 & 468  \\
    267      & 211 & 369 & 369 \\
    468      & 369 & 648 & 648 \\
    468      & 369 & 648 & 648 \\
\end{bmatrix}
,
\begin{bmatrix}
    -338       & -267 & -468 & -468  \\
    -267      & -211 &-369 & -369 \\
    -468      & -369 & -648 & -648 \\
    -468      & -369 & -648 & -648 \\
\end{bmatrix}
\]

As shown in ~\figurename~\ref{SE1block}, the private fragment we selected is the $2^{nd}LL$ DWT coefficients and all rest coefficients are fragmented into two public and protected fragments.


The storage design for the three fragments could be flexible. If the avalanche effect~\cite{webster1985design} must be a concern (communication channel is unreliable and transmission error rate is high), the private fragment and $1^{st}$ public and protected fragment should be stored locally to avoid avalanche effect. In such case, the storage space requirement locally for one block is 164-bits storage in total (40-bits for private and 124 bits for $1^{st}$ public and protected fragment) and cloud storage usage is 480 bits
(the $2^{nd}$ protected and public fragment). 
However, if the channel is reliable and error transmission are rarely to be seen, the $1^{st}$ public and protected fragment can also be put on clouds so the local storage is optimized which is important for a smart phone use case. Anyway, the total storage space usage is 644-bits for one block (initial 512 bits) which is about 26\% more but in both cases most of the data can be stored on clouds.

In summary, the preprocessing step is the DWT-2D based on \lq\lq{}Le Gall 5/3\rq\rq{} filter which is designed to be an integer-to-integer map, such that the DWT is lossless. As a result, on one hand, any rounding error is avoided; on the other hand, the extra storage space usage caused by the $int$ to $float$ conversion does not exist. Moreover, in our design, we consider any kind of data type as $int$ with bit-length of 8 bits. That is to say, no matter what kind of original data type it is, we process the data by reading one byte one time and deal with it as an 8-bit integer. Then the input bytes will form an "image" (2D matrix) of a configurable size ready for the whole SE process. In this process, the storage method of output data is carefully designed to provide integrity for any kind of input data.

\section{Security analysis}

A secure encryption algorithm ought to resist a various array of classic attacks \cite{nyberg1995provable, cho2011securing}. In this section, different security tests on the proposed scheme are performed to establish its high level of security.


The basic assumption is that the selected private fragment of data can easily be secured by using AES-128 (also, it is easy to replace AES-128 with any other encryption algorithms as in ~\figurename~\ref{SE1block}), so the security property of this private fragment is not analyzed here either it is stored locally or for further securing sharing. 

To validate a vast deployment (robustness) of the proposed method, the public and protected fragment which is stored on clouds in our use case are analyzed in terms of security performance to verify whether it reaches the required cryptographic performance. In \figurename \ref{SE1block}, the design is to put the two public and protected fragments on clouds. Thus, only these two fragments should be analyzed. However, according to our work, the security property of the $1^{st}$ public and protected fragment is very similar to the $2^{nd}$ public and protected fragment so we only present the results for the $2^{nd}$ public and protected fragment here.

In the following, we present the figures for the security analysis and all statistical results for an image, three kinds of video files, and English texts can be found in \tablename~\ref{statistical1_lenna}, \tablename~\ref{statistical1_video}, and \tablename~\ref{statistical1_text} respectively. As long as the video and text files are larger than single image files, the statistical results for the videos and texts are the average one of 100 randomly picked chunks (chunk size 1024KB) inside the video file contents.
Some criteria like PSNR and SSIM just suit for measuring images while not for text files.

Here the analysis measures on original data chunk and public and protected fragment used following are given: 1. \textit{Uniformity analysis} is given by calculating the Probability Density Function (PDF); 2. \textit{Information entropy analysis} is given by calculating the Entropy; 3. \textit{Correlation analysis} is given by calculating correlation coefficients; 4. \textit{Difference analysis} is to test the difference in bit level and also the Normalized Mutual Information (NMI); 5. \textit{Sensitivity analysis} is given by calculating the sensitivity when input plain-text and key changes; 6. \textit{Visual Degradation analysis} is only for bitmaps by calculating Peak Signal-to-Noise Ratio (PSNR) and Structural Similarity (SSIM); 7. \textit{Errors propagation} is also discussed.

\begin{table}[htbp!] 
\centering
\caption{Statistical results of sensitivity for public and protected fragment (stored in Cloud) for Lenna image.}
\label{statistical1_lenna}
\begin{tabular}{ c c c c c c}
\hline
\multicolumn{6}{c}{$\;\;\;\;\;\;\;\;\;\;\;\;\;\;\;$ Statistical results for images $\;\;\;\;\;\;\;\;\;\;\;\;\;\;\;$ }                 \\ \hline
\multicolumn{2}{c}{}   & Min  &  Mean &  Max & Std  \\ \hline
\multicolumn{2}{c}{PSNR} &  9.1943& 9.2303& 9.2616& 0.0093\\   
\multicolumn{2}{c}{SSIM} &   0.03& 0.0359& 0.0412& 0.0016\\ 
\multicolumn{2}{c}{$Dif$} &49.8886&49.9990&50.1077& 0.0351\\
\multicolumn{2}{c}{$KS$} &49.8943&50.0011&50.1280& 0.0347\\  
\multicolumn{2}{c}{$\rho2$} & 0.0189& 0.0193& 0.0196& 0.0001\\  
\multicolumn{2}{c}{$\rho-h $} &  -0.0614& 0.0002& 0.0492& 0.0156\\
\multicolumn{2}{c}{$\rho-v$} & -0.0515& 0.0001& 0.0522& 0.0154\\ 
\multicolumn{2}{c}{$\rho-d$} & -0.0448& 0.0005& 0.0580& 0.0157\\ 
\multicolumn{2}{c}{NMI} & 0.0189& 0.0193& 0.0197& 0.0027\\ \hline
\end{tabular}
\end{table}

\begin{table}[htbp!] 

\centering
\caption{Statistical results of sensitivity for public and protected fragment (stored in Cloud) for a English text (ASCII coding).}
\label{statistical1_text}
\begin{tabular}{c c c c c c}
\hline
\multicolumn{6}{c}{$\;\;\;\;\;\;\;\;\;\;\;\;\;\;\;$ Statistical results for texts$\;\;\;\;\;\;\;\;\;\;\;\;\;\;\;$ }                 \\ \hline
\multicolumn{2}{c}{}   & Min  &  Mean &  Max & Std  \\ \hline
\multicolumn{2}{c}{$Dif$} &49.7008&50.0042&50.3350& 0.0992\\
\multicolumn{2}{c}{$KS$} &49.7417&49.9978&50.4112& 0.1030\\  
\multicolumn{2}{c}{$\rho$} & -0.0145& 0.0002& 0.0192& 0.0055\\  
\multicolumn{2}{c}{$NMI$} &  0.0669& 0.0685& 0.0701& 0.0005\\\hline
\end{tabular}
\end{table}

\begin{table}[htbp!] 

\centering
\caption{Statistical results of sensitivity for public and protected fragment (stored in Cloud) for videos.}
\label{statistical1_video}
\begin{tabular}{c c c c c c}
\hline
\multicolumn{6}{c}{$\;\;\;\;\;\;\;\;\;\;\;\;\;\;\;$ Statistical results for MP4$\;\;\;\;\;\;\;\;\;\;\;\;\;\;\;$ }                 \\ \hline
\multicolumn{2}{c}{}   & Min  &  Mean &  Max & Std  \\ \hline
\multicolumn{2}{c}{$Dif$} &49.8211&49.9918&50.1684& 0.0632\\
\multicolumn{2}{c}{$\rho$} & -0.0098& -0.0004& 0.0078& 0.0031\\  
\multicolumn{2}{c}{$NMI$} &  0.1005& 0.1145& 0.1198& 0.0018\\\hline
\multicolumn{6}{c}{$\;\;\;\;\;\;\;\;\;\;\;\;\;\;\;$ Statistical results for MKV$\;\;\;\;\;\;\;\;\;\;\;\;\;\;\;$ }                 \\ \hline
\multicolumn{2}{c}{}   & Min  &  Mean &  Max & Std  \\ \hline
\multicolumn{2}{c}{$Dif$} &49.9985&49.8323&50.2008& 0.0731\\
\multicolumn{2}{c}{$\rho$} & -0.0090& 0.0001& 0.0085& 0.0039\\  
\multicolumn{2}{c}{$NMI$} &  0.1010& 0.1027& 0.1035& 0.0067\\\hline
\multicolumn{6}{c}{$\;\;\;\;\;\;\;\;\;\;\;\;\;\;\;$ Statistical results for RMVB$\;\;\;\;\;\;\;\;\;\;\;\;\;\;\;$ }                 \\ \hline
\multicolumn{2}{c}{}   & Min  &  Mean &  Max & Std  \\ \hline
\multicolumn{2}{c}{$Dif$} &49.8344&49.9994&50.1013& 0.0792\\
\multicolumn{2}{c}{$\rho$} & -0.0096& -0.0005& 0.0112& 0.0037\\  
\multicolumn{2}{c}{$NMI$} &  0.2616& 0.2745& 0.2773& 0.0022\\\hline
\end{tabular}
\end{table}

\newpage


\subsection{Uniformity Analysis}

The encrypted data should possess certain random properties such as uniformity, which is essential to resist against frequency attacks. Accordingly, the Probability Density Function(PDF) of the public and protected fragment should be as uniform as possible. This means that each symbol (pixels in image case) has an occurrence probability close to $\frac{1}{n}$, where $n$ is the number of symbols ($\frac{1}{256}=0.0039$ in byte level). We start by analyzing the image data and then other data to prove that the proposed method can attain the uniformity independently for its public and protected fragment.

\begin{figure}[htbp!] 
\centering
\includegraphics[width=0.7\textwidth]{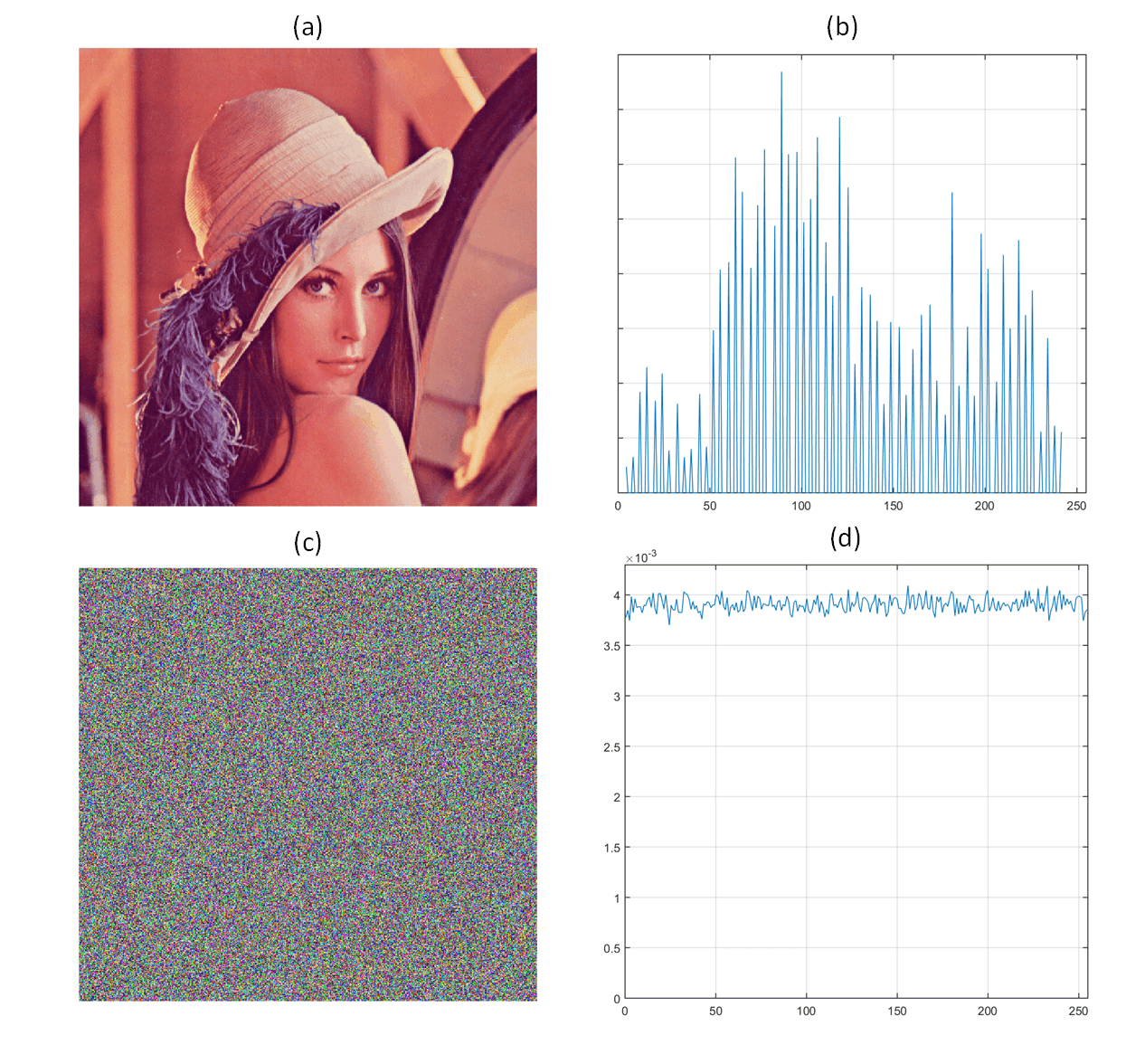}
\vspace{-1em}
\caption{(a) Original Lenna, (b) PDF of original Lenna512, (c) Public and protected fragment, (d) PDF of public and protected fragment.}
\label{fig:lenapdf}
\end{figure}

The original plain image Lenna and its corresponding PDF are shown in \figurename ~\ref{fig:lenapdf}-(a),(b). While,  in~\figurename~\ref{fig:lenapdf}-(c),(d), the corresponding fragment stored in cloud (c) to their corresponding PDF (d) is shown, respectively. It can be observed that the PDF of the public and protected fragment is close to uniform distribution since the probability of different symbols in the PDF figure are very different with the original one and tends to be uniform.

\begin{figure}[htbp!] 
\centering
\includegraphics[width=0.7\textwidth]{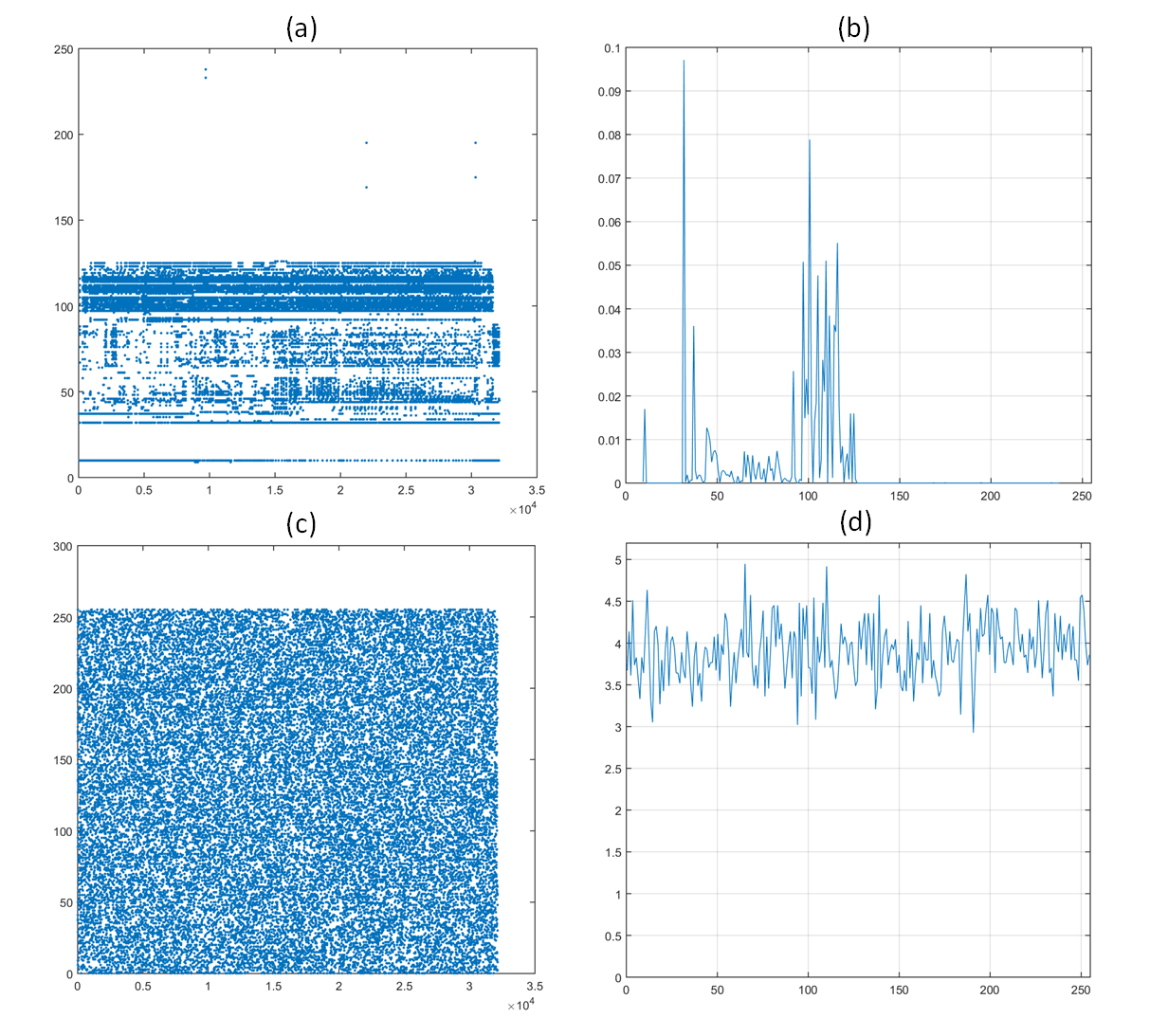}
\vspace{-1em}
\caption{Randomly chosen original text byte representation (a) and its PDF (b), Corresponding protected and public fragment (c) with its PDF (d).}
\label{fig:pdftext}
\end{figure}
\begin{figure}[htbp!] 
\centering
\includegraphics[width=0.66\textwidth]{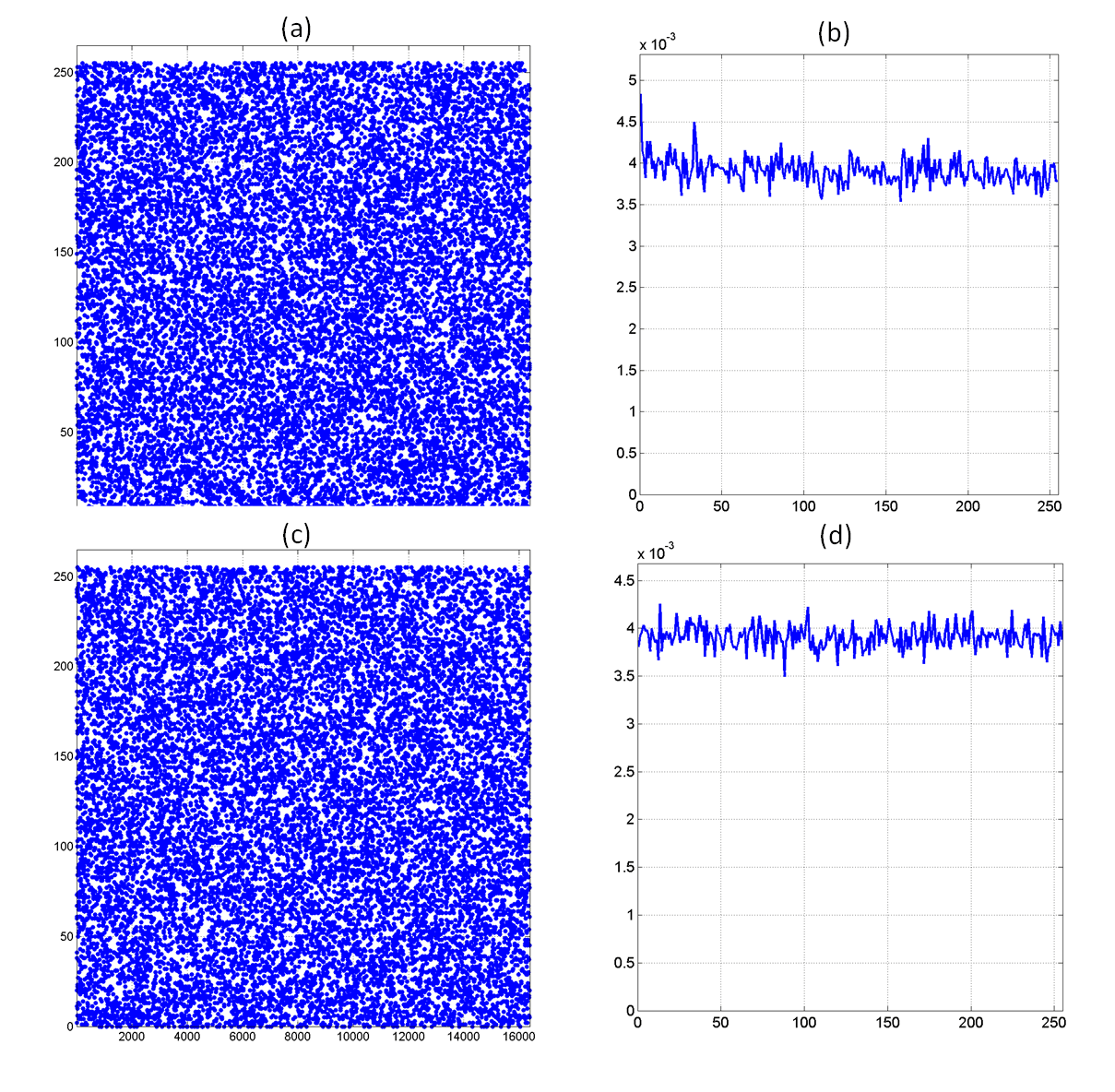}
\vspace{-1.5em}
\caption{Randomly chosen original MP4 file byte representation (a) and its PDF (b), Corresponding protected and public fragment (c) with its PDF (d).}
\label{mp4pdf}
\end{figure}
\begin{figure}[htbp!] 
\centering
\includegraphics[width=0.66\textwidth]{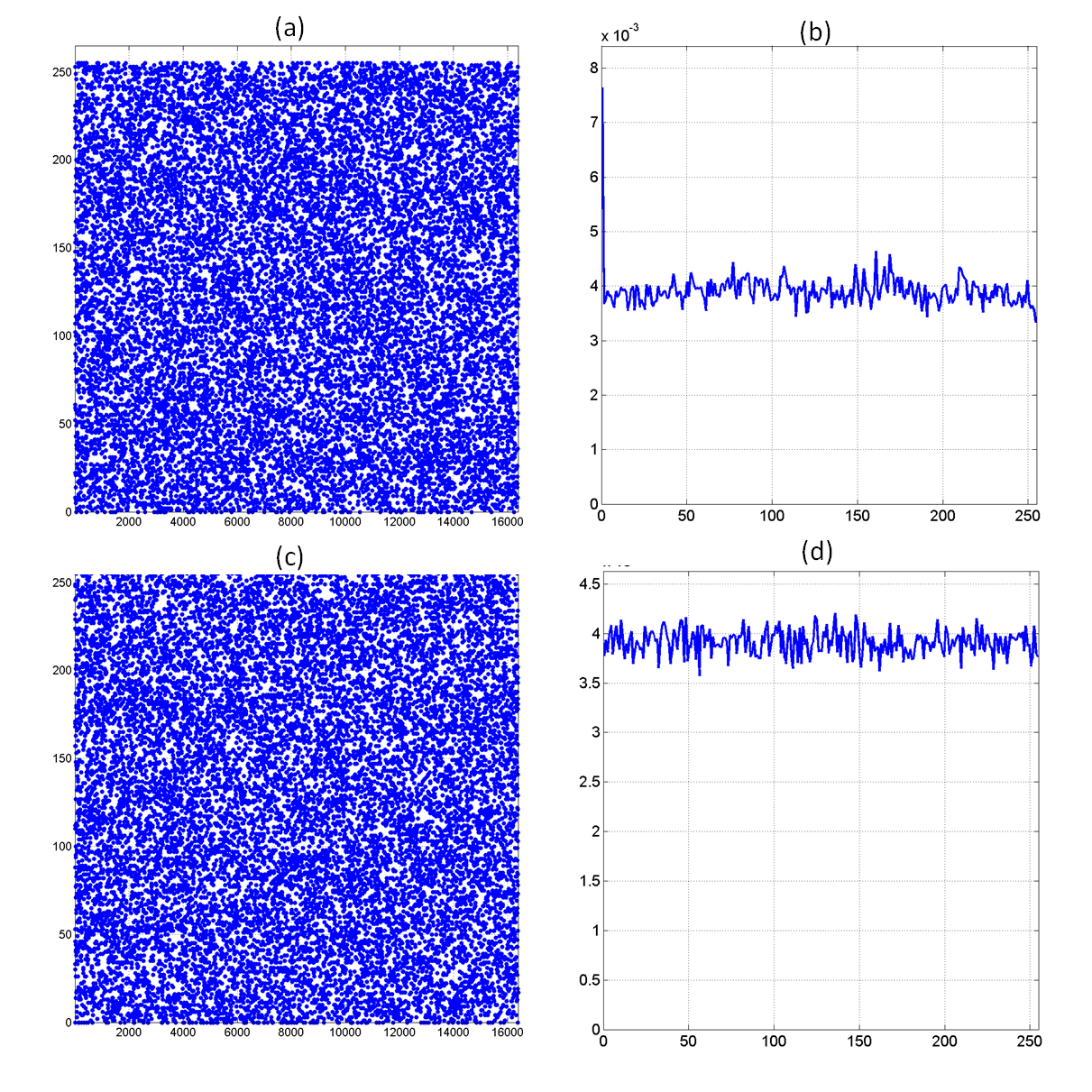}
\vspace{-1.5em}
\caption{Randomly chosen original MKV file byte representation (a) and its PDF (b), Corresponding protected and public fragment (c) with its PDF (d).}
\label{mkvpdf}
\end{figure}
\begin{figure}[htbp!] 
\centering
\includegraphics[width=0.66\textwidth]{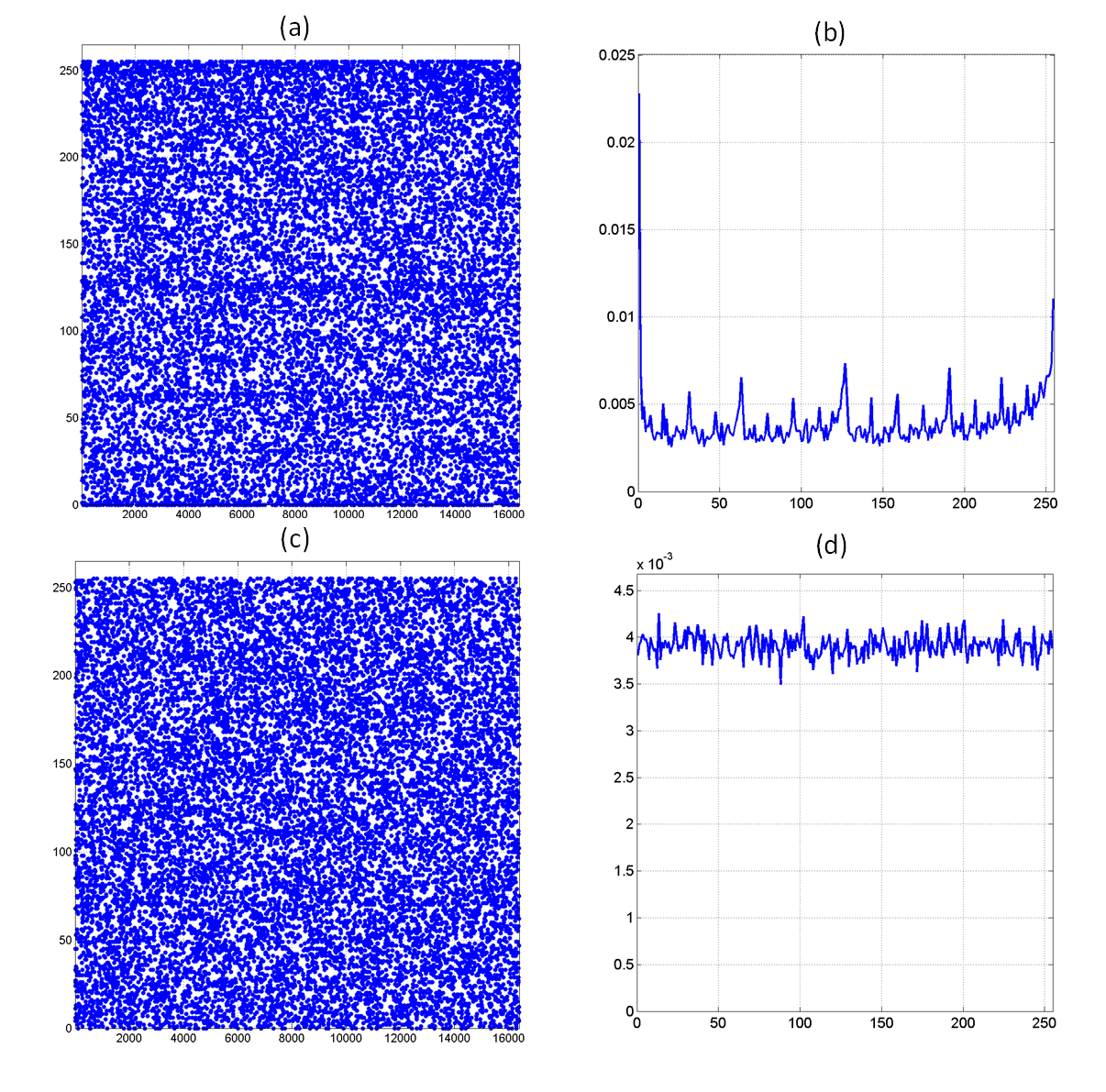}
\vspace{-1.5em}
\caption{Randomly chosen original RMVB file byte representation (a) and its PDF (b), Corresponding protected and public fragment (c) with its PDF (d).}
\label{rmvbpdf}
\end{figure}

Also, in \figurename~\ref{fig:pdftext}, the byte representation (read the data chunk byte by byte and form the matrix with each element is the value of the byte before DWT-2D) of an original chosen text file is presented in (a) and its corresponding PDF in (b)  in addition to its corresponding fragment byte representation that is stored in cloud (c) and with its corresponding PDF (d). The obtained result indicates that the public and protected fragment of the text file posses also a uniform distribution. The similar results can be observed in \figurename \ref{mp4pdf}, \figurename \ref{mkvpdf}, and \figurename \ref{rmvbpdf}.

From these results, we have shown that the distribution of the public and protected fragment tends to the uniform one no matter of the input data type. For the video cases, as long as the input chunks are video file contents which are already compressed and encoded, the original byte representations have no visual information. But the PDF of the public and protected fragments correspondingly have shown a tend of uniform distribution.

Moreover, to validate this result, an entropy test is realized in the sub-matrix level of size $8\times 8$ (same size of the input block).

\newpage
\subsection{Information Entropy Analysis}

The information entropy of a data sequence $M$ is a parameter that measures the level of uncertainty in a random variable~\cite{zhang2011novel} and is expressed in bits, defined using equation (5.6):

\begin{equation}
H(m)= -\sum_{i=1}^{n} p(m_i)\log_2 \frac{1}{p(m_i)}
\end{equation}

where $p(m_{i})$ denotes the probability of symbol $m_{i}$. It is easy to calculate for a random source emitting $2^{N}$ symbols, the entropy should be $N$. In this design, as the data are always seen as 8-bit per element, the pixel data have $2^{8}$ possible values. As such, the entropy for a "true random" information source must be 8. For the bitmap case, the entropy of the public and protected fragments for different images are always more than 7.999 which proves high randomness.

\begin{figure}[htbp!] 
\centering
\includegraphics[width=0.8\textwidth]{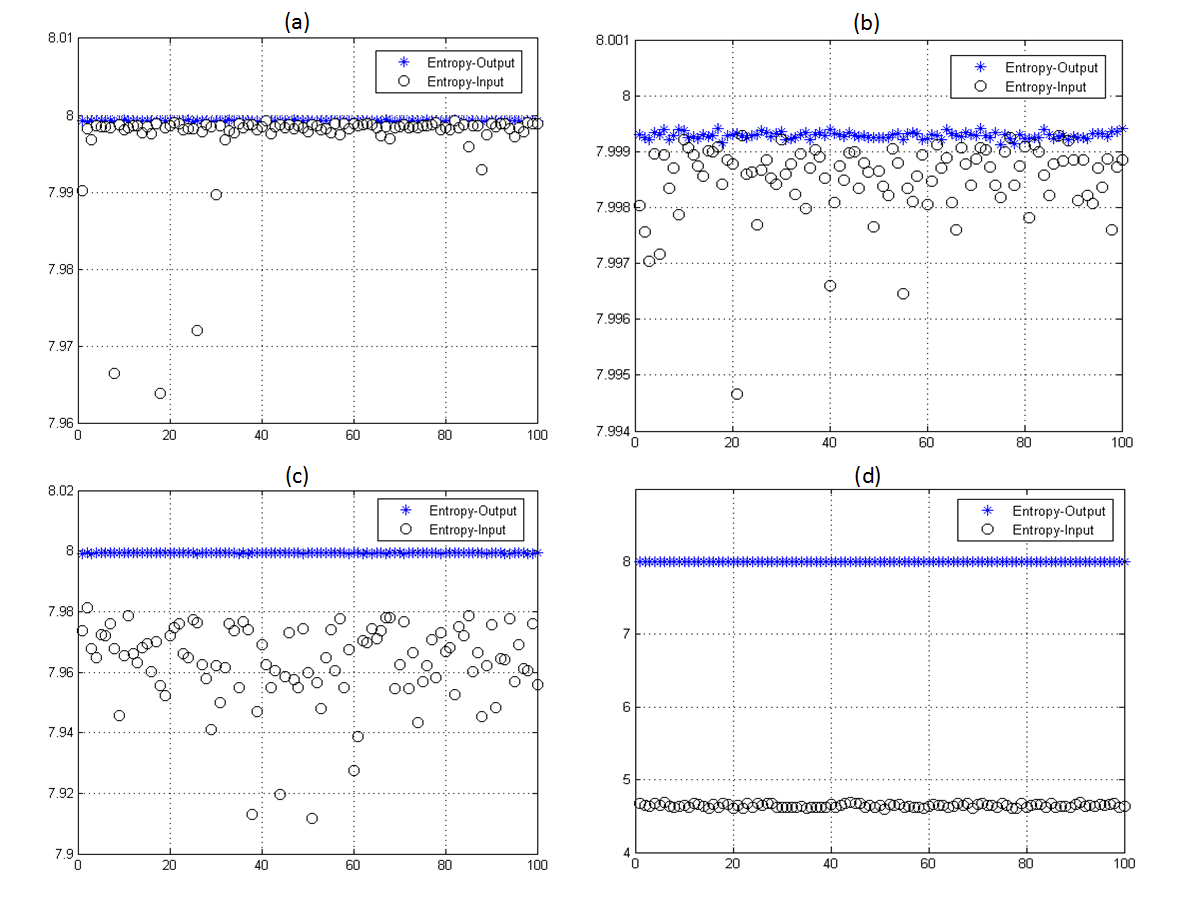}
\caption{Entropy test results distribution for 100 random chunks for videos and texts: (a) mkv, (b) mp4, (c) rmvb, and (d) text.}
\label{fig:entropy}
\end{figure}

In this subsection, the entropy tests are done on the public and protected fragments for not only bitmap files but also for English text files and three different video formats. For each file formats, 100 data chunks (each one is 1MB) are randomly chosen for the entropy test. As shown in \tablename \ref{entropy_fig} and \figurename \ref{fig:entropy}-(d), the public and protected fragment of text chunks always have high randomness with entropy values are always between 7.9992 to 7.9995.






\begin{table}[htbp!] 

\centering
\caption{Entropy test for 100 random chunks for videos and texts.}
\label{entropy_fig}
\begin{tabular}{c c c c c c}
\hline
\multicolumn{6}{c}{$\;\;\;\;\;\;\;\;\;\;\;\;\;\;\;$ Entropy test results for videos and texts.$\;\;\;\;\;\;\;\;\;\;\;\;\;\;\;$ }                 \\ \hline
\multicolumn{2}{c}{}   & Min  &  Mean &  Max & Std  \\ \hline

\multicolumn{2}{c}{$text(original)$} &4.5961&4.6423&4.6938& 0.0239\\
\multicolumn{2}{c}{$text(protected)$} &7.9992&7.9993&7.9995& 0.0000\\
\hline

\multicolumn{2}{c}{$mkv(original)$} & 7.96399 & 7.99726 & 7.99928 & 0.0055\\
\multicolumn{2}{c}{$mkv(protected)$} &7.99916&7.99930&7.99945& 0.00006\\
\hline

\multicolumn{2}{c}{$rmvb(original)$} &7.91174&7.96303&7.98120& 0.13383\\
\multicolumn{2}{c}{$rmvb(protected)$} &7.99914&7.99930&7.99944& 0.00006\\
\hline

\multicolumn{2}{c}{$mp4(original)$} &7.99467&7.99851&7.99930& 0.0006\\
\multicolumn{2}{c}{$mp4(protected)$} &7.99912&7.99930&7.99941& 0.00005\\

\hline

\end{tabular}
\end{table}


As long as video files are already compressed, encoded and formatted, the entropy of the original video data chunks are already close to 8. However, in \tablename \ref{entropy_fig} and \figurename \ref{fig:entropy}-(a),(b),(c), there are still improvements of the randomness for the three video formats. Therefore, the proposed scheme can achieve a very low entropy effect that can resist an attack based on entropy analysis for the file formats tested.


\subsection{Test Correlation between Original and protected and public fragments}

Lower correlation between original data and public and protected fragment is an important factor that allows validating the independence between them. Having a correlation coefficient close to zero means that the high degree of randomness is obtained. The correlation coefficient $r_{xy}$ is calculated using the following equations (5.8):

\begin{equation}
r_{xy}={}{}\frac{cov (x,y)} {\sqrt{D(x)\times{D(y)}}}
\end{equation}
where 
\begin{eqnarray}
&& \; E(x)=\frac{1}{N}\times \sum_{i=1}^N x_i \nonumber\\
&&{} \;D(x)=\frac{1}{N}\times \sum_{i=1}^N (x_i-E(x))^2\nonumber \\
&& \; cov (x,y)=\frac{1}{N}\times \sum_{i=1}^N (x_i-E(x))(y_i-E(y))  \nonumber 
\end{eqnarray}

In this test, we use image, text and video files as input for analyzing correlation. Indeed, the variation of coefficient correlation between original data and the public and protected fragment is obtained by applying the upper equations and the result is shown in \tablename~\ref{statistical1_lenna}, \tablename~\ref{statistical1_text}, and \tablename~\ref{statistical1_video} (see value distribution of $\rho^{2}$ for image case and $\rho$ for text and video case). The obtained result indicates that the coefficient correlation varies in a small interval very close to 0. This means that low correlation coefficient is attained by employing the proposed scheme and consequently the independence between the original data and the fragment is attained. 

\begin{figure}[htbp!] 
\centering
\includegraphics[width=1\textwidth]{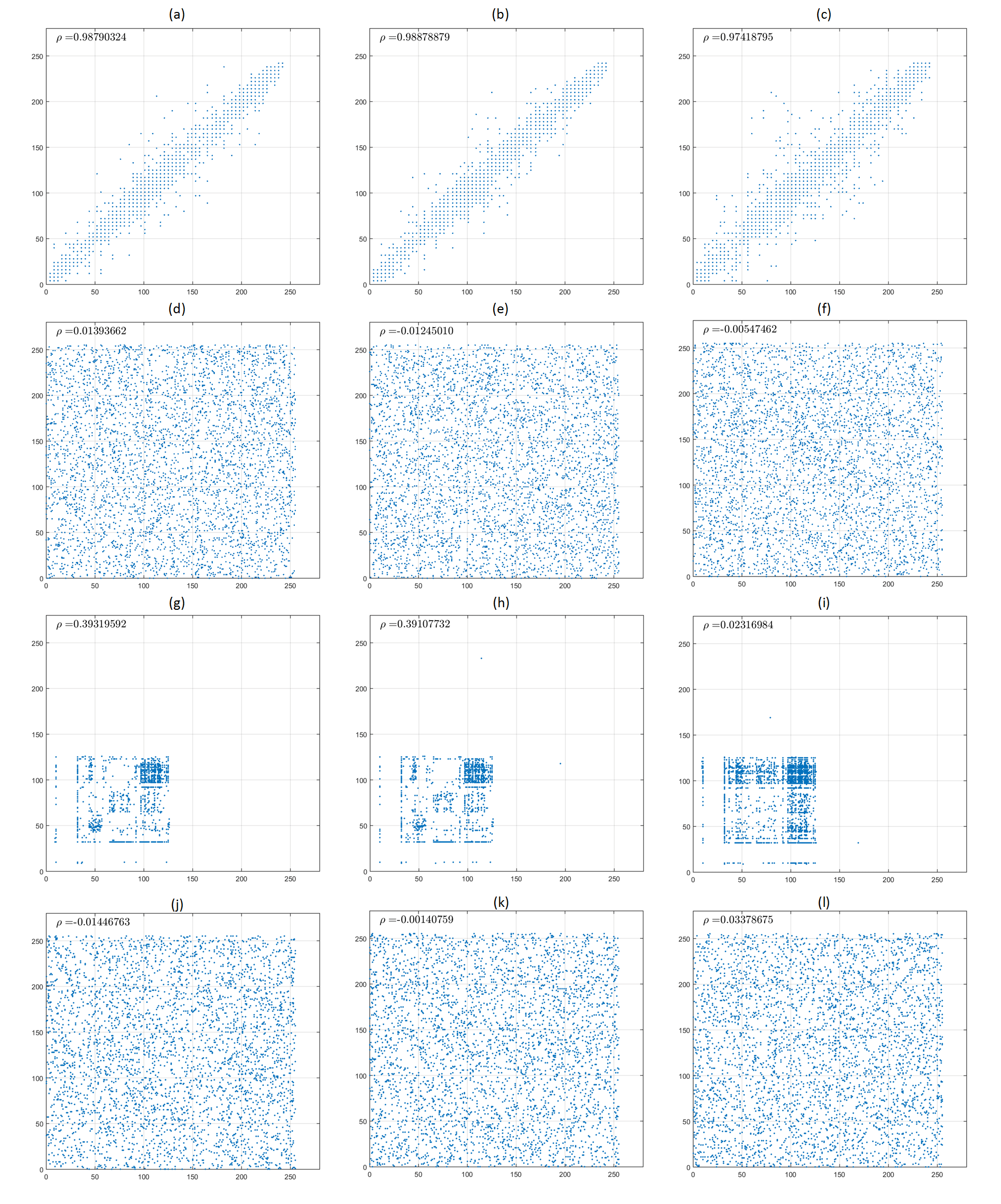}
\caption{Correlation distribution in adjacent pixels in original Lenna: (a) horizontally, (b) vertically, (c) diagonally.\\Correlation in adjacent pixels in the public and protected fragment:(d) horizontally, (e) vertically, (f) diagonally.\\Correlation distribution in adjacent pixels in text:(g) horizontally, (h) vertically, (i) diagonally.\\ Correlation in adjacent pixels in the public and protected fragment: (j) horizontally, (k) vertically, (l) diagonally.}
\label{Corr_all}
\end{figure}

Additionally, to validate that the spacial redundancy is removed ~\cite{norouzi2014novel,rhouma2008cryptanalysis}, for the image case, the correlation between pixels of original image and the public and protected fragment are performed. This test selects randomly $N=4096$ pairs of two adjacent pixels in horizontal, vertical, and diagonal direction. The obtained results are presented in \figurename ~\ref{Corr_all}, for the original (a)-(c) and the fragment (d)-(f) in horizontal, vertical and diagonal direction, respectively (same for text file from (g) to (l)). The result in this figure indicates clearly the high correlation between adjacent pixels in original image (correlation coefficient close to 1). While, for the public and protected fragment, the correlation coefficients become very low (close to 0) which clearly shows that the proposed scheme reduces severely the spatial redundancy. 

Moreover, the variation of the correlation coefficient  between adjacent pixels of public and protected fragment of Lenna image versus 1000 random keys are shown in \tablename~\ref{statistical1_lenna} ( $\rho-h$, $\rho-d$, $\rho-v$ respectively). The results are close to 0, which confirms that spatial redundancy is almost eliminated and very little detectable relation can be found in the public and protected fragment for both image and text case. Similar results are obtained using text file as input (see \figurename ~\ref{Corr_all} (g)-(l)). For the video cases, the results in \tablename~\ref{statistical1_video} are the average values for 100 randomly picked chunks.

\subsection{Difference Between input Data and the public and protected fragment}

\begin{figure}[!htbp]
\centering
\includegraphics[width=0.46\textwidth]{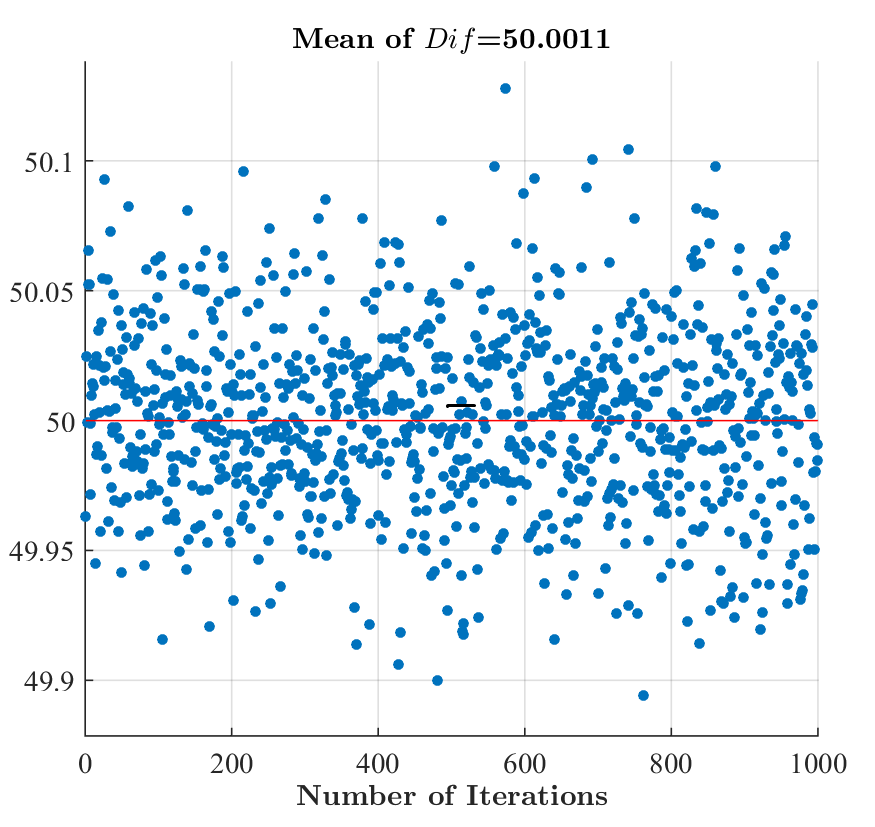}
\includegraphics[width=0.50\textwidth]{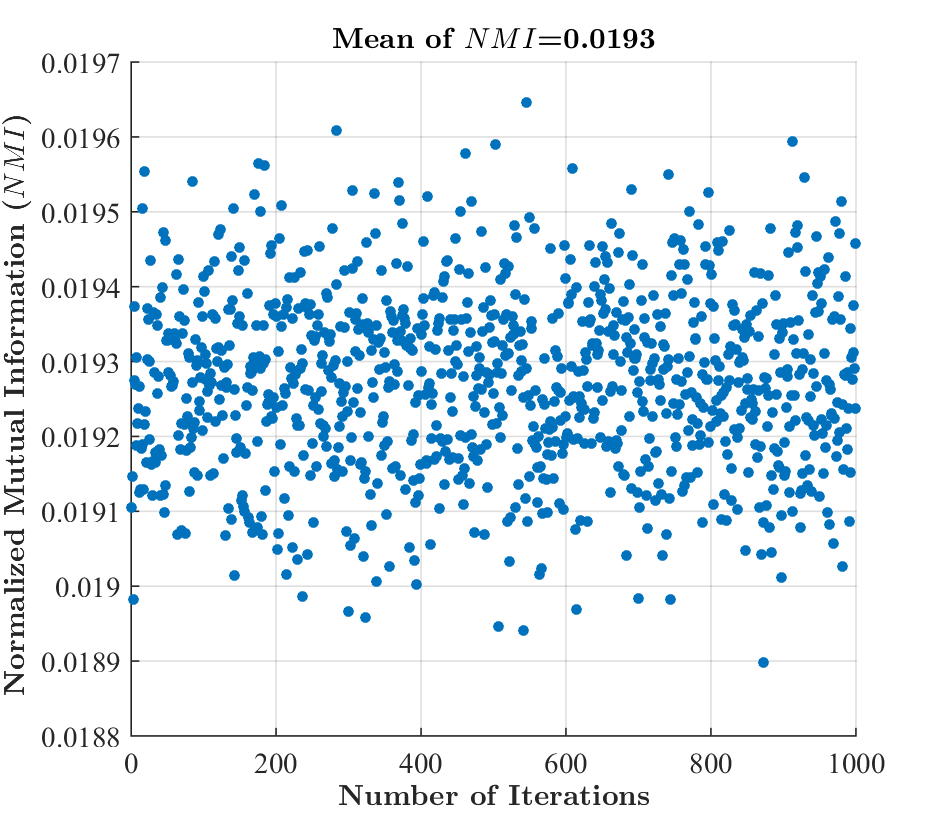}

\caption{Difference (a) and NMI (b) between original Lenna and the public and protected fragment versus 1000 random different keys.}
\label{fig:diff}
\end{figure}

The public and protected fragment must be statistically different from the original data (50\%) in bit level. The proposed scheme has achieved a high value of difference before and after process for all data formats tested. For example, the plain image Lenna was tested and the obtained result in \figurename ~\ref{fig:diff}-(a) shows that 50\% of bits is being changed between the public and protected fragment and the plain image. Additionally, similar result is obtained for text and video files, and statistical value is shown in \tablename~\ref{statistical1_text} and \tablename~\ref{statistical1_video}(see value distribution of $Dif$ for all cases).

 
To confirm this result, we also applied the Normalized Mutual Information (NMI)~\cite{veyrat2009mutual} between the original data blocks and public and protected fragments and the obtained results (for 1000 random secret keys: image case in \figurename ~\ref{fig:diff}-(b), \tablename~\ref{statistical1_lenna} and text case in \tablename~\ref{statistical1_text}; for 100 randomly picked data chunks: video case in \tablename~\ref{statistical1_video}) shows that NMI value is always close to 0. Consequently, this indicates that no detectable information can be extracted from the public and protected fragment.

\subsection{Sensitivity Test}

Differential attacks are based on studying the relation between two encrypted data resulting from a slight change like usually one different bit in the original plain-image or in the key. A successful sensitivity test shows how much a slight change will affect the cipher data. In other words, the higher the ciphered data changes when slight change happens in input, the better sensitivity of the encryption algorithm is. Here we analyze different types of sensitivity.

For the \textbf{Plain-text Sensitivity}, it is designed that in current version, the very similar blocks will have the very different public and protected fragments due to the randomness introduced by SHA algorithms. In fact, as long as most file transmitted on Internet are compressed and formatted, many same blocks within one chunk are rare to see. For this specific case, a counter (nouce) could be added as the input of the SHA algorithm in \figurename \ref{SE1block} to generate different output for the same input blocks.


Concerning the {\textbf{Key Sensitivity}} tests, it is one of the most important tests and permits to quantify its sensitivity against any slight change in the secret key.
In fact, for the private fragment, the encryption algorithm used (AES-128) could meet such sensitivity requirement. For the public and protected fragments, to study the key sensitivity, two secret keys are used : $SK_1$ and $SK_2$ that differ in only one random bit. The two plain images are processed separately and the Hamming distance of the corresponding public and protected fragments $C_1$ and $C_2$ is computed and also for the chosen text file (same methods used in ~\cite{fawaz2016efficient}), and illustrated as ~\tablename ~\ref{statistical1_lenna} (see $KS$ for both cases) versus 1000 tests. 


\begin{figure}[!htbp]
\centering
\includegraphics[width=0.45\textwidth]{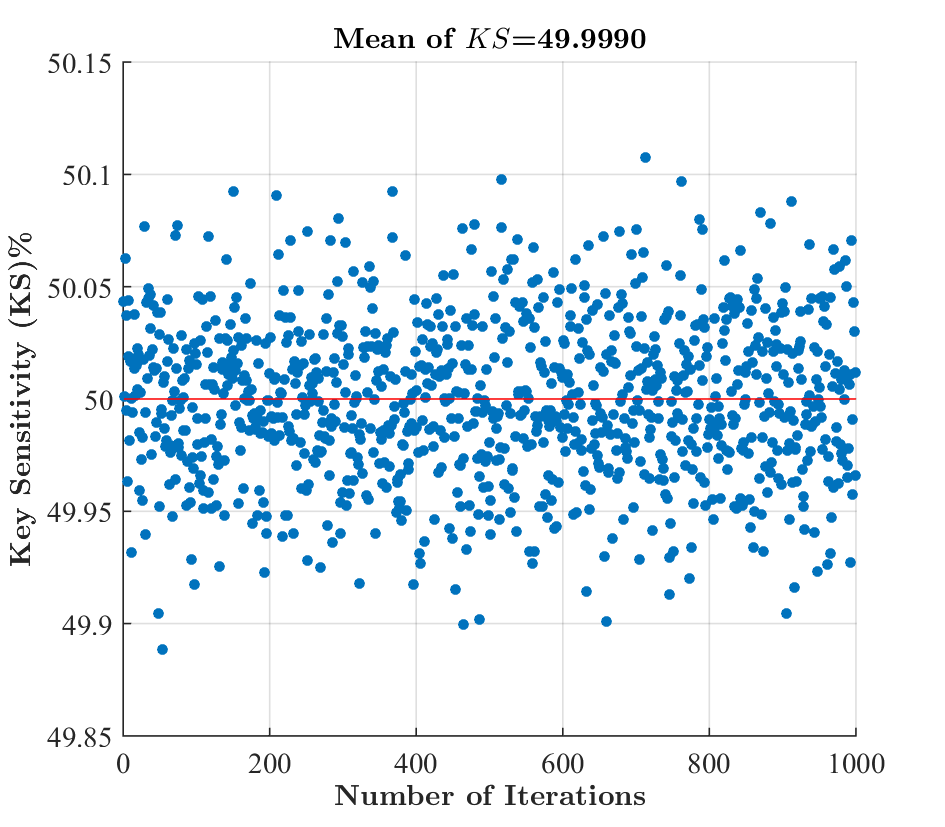}
\includegraphics[width=0.45\textwidth]{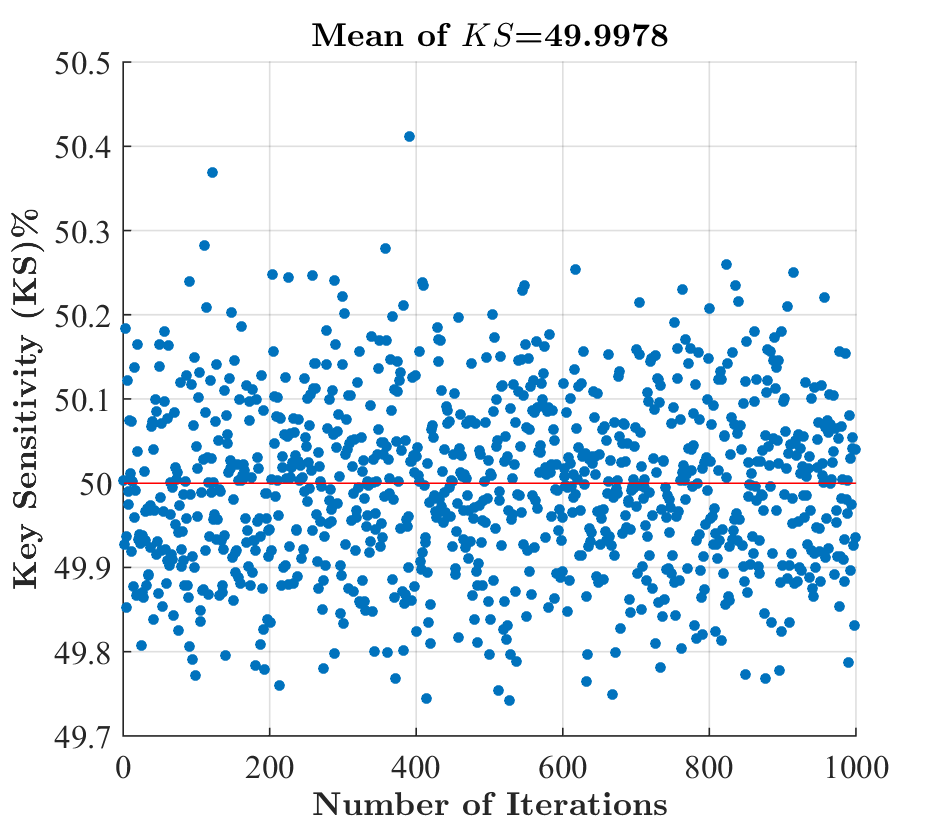}

\caption{Sensitivity tests for Lenna and text versus 1000 random different keys.}
\label{fig:ks}
\end{figure}

It is seen that the obtained values are always close to the optimal value (about $50\%$ bits changes when 1 bit change in the key) for both input data as shown in \figurename \ref{fig:ks}. This indicates that the proposed method ensures high sensitivity against any tiny change in the secret key. Similar results are obtained for the three video files used.

\subsection{Visual Degradation for images}

This test is specific for image that permits to quantify the visual degradation that is reached by employing a protection scheme. In fact, the degradation operated on the original image must be done in way that the visual content presented in the protected image must not be recognized. Two well known parameters are studied to measure the encryption visual quality which are Peak Signal-to-Noise Ratio (PSNR) \cite{huynh2008scope} and Structural Similarity (SSIM) \cite{wang2004image}.

\begin{figure}[!htbp]
\centering
\includegraphics[width=0.45\textwidth]{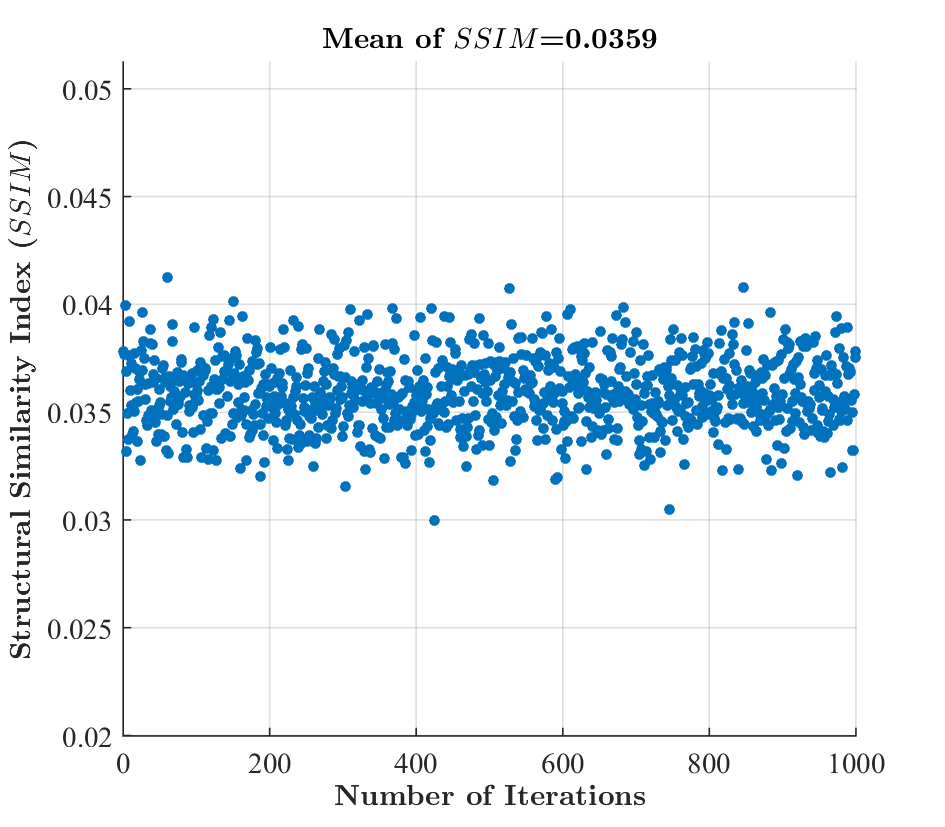}
\includegraphics[width=0.45\textwidth]{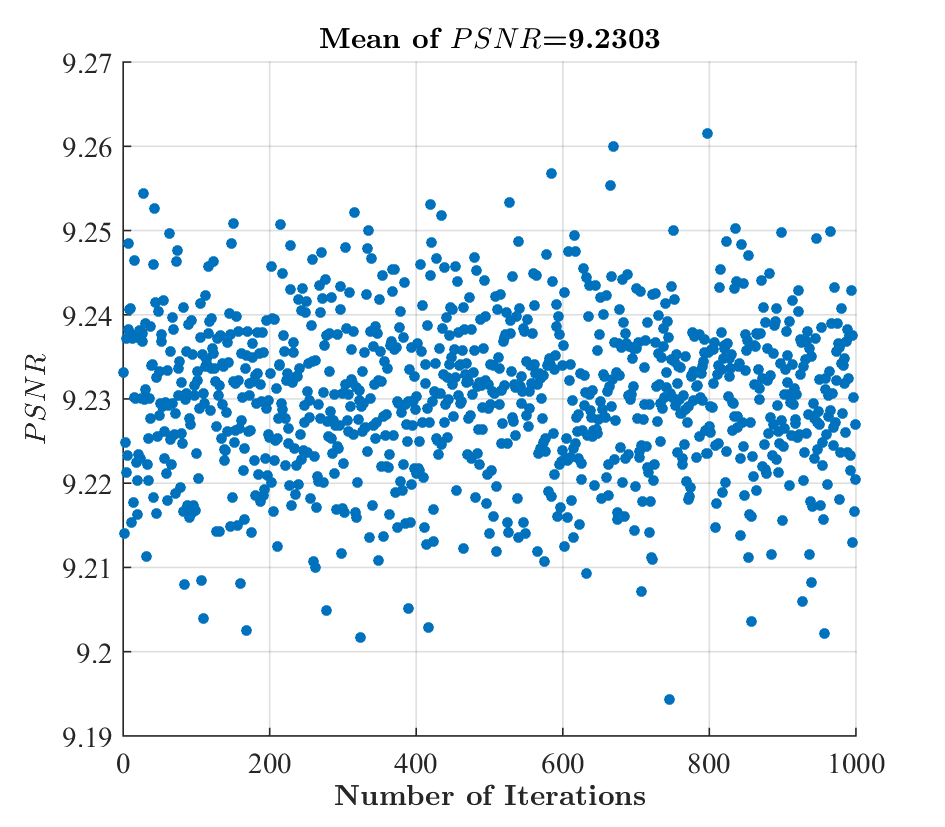}

\caption{$PSNR$ and $SSIM$ variation between original Lenna image and the corresponding public and protected fragment versus $1000$ different keys.}
\label{fig:PSNR and MSSIM}
\end{figure}

PSNR is derived from the Mean Squared Error (MSE), while MSE represents the cumulative squared error between two images. A low PSNR value \cite{huynh2008scope} indicates that there is a high difference between the original image and the public and protected fragment. 

Concerning SSIM \cite{li2002cryptanalysis}, it is defined after the Human Visual System (HVS) has evolved so that we can extract the structural information from the scene. SSIM is in the interval [0,1] and a value of 0 means that there is no correlation between the original image and the public and protected fragment, while a value close to 1 means that the two images are approximately the same. PSNR and SSIM are measured between the original Lenna image and its public and protected fragment for $1000$ different keys and corresponding value distribution presented in ~\figurename~\ref{fig:PSNR and MSSIM}, respectively. The mean PSNR value is $9.23$ dB which validates that the proposed scheme provides a high difference in visual between the original image and its public and protected fragment. Also, the SSIM value did not exceed $0.036$, which means that a high and hard visual distortion is obtained.

As a conclusion, the proposed scheme ensures a hard visual degradation. This means that no useful visual information or structure about the original image could be revealed from the protected and public fragments.

\subsection{Propagation of errors}

Indeed, an important criteria that should be ensured for any protection scheme is the error propagation while data is transmitted.
The interference and noise in the transmission channel might cause errors. Bit error means that a substitution of '0' bit into '1' bit or vice versa. This error may propagate and lead to the destruction of decrypting data, which is a big challenge since a trade-off between avalanche effect and error propagation are shown in~\cite{massoudi2008overview}.
In this proposal, if a bit error takes place in sub-matrix of the public and protected fragment, the error will propagate randomly only in its corresponding sub matrix and will not affect its consecutive corresponding neighbour sub-matrix. Moreover, as we discussed before, in~\figurename~\ref{SE1block}, the $1^{st}$ public and protected fragment can also be stored locally so when the communication channel is unreliable and transmission error occurs, the $2^{nd}$ public and protected fragment are the only one that is affected. In this case, the defragmenting process is the XORing between the correct SHA-512 result of $1^{st}$ public and protected fragment and $2^{nd}$ public and protected fragment with errors. Thus, the decrypting $2^{nd}$ public and protected fragment will have 1 bit error if there is 1 bit error in the transmitted $2^{nd}$ public and protected fragment. As a result, we can conclude that in such communication channel, this design of dispersion is efficient to prevent the error propagation.

\subsection{Cryptanalysis Discussion: Resistance against well-known types of attacks}

In this subsection, typical published cryptanalytic cases are considered and a brief analysis of the proposed scheme against several cryptanalytic attacks is provided from a cryptanalysis viewpoint. The proposed method is considered to be public and the attacker has complete knowledge to all steps but no knowledge about the secret key.

The strength of the proposed scheme against attacks is based on the existing cipher systems we deployed. 

For the private fragment, AES has key space that can be $2^{128}$, $2^{192}$ or $2^{256}$, which is sufficiently large to make the brute-force attack almost infeasible. Furthermore, differential and linear attacks would become ineffective. For the public and protected fragments, SHA-256 and SHA-512 guarantee the randomness. In fact, any change in any bit of the secret key causes a significant difference in the produced public and protected fragment as seen in ~\tablename ~\ref{statistical1_lenna}. Hence, a key is used for every block (shown in \figurename~\ref{SE1block}) and as the difficulty of cipher-text-only attack is equal to one of the brute force attacks, it becomes impossible for a cipher-text-only attack to retrieve useful information from the public and protected fragment in our scheme.

For further cases like single plain-text failure and accidental key disclosure, Initialization Vector (IV) could be introduced to generate dynamic keys for each of the chunk. In such case, it is very difficult for an attacker to recover the dynamic secret key that is changed for every input chunk.

\begin{figure}[htbp!] 
\centering
\includegraphics[width=1\textwidth]{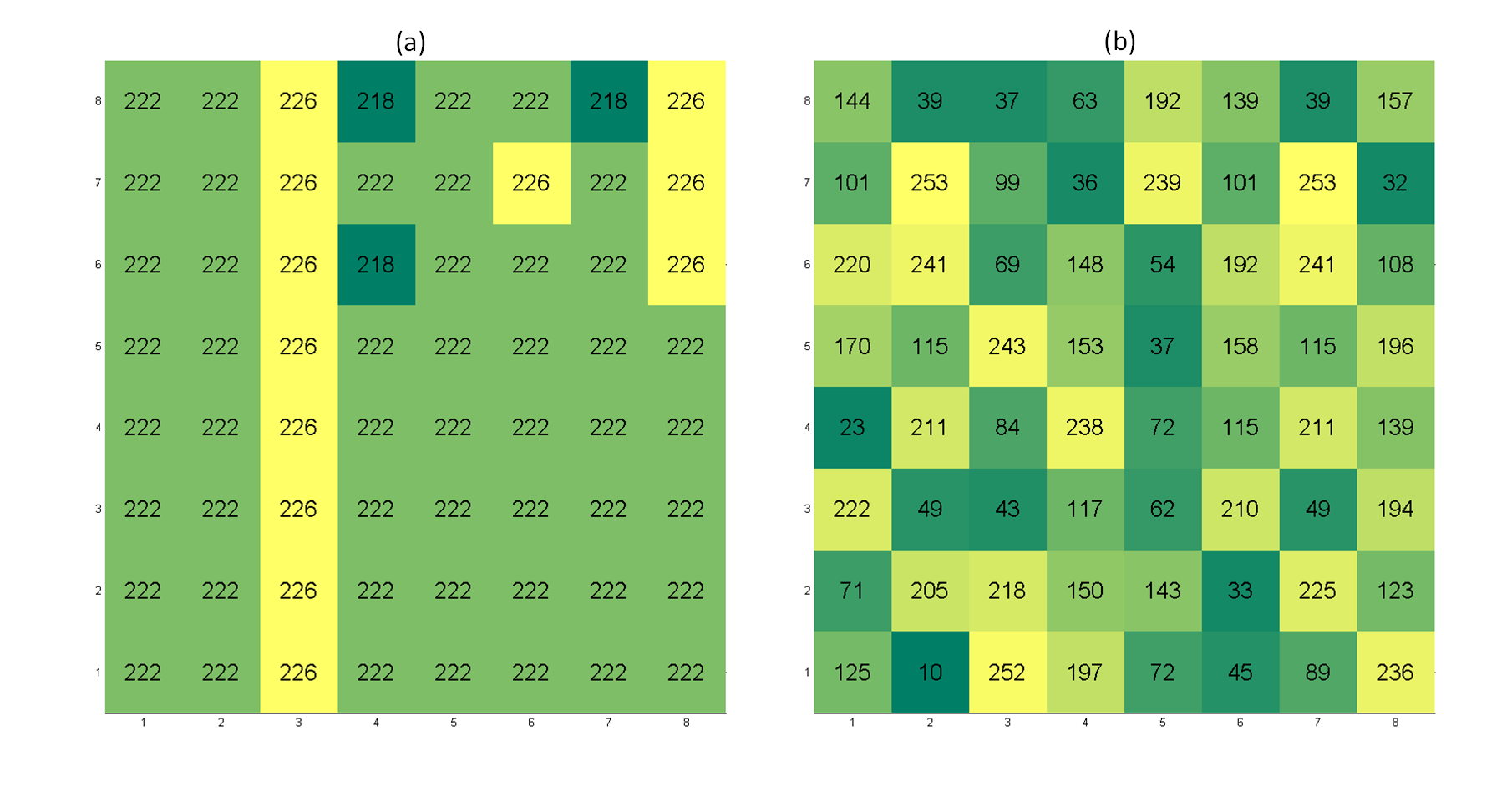}
\caption{(a) $8 \times 8$ cropped plain matrix with its corresponding gray scale matrix, (b) public and protected fragment of this matrix using the proposed scheme with its gray scale value.}
\label{fig:crop}
\end{figure}

With regard to resisting the statistical attacks, the proposed approach achieves that the plain-text are changed in positions and values, which means that the confusion and diffusion properties are ensured in addition. An example is illustrated in \figurename~\ref{fig:crop}, where a $8\times 8$ matrix of the original Lenna image and its public and protected fragment are illustrated by values. This result demonstrates that all values are changed. Therefore, the randomness property is ensured and this consequently permits to prevent the reverse-attack algorithm.

More importantly, the spatial redundancy between adjacent elements of input plain data are removed and a high randomness degree of the whole fragment are proved. Different statistical tests such as the entropy analysis, probability density function, correlation tests are applied to validate the independence and uniformity property. Consequently, these results indicate that no useful information can be detected from the public and protected fragment. This validates the robustness of the proposed scheme and their high resistance to statistical attacks.

Moreover, key sensitivity analysis demonstrates the efficiency of the proposed scheme against related key attacks, while any change in any one bit of key provide a different (50\%) public and protected fragment.



\section{Benchmark with two computer architectures}

In this section, we evaluate the performance of the whole protection process. As we are considering the allocation of the calculations on a PC platform, the hardware resource we have are a CPU and a GPGPU. However, the very different calculation capacities of GPGPU change the whole execution time of SE~\cite{mittal2015survey}~\cite{owens2008gpu}. So, the performance is evaluated in two typical use cases that are a laptop equipped with a low-end GPU and a desktop equipped with a high-end GPU. 

\begin{table}[htbp!] 
\centering
\caption{Performance evaluation for every calculation tasks of SE for two platforms.}

\label{perf}
\begin{tabular}{c c c c c c}
\hline
 & Input chunk size (byte) & $1024\times1024$ & $2048\times2048$ & $3200\times3200$ & $4800\times4800$ \\ \hline
\multirow{4}{*}{\begin{tabular}[c]{@{}l@{}}Laptop\\ Scenario\end{tabular}}  & GPU time (DWT-2D)  & 1.39ms           & 4.87ms           & 12.6ms           & 24.1ms           \\ \cline{2-6} 
                                                                            & GPU time (SHA-256) & 0.33ms           & 1.31ms           & 3.3ms            & 7.3ms            \\ \cline{2-6} 
                                                                            & GPU time (SHA-512) & 1.45ms           & 5.8ms            & 14.2ms           & 31.8ms           \\ \cline{2-6} 
                                                                            & CPU time (AES-128) & 0.29ms           & 1.14ms           & 3.06ms           & 6.67ms           \\ \hline
\multirow{4}{*}{\begin{tabular}[c]{@{}l@{}}desktop\\ Scenario\end{tabular}} & GPU time (DWT-2D)  & 0.34ms           & 0.79ms           & 1.7ms            & 3.3ms            \\ \cline{2-6} 
                                                                            & GPU time (SHA-256) & 0.05ms           & 0.13ms           & 0.29ms           & 0.63ms           \\ \cline{2-6} 
                                                                            & GPU time (SHA-512) & 0.13ms           & 0.69ms           & 1.58ms           & 3.2ms            \\ \cline{2-6} 
                                                                            & CPU time (AES-128) & 0.23ms           & 0.96ms           & 2.37ms           & 5.3ms            \\ \hline
\end{tabular}

\end{table}

\begin{figure}[htbp!] 
\centering
\includegraphics[width=1\textwidth]{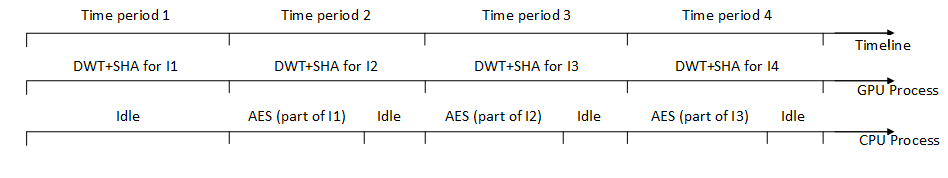}

\caption{Time overlapping architecture of the implementation.}
\label{fig:overlap}

\end{figure}

The key decision of the design is to distribute the calculation tasks between the GPU and the CPU. As pointed in Section 2.3.2, the DWT-2D, SHA-256 and SHA-512 can benefit from the GPU acceleration, so the design is based on the parallel execution of CPU with GPU while the GPU will take calculation tasks of DWT-2D, SHA-256 and SHA-512 in the process, CPU takes AES-128 for only private fragment. The initial plan for both low-end and high-end GPU cases is to keep the GPU busy and CPU would have time space for other tasks (~\figurename ~\ref{fig:overlap}).

For the laptop, there is an Intel I7-3630QM CPU and a Nvidia Nvs 5200M GPU. For the desktop, we have a CPU of Intel I7-4770K and a GPU of Nvidia Geforce gtx 780. In order to verify our initial plan and allocate the right calculation tasks to the right chip, we evaluate each of the tasks on laptop and desktop and results are shown in ~\tablename ~\ref{perf}. For different size of input data chunk, the execution time of GPU for the second input data chunk can always overlap the execution time of CPU for the selected DWT-2D coefficients of the first input data chunk. 

From the two use cases we evaluated, the overlay design in \figurename ~\ref{fig:overlap} works. And the speed of the whole SE process relies on how fast the GPU can process its calculation tasks on input data as long as there are many chunks as input. That is to say, in these two scenarios, the time consumed by GPU is evaluated as the benchmark for our SE method. The calculation speed of our scheme evaluated for this laptop scenario is about 360 MB/s and for this desktop scenario is about 2.8-3.2GB/s.

\section{Discussion for benchmark}

As shown in Table 3.1 in Chapter 3, the desktop GPU we used contains 2304 CUDA cores compared with the 96 CUDA cores on the laptop GPU. It is easy to conclude calculation speed of the desktop GPU is much faster than the laptop GPU which is a common scenario for different GPUs. As a consequence, Speed of the SE method is very different for the two PC scenarios as shown before. This very difference exists on the two typical hardware cases: low-end GPUs normally for laptops and high-end GPUs for desktops or gaming PCs.

\begin{figure}[htbp!]
\centering
\includegraphics[width=1\textwidth]{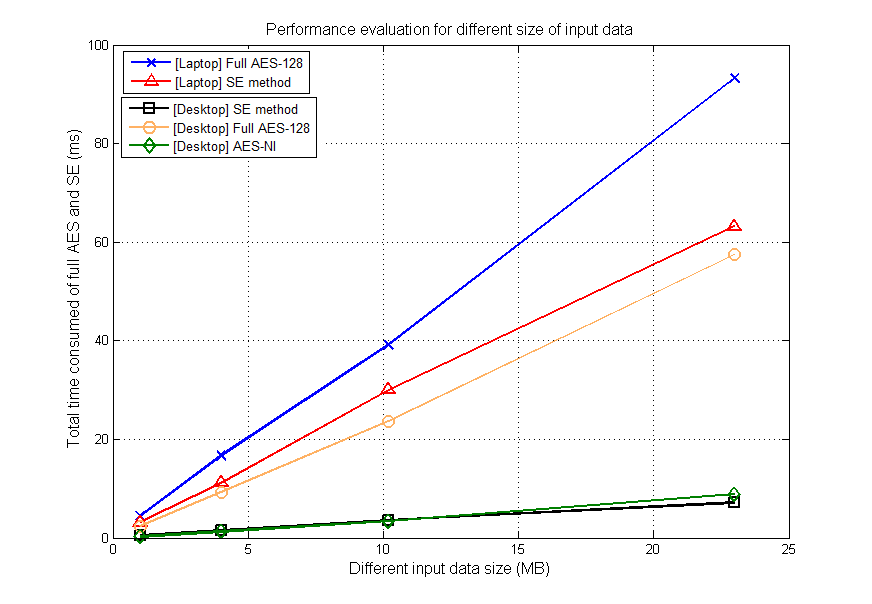}
\caption{Performance evaluations for SE compared with full AES on laptop and desktop scenarios.}
\label{fig:benchmarkall}
\end{figure}

In fact, as long as GPU architectures are rapidly evolving, the hardware configuration of a GPU may strongly influence software applications architectural choices necessary to derive the best possible implementation. This point could even invert the results of our evaluation since a large number of cores could very well favor the GPU calculation tasks that finally change the overlay design. Also as pointed by~\cite{gregg2011data}, when a GPU calculation capacity is very high, bottleneck of the process is the memory transfer between the GPU memory and the host memory instead of calculation speed itself. This is mainly due to the limitation of transmission speed through the PCIE bus (Peripheral Component Interconnect Express \cite{budruk2004pci}) between host memory and GPU memory. This problem can be avoided by arranging more calculation tasks on GPU instead of too much I/O usage or overlapping the transmission by execution time.

Here we presented the comparison of the SE method with the traditional CPU-only AES-128 speed (\cite{dai2007crypto++}) in ~\figurename ~\ref{fig:benchmarkall}. It is worth noticed that~\cite{bogdanov2015comb} pointed out that the CPU-only AES could also be very fast with the support of the New Instructions extension (NI) brought by Intel. This AES-NI could accelerate the AES on CPU for more than 5 times and achieve almost 3GB/s on a NI-enable CPU (shown in  ~\figurename ~\ref{fig:benchmarkall}) which is almost the same speed as our method based on GPGPU. However, as pointed out in the Nvidia white paper \cite{nvidiawhite1080}, the Nvidia GeForce 1080 GPU (released in 2016) is already three times faster than the GPU used in this thesis (Nvidia GeForce 780, released in 2013), it is only fair to say the performance gain could be larger with employing the start-of-the-art GPGPU. Such rapid evolving in hardware is not seen in recent years' CPU manufacturing. In conclusion, our design on GPU could always achieve better performance than AES-NI on CPU with fair hardware.

\section{Fragment transmission}

\begin{figure}[htbp!] 
\centering
\includegraphics[width=1\textwidth]{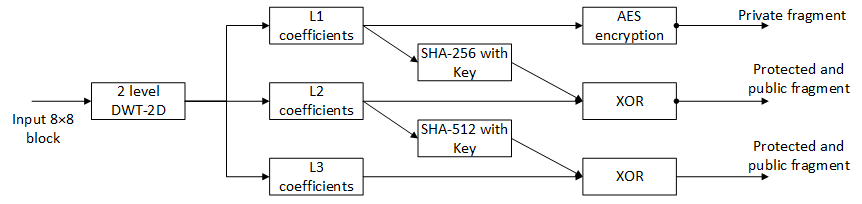}
\caption{Three fragments from one data chunk for further securing data sharing.}
\label{datasharing}
\end{figure}

This data protection method can be also used for secure data transmission and sharing between end-users. This design could generates three fragments for data sharing for each of the data chunk as shown in \figurename \ref{datasharing}. For the \textit{private fragment}, the encryption algorithm used in this particular design is AES-128 (could be easily replaced by AES-NI with a NI-enabled CPU or any other encryption algorithms). This \textit{private fragment} is the only fragment that is directly transmitted between end-users. The other two \textit{protected and public fragments} are transmitted through the public cloud servers without leaking information. The storage space that the \textit{private fragment} takes is $7.8\%$ of the original data size, while the \textit{$1^{st}$ and $2^{nd}$ protected and public fragments} take $24.2\%$ and $93.8\%$ of the original data storage space respectively.

Although the total data storage space usage is about 25\% larger than the original one, the local storage space usage is only 7.8\% which highly reduces the local storage usage and exploit the convenience brought by the Cloud servers with both efficiency and security.



\begin{figure}[htbp!] 
\centering
\includegraphics[width=1\textwidth]{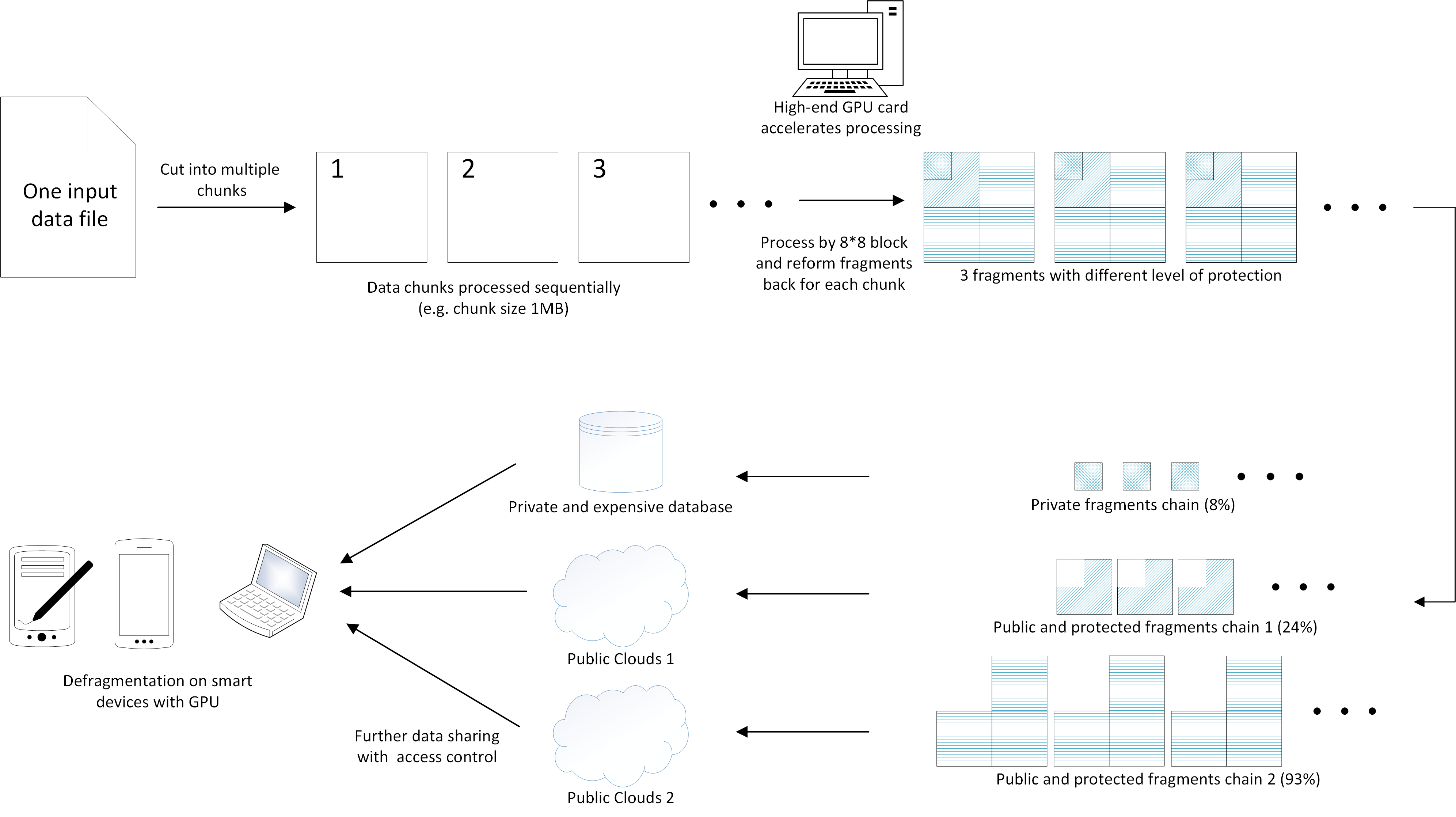}
\caption{Use case: secure data sharing between end-users through different Cloud servers based on fragmentation and dispersion.}
\label{transmission}
\end{figure}

\figurename \ref{transmission} gives an example for using this method for data sharing between end-users. The sender processes the fragmentation and protection before sending to the Clouds. The \textit{$1^{st}$ and $2^{nd}$ protected and public fragments} will be sent to different Cloud servers and be available for any receiver to download at anytime. The \textit{private fragment} will be sent through direct communication channel between senders and receivers. 

The first advantage of this design is to largely reduce the local storage space usage while providing security and privacy for the end-user. This is specially useful in some use cases like when the end-user device is a mobile phone (local storage space is limited and expensive compared with the free large Cloud storage space). The second one is to efficiently process the fragmentation and protection solely on end-user's device which avoid any plain data transmission on insecure channel and 
could avoid using specific service provided by server end like BlackBerry Enterprise Server (BES) \cite{chen2002enterprise}. More importantly, as pointed out by \cite{cheng2011using} and \cite{zhao2015fast}, DWT could be accelerated by mobile GPU which allows possible usage of this scheme in future when GPGPUs are largely deployed on smart phones.

\chapter{Conclusion and future work}

\ifpdf
    \graphicspath{{Chapter6/Figs/Raster/}{Chapter6/Figs/PDF/}{Chapter6/Figs/}}
\else
    \graphicspath{{Chapter6/Figs/Vector/}{Chapter6/Figs/}}
\fi

In this thesis, a data protection scheme combining fragmentation, encryption and dispersion is presented based on improvement of selective encryption algorithms. In the past two decades, most SE algorithms, were initially dedicated to SE specific multimedia. They are based upon one of the steps for formatting compressing the content (transformation, encoding, and packeting). 

From previous works, SE methods provides mainly more efficiency compared with traditional full encryption methods by protecting only a part of original data. However, this thesis points out that SE methods do not always provide efficiency considering the rapid evolution of both algorithms and hardware (pointed out in Chapter 3). Thus, an architecture according to existing different hardware configurations (the recent evolvement of hardware should be considered to fit in the frame and the scheme has the adaptivity for very different environments) is discussed including a flexible software architecture (the algorithms deployed could be easily replaced by the new ones while the main framework of the scheme remains the same).

In Chapter 4, two levels of Bitmap protection schemes are presented both using DCT $8\times8$ preprocessing and GPGPU acceleration. We defined a first level of lightweight protection with a very fast speed and a second level of strong protection with a good protection quality. For the second level of protection, many detailed designs are implemented for less information loss and avoiding recursive rounding error which improved the previous bitmap-related SE methods. Fragmentation is used in this scheme with an additional optimized memory allocation and transmission.


In Chapter 5, an agnostic SE architecture is presented based on lossless DWT $8\times8$ preprocessing. The initial motivation of this architecture is to solve the question of integrity for bitmap images. Then we realized that this method could deal with text format. And we verified that in fact it is agnostic with regard to the format of the information to be protected at the difference of any SE method that we have been able to see in the current literature.


The proposed architecture employs the AES encryption algorithm to protect the private fragment that will be stored locally in the use case. It can be easily replaced by using other encryption algorithms like AES-NI instead. For the other fragments, SHA algorithms are used to produce a key-stream that will be employed to encrypt the public and protected fragment by mixing them. The SHA will guarantee the randomness of the key-streams generated even from the very similar neighbour $8\times8$ blocks which can provide the needed protection for the public and protected fragments.


More importantly, GPGPU is employed to reduce the overhead of applying the optimization DWT-2D operation, and sometimes both the AES and SHA operations. The architecture is flexible for different hardware configurations with the pure GPGPU experimentation and CPU with GPGPU overlay design. In order to validate that the proposed design can ensure the required goals, a benchmark was realized between the proposed experimentation and a full AES encryption for different kinds of data on two very different hardware platforms. And several experimental and theoretical security analysis were realized to prove the security, robustness and resistance against error propagation.

Therefore, the proposed solution can be considered as an agnostic selective encryption algorithm candidate that can be applied for most computer distributed systems in particular, we can use this method to store large amount of data in public clouds in a secure manner.

For future work, we propose considering to reconsider the design for redefining the important private fragment. According to recent work, the low frequency coefficients in transformations (like DWT $8\times 8$ used in this thesis) are not necessarily more "important" than the high frequency coefficients. In fact, for agnostic fashion data protection schemes, a practical way to measure the importance of transformation coefficients is to calculate the influence of input values to the coefficients. For instance, some low frequency coefficients are related to only a small subset of input values (the change of rest input values will not lead to the change of these low frequency coefficients) while some high frequency coefficients can be related to more input values. This fact would give us a hint about how to define more "important" fragments by choosing those coefficients that are related to all input values ignoring how low or high frequencies are.

Regarding the implementation aspect, the non-general-purpose mobile GPU platforms should be experimented to fit the architecture of the proposed agnostic SE method to achieve both performance efficiency and energy saving purposes on today's smart phones. And, of course, more kinds of personal computer GPGPU platforms (such as multiple CPUs or GPGPUs platforms or newest generation hardware) should be experimented with deriving a general solution for the allocation between CPUs and GPGPUs. As pointed out in former discussion, the evolution of hardware configuration never stops and sometimes the deployment of new hardware would totally change the software design.

At last, there are some future work regarding fragment dispersion and transmission with an environment based on user device and public and clouds. More standards like availability, data recovery should be taken into consideration for a more adaptable architecture.

\clearpage

\begin{spacing}{0.9}


\bibliographystyle{apalike}
\cleardoublepage
\bibliography{References/references} 



\end{spacing}





\printthesisindex 

\end{document}